\documentclass[prd,nofootinbib,twocolumn,superscriptaddress,preprintnumbers,
balancelastpage,longbibliography]{revtex4-1}

\usepackage{graphicx}
\usepackage{color}
\usepackage{hepunits}
\usepackage{amsmath}
\usepackage{booktabs}
\usepackage{array}
\usepackage{makecell}
\usepackage[framemethod=TikZ]{mdframed}
\usepackage[colorlinks=true
,urlcolor=blue
,anchorcolor=blue
,citecolor=blue
,filecolor=blue
,linkcolor=red
,menucolor=blue
,linktocpage=true
,pdfproducer=medialab
,pdfa=true
]{hyperref}
\usepackage[utf8]{inputenc}
\usepackage[english]{babel}

\newcolumntype{C}[1]{>{\centering\let\newline\\\arraybackslash\hspace{0pt}}m{#1}}

\newcommand{\overbar}[1]{\mkern 2mu\overline{\mkern-2mu#1\mkern-2mu}\mkern 2mu}

\begin{document}
\title{Revealing the Dark Matter Halo with Axion Direct Detection}
\author{Joshua W. Foster}
\affiliation{Leinweber Center for Theoretical Physics, Department of Physics, University of Michigan, Ann Arbor, MI 48109}
\author{Nicholas L. Rodd}
\affiliation{Center for Theoretical Physics, Massachusetts Institute of Technology, Cambridge, MA 02139}
\author{Benjamin R. Safdi}
\affiliation{Leinweber Center for Theoretical Physics, Department of Physics, University of Michigan, Ann Arbor, MI 48109}

\preprint{LCTP-17-08, MIT-CTP 4964}

\begin{abstract}
The next generation of axion direct detection experiments may rule out or confirm axions as the dominant source of dark matter.  
We develop a general likelihood-based framework for studying the time-series data at such experiments, with a focus on the role of dark-matter astrophysics, to search for signatures of the QCD axion or axion like particles.
We illustrate how in the event of a detection the likelihood framework may be used to extract measures of the local dark matter phase-space distribution, accounting for effects such as annual modulation and gravitational focusing, which is the perturbation to the dark matter phase-space distribution by the gravitational field of the Sun.
Moreover, we show how potential dark matter substructure, such as cold dark matter streams or a thick dark disk, could impact the signal.
For example, we find that when the bulk dark matter halo is detected at 5$\sigma$ global significance, the unique time-dependent features imprinted by the dark matter component of the Sagittarius stream, even if only a few percent of the local dark matter density, may be detectable at $\sim$2$\sigma$ significance.
A co-rotating dark disk, with lag speed $\sim$50 km$/$s, that is $\sim$20\% of the local DM density could dominate the signal, while colder but as-of-yet unknown substructure may be even more important.
Our likelihood formalism, and the results derived with it, are generally applicable to any time-series based approach to axion direct detection.
\end{abstract}

\maketitle

\section{Introduction}\label{sec:Introduction}

The local distribution of dark matter (DM) leaves a unique fingerprint on an emerging signal at axion direct detection experiments.
While it has long been recognized that the local phase-space distribution of DM may be partially uncovered with direct-detection experiments searching for heavy DM candidates with masses $m_{\rm DM} \gtrsim \text{MeV}$ (for a recent review, see~\cite{Green:2017odb}), the role of the DM distribution at axion direct detection experiments, where $m_{\rm DM} \lesssim \, \, \text{meV}$, remains less explored. 
In this work, we develop a likelihood-function-based analysis framework for analyzing the output of axion DM direct detection experiments.
Using this framework, we explore in detail the impact of the DM phase-space distribution on the experimental sensitivity to the axion; in the presence of a signal, we show that many aspects of the full time-dependent phase-space distribution can be uncovered.

The need for understanding how the DM phase-space distribution is manifest in axion direct detection experiments has taken on a new sense of urgency recently due to a multitude of new experimental efforts.
In addition to the long-running ADMX experiment~\cite{Asztalos:2001tf,Shokair:2014rna,PhysRevLett.104.041301}, there has been a raft of new ideas for directly detecting axion DM, including ABRACADABRA~\cite{Kahn:2016aff}, CASPEr~\cite{Budker:2013hfa}, CULTASK~\cite{Chung:2016ysi}, DM Radio~\cite{Chaudhuri:2014dla,Silva-Feaver:2016qhh}, MADMAX~\cite{TheMADMAXWorkingGroup:2016hpc,Majorovits:2016yvk,Millar:2016cjp,Ioannisian:2017srr}, HAYSTAC~\cite{Brubaker:2016ktl,Kenany:2016tta,Brubaker:2017rna}, nEDM~\cite{Stadnik:2013raa,Abel:2017rtm}, ORGAN~\cite{McAllister:2017lkb}, QUAX~\cite{Ruoso:2015ytk,Barbieri:2016vwg,Crescini:2016lwj}, TASTE~\cite{Anastassopoulos:2017kag}, and more~\cite{Baker:2011na,Graham:2011qk,Graham:2013gfa,Horns:2012jf,Sikivie:2013laa,Horns:2013ira,Beck:2013jha,Beck:2014aqa,Hong:2014vua,Roberts:2014dda,Roberts:2014cga,Stadnik:2014tta,Hill:2015kva,Hill:2015vma,McAllister:2015zcz,Yokoi:2016umv,Alexander:2017zoh,Cao:2017ocv,McAllister:2017ern,Graham:2017ivz,Safronova:2017xyt,Beck:2017sle,Arvanitaki:2017nhi}.
Our statistical framework allows us to better quantify limits and detection thresholds for the proposed experiments.
Moreover, it also shows how various features of the DM distribution, for example annual modulation, gravitational focusing, and potential substructure such as local DM streams, can affect the sensitivity of these experiments and how they can be searched for in the data.

The resurgence of effort towards detecting axion DM is driven by a combination of factors, including the increasing tension that heavier DM candidates are facing from null searches, technological advancements that make axion searches more feasible, and new ideas for how to detect axion DM in the laboratory.
However, axion DM is also a focus point due to its strong theoretical motivation. 
The quantum chromodynamics (QCD) axion was originally invoked to  solve the strong CP problem of the neutron electric dipole moment~\cite{Peccei:1977hh,Peccei:1977ur,Weinberg:1977ma,Wilczek:1977pj}.
It was later realized that the QCD axion behaves like cold DM for cosmological and astrophysical purposes~\cite{Preskill:1982cy,Abbott:1982af,Dine:1982ah}.
The axion interacts with the electromagnetic sector through the following operator:
\begin{equation}
\mathcal{L}_a = - {1 \over 4} g_{a \gamma \gamma} a F_{\mu \nu} \tilde F^{\mu \nu} \,,
\label{eq:AxionEMCoupling}
\end{equation}
where $F_{\mu \nu}$ is the electromagnetic field strength, $a$ is the axion field, and $g_{a \gamma \gamma}$ is the coupling.\footnote{Throughout this work we will consider exclusively the electromagnetic coupling, but the framework we introduce can be straightforwardly extended to nucleon couplings.}
We may parametrize the coupling as $g_{a \gamma \gamma} = g \alpha_{\rm EM} / (2 \pi f_a)$, where $f_a$ is the axion decay constant, $\alpha_{\rm EM}$ is the electromagnetic fine structure constant, and $g$ is a model dependent parameter, which takes a value $-1.95$ ($0.72$) for the KSVZ~\cite{Kim:1979if,Shifman:1979if} (DFSZ~\cite{Dine:1981rt,Zhitnitsky:1980tq}) QCD axion, although the space of models covers an even broader range (see, {\it e.g.},~\cite{Kim:1998va}).
The axion decay constant determines the axion mass through the coupling of the axion to QCD:
\begin{equation}
m_a \approx {f_\pi m_\pi \over f_a} \,,
\end{equation}
which is given in terms of the pion mass and decay constant, $m_\pi$ and $f_\pi$, respectively.
Depending on the detailed cosmological scenario, the QCD axion may make up all of the DM for axion masses roughly in the range  $\sim 10^{-12}$ eV to $\sim 10^{-5}$ eV (see~\cite{Marsh:2015xka} for a review).
Lower masses are disfavored by requiring the axion decay constant, which is the scale of new physics that generates the axion, to be sub-Planckian.
At higher masses it becomes more difficult to generate the required abundance of DM through the misalignment mechanism and the decay of topological defects (see, {\it e.g.},~\cite{Patrignani:2016xqp}).
In addition to the QCD axion, it is also possible to have more general axion-like DM particles that still couple to electromagnetism, but not to QCD.
The mass of these axion-like particles is a free parameter, since there is no contribution from QCD; however, axion-like particles do not address the strong CP problem.

Most axion direct detection experiments exploit the fact that axion DM may be described by a coherently-oscillating classical field $a$ that acts as a source of $F_{\mu \nu} \tilde F^{\mu \nu}$.
The oscillation frequency of $a$ is set by its mass $m_a$, while the coherence of the oscillations is set by the local DM velocity distribution.
Locally, we expect the velocity dispersion of the bulk DM halo to be $\sim$$10^{-3}$ in natural units, which leads to the expectation that the axion coherence time is $\tau \sim 10^6 \times (2 \pi / m_a)$.
Consequently, the axion sources a coherent signal that experiments can repeatedly sample by taking time-series data sensitive to the possible interactions of the axion.
For example, in ADMX, which is the only experiment so far to constrain part of the QCD axion parameter space,\footnote{This, of course, depends on the exact definition of what constitutes a QCD axion.
Recent studies have suggested the window could be broader than what we discuss in this work, see, {\it e.g.},~\cite{DiLuzio:2016sbl,Agrawal:2017cmd}.
Under such extended definitions, results from the HAYSTAC experiment may already probe the QCD parameter space~\cite{Brubaker:2016ktl}.} the coherent axion background sources electromagnetic modes in a resonant cavity.
The experiment tunes the resonant frequency of the cavity to scan over different possible masses.  Most axion experiments make use of high-$Q$ oscillators or cavities to build up the otherwise small signal.
However, some experiments, such as ABRACADABRA and MADMAX, can operate in a broadband mode that allows multiple masses to be searched for simultaneously, albeit with slightly reduced sensitivity.

Resonant experiments, such as ADMX, typically analyze their data by comparing the power output from the resonator, measured across the frequency bandwidth of the signal as determined by the coherence time, to the expectation under the null hypothesis using, for example, the Dicke radiometer equation~\cite{Dicke:1946}, supplemented with Monte Carlo simulations as described in~\cite{Asztalos:2001tf,Peng:2000hd}.
In this work, we present a likelihood-function based approach to analyzing the data at resonant and broadband axion experiments that takes as input the Fourier components of the time-series data, with frequency spacing potentially much smaller than the bandwidth of the signal.
We show that the velocity distribution of the local halo is uniquely encoded in the spectral shape of the Fourier components, within the frequency range set by the coherence time, and that it may be extracted from the data in the event of a detection.

We present an analytic analysis of the likelihood function using the Asimov dataset~\cite{Cowan:2010js}, which also allows us to calculate the sensitivity of axion experiments to DM substructure such as cold DM streams and a co-rotating dark disk.
For example, we show that soon after the discovery of axion DM from the bulk DM halo, the DM component of the Sagittarius stream, which has been extensively discussed in the context of electroweak-scale direct detection~\cite{Freese:2003tt,Purcell:2012sh,Savage:2006qr,OHare:2014nxd}, should become visible in the data through the likelihood analysis.
Moreover, we may use the formalism to accurately predict exclusion and discovery regions analytically.

Most previous studies of axion direct detection have not addressed the question of how to extract measures of the local phase-space distribution from the data.
In~\cite{Millar:2017eoc}, it was demonstrated that effects of the non-zero axion velocity will need to be accounted for in future versions of the MADMAX experiment.
Ref.~\cite{Sloan:2016aub} recently performed simulations to show how the sensitivity of ADMX changes for different assumptions about the velocity distribution, such as the possibility of a co-rotating dark disk or cold flows from late infall, using the analysis method used by ADMX in previous searches (see, for example,~\cite{Asztalos:2001jk,Asztalos:2003px}).
In~\cite{Ling:2004aj} (see also~\cite{Vergados:2016rlh}) it was pointed out that the width of the resonance should modulate annually due to the motion of the Earth around the Sun, which slightly shifts the DM velocity distribution.
Recently,~\cite{OHare:2017yze} took an approach similar to that presented in this work and considered a likelihood-based approach to annual modulation and reconstructing the halo velocity distribution.
We extend this approach to accurately account for the statistics of the axion field, to include previously-neglected but important phenomena such as gravitational focusing~\cite{Lee:2013wza} induced by the Sun's gravitational potential, and to analytically understand, using the Asimov formalism~\cite{Cowan:2010js}, the effect of DM substructure.

We organize the remainder of this work as follows.
To begin with, in Sec.~\ref{sec:likelihood} we derive a likelihood for axion direct detection.
The result is derived for both broadband and resonant experimental configurations.
Section~\ref{sec:sensitivity} determines the expected limit and detection thresholds from this likelihood.
In Sec.~\ref{sec:BulkHalo} we discuss our results in the context of an axion population following a time independent bulk halo.
Finally, Sec.~\ref{sec:DMDistribution} extends the discussion of the axion phase space to include annual modulation, gravitational focusing, and the possibility of local DM substructure such as cold streams.
We note that the analysis framework presented in this work is also provided in the form of publicly available code and can be accessed at \url{https://github.com/bsafdi/AxiScan}.

\newpage
\section{A Likelihood for Axion Direct Detection}\label{sec:likelihood}

In this section we derive a likelihood that describes how the statistics of the local DM velocity distribution are transformed into signals at axion direct detection experiments.
The main result that will be used throughout the rest of the paper is the likelihood presented in~\eqref{eq:AxionLikelihood}; however, there will be several intermediate steps.
In particular, in the first subsection we show how to write the local axion field as a sum over Rayleigh-distributed random variables, as specified in~\eqref{eq:alocal}.
In the following subsection we will show that when coupled to an experiment sensitive to the axion, if data is taken in the form of a power spectral density (PSD), it will be exponentially distributed, as given in~\eqref{eq:PSDasExponential}.
In the main body we will only derive the distribution of the signal, but in App.~\ref{app:SigBkgDist} we will show that the background only, and signal plus background distributions, are both exponentially distributed also.
Combining these, we then arrive at a form for the likelihood function.

In the initial derivation of the likelihood we will focus on how our formalism applies to a broadband experiment.
However, the modification to a resonant framework is straightforward and we present the details in the final subsection.

\subsection{The Statistics of the Local Axion Field}

Our goal in this section is to build up the local axion field from the underlying distribution of fields describing individual axions.
Thus as a starting point let us consider an individual axion-like particle, which we think of as a non-relativistic classical field.\footnote{Individual axion-like particles should technically be described as quantum objects not classical fields.
Nevertheless the local occupancy numbers of these quantum particles is enormous.
For example, taking axion dark matter with $m_a \sim 10^{-10}$ eV, the number of axions within a de Broglie volume is $\sim$$10^{36}$.
Accordingly the distinction is unimportant since formally when we say single particles we really mean a collection of particles in the same state with high enough occupancy number such that the ensemble is described by a classical wave.
For simplicity, however, we refer to these classical building blocks as ``particles."
}
If we assume that there are $N_a$ such particles locally that make up the local DM density $\rho_{\rm DM}$, then we can write down the field describing an individual particle as
\begin{equation}
a_i(v,t) = \frac{\sqrt{2\rho_{\rm DM}/N_a}}{m_a} \cos \left[ m_a \left( 1 + \frac{v_i^2}{2} \right) t + \phi_i \right]\,,
\label{eq:individualaxion}
\end{equation}
where $i \in 1, 2, \ldots, N_a$ is an index that identifies this specific axion particle, $m_a$ is the axion mass, $v_i$ is the velocity of this axion, and $\phi_i \in [0,2\pi)$ is a random phase.
The phase coherence of the full axion field constructed from the sum each of these particles is dominated by the common mass they share and to a lesser extent by velocity corrections which are drawn from a common DM velocity distribution.
Beyond this we take the fields to be entirely uncorrelated, which is represented by the random phase.
Axion self interactions could induce additional coherence.  However, given the feeble expected strength of these interactions we assume such contributions are far subdominant to those written.

From here to build up the full axion distribution we need to sum~\eqref{eq:individualaxion} over all $i$.
We proceed, though, through an intermediate step that takes advantage of the fact that there will be many particles with effectively indistinguishable speeds.
As such let us partition the full list of $N_a$ particles into subsets $\Omega_j$, which contain the $N_a^j$ particles with speeds between $v_j$ and $v_j + \Delta v$, where $\Delta v$ is small enough that we can ignore the difference between their speeds.
In this way the contribution from all particles in subset $\Omega_j$ is given by
\begin{equation}
a_j(t) = \sum_{i \in \Omega_j} \frac{\sqrt{2\rho_{\rm DM}}}{m_a\sqrt{N_a}} \cos \left[ m_a \left( 1 + \frac{v_j^2}{2} \right) t + \phi_i \right]\,.
\end{equation}
Note that it is only the random phase that differs between elements of the sum:
\begin{equation}\begin{aligned}
&\sum_{i \in \Omega_j} \cos \left[ m_a \left( 1 + \frac{v_j^2}{2} \right) t + \phi_i \right] \\
= &{\rm Re} \left\{ \exp \left[ i m_a \left( 1 + \frac{v_j^2}{2} \right) t \right] \left( \sum_{i \in \Omega_j} \exp \left[ i \phi_i \right] \right) \right\}\,.
\end{aligned}\end{equation}
To proceed further, we recognize that the sum over phases is equivalent to a 2-dimensional random walk; this allows us to write
\begin{equation}\begin{aligned}
\sum_{i \in \Omega_j} \exp \left[ i \phi_i \right] = \alpha_j e^{i\phi_j}\,,
\label{eq:RayleighApprox}
\end{aligned}\end{equation}
where $\phi_j \in [0,2\pi)$ is again a random phase and $\alpha_j$ is a random number describing the root-mean-squared distance traversed in a 2-dimensional random walk of $N_a^j$ steps.
These distances are governed by the Rayleigh distribution, which takes the form
\begin{equation}
P[\alpha_j] = \frac{2\alpha_j}{N_a^j} e^{-\alpha_j^2/N_a^j}\,.
\end{equation}
For future convenience, we remove $N_a^j$ from the distribution by rescaling \mbox{$\alpha_j \to \alpha_j \sqrt{N_a^j/2}$}, so that we can complete our result for this velocity component as follows:
\begin{align}
a_j(t) &= \alpha_j \frac{\sqrt{\rho_{\rm DM}}}{m_a} \sqrt{\frac{N_a^j}{N_a}} \cos \left[ m_a \left( 1 + \frac{v_j^2}{2} \right) t + \phi_j \right]\,, \notag \\
P[\alpha_j] &= \alpha_j e^{-\alpha_j^2/2}\,.
\label{eq:ajandalphajdist}
\end{align}

The final step to obtain the full local axion field is to sum over all $j$.
Before doing so, however, we note the important fact that the speeds, $v_j$, are being drawn from the local DM speed distribution, $f(v)$.
A simple ansatz for $f(v)$ is given by the standard halo model (SHM):\footnote{We note in passing that data from the \textit{Gaia} satellite is likely to lead to updates to this simple model~\cite{Herzog-Arbeitman:2017fte,Herzog-Arbeitman:2017zbm}.
Further, there is also likely a cut-off at the Galactic escape velocity, $\sim$$550$ km/s, though this will not play an important role in the analyses in this work.}
\begin{equation}\begin{aligned}
f_{\rm SHM}(v \vert v_0, v_{\rm obs}) &= \frac{v}{\sqrt{\pi} v_0 v_\text{obs}} e^{-(v+v_\text{obs})^2 / v_0^2} 
\\ &\times  \left( e^{4 v v_{\rm obs} / v_0^2} - 1 \right)\,,
\label{eq:SHM}
\end{aligned}\end{equation}
where in conventional units $v_0 \approx 220$ km/s is the speed of the local rotation curve, and $v_{\rm obs} \approx 232$ km/s is the speed of the Sun relative to the halo rest frame.\footnote{When manipulating the velocity distribution, we will often work in natural units.}
As shown in Sec.~\ref{sec:DMDistribution}, small variations on this simple model can induce large changes to the expected experimental sensitivity, but $f_{\rm SHM}(v)$ is likely to approximately describe the bulk of the local DM speed distribution and so gives a good initial proxy for $f(v)$.
As a first use of $f(v)$, we can rewrite $N_a^j$ in terms of $f(v)$, as from the definition of $j$ we have $N_a^j = N_a f(v_j) \Delta v$.
With this we arrive at the main goal of this section, a form for the local axion distribution:
\begin{equation}\begin{aligned}
a(t) =\,&\frac{\sqrt{\rho_{\rm DM}}}{m_a} \sum_j\,\alpha_j \sqrt{f(v_j) \Delta v} \\ \times &\cos \left[ m_a \left( 1 + \frac{v_j^2}{2} \right) t + \phi_j \right]\,,
\label{eq:alocal}
\end{aligned}\end{equation}
where note the sum over $j$ is effectively a sum over velocities, and again we emphasize that each $\alpha_j$ is a random number drawn from the distribution given in~\eqref{eq:ajandalphajdist}.

\subsection{Coupling the Axion to a Broadband Experiment}\label{sec:Broadband}

We now discuss how to quantify the coupling of the DM axion field to an experiment sensitive to the coupling in~\eqref{eq:AxionEMCoupling}, using the form of the local axion field given in~\eqref{eq:alocal}.
Then, we write down a likelihood function that may be used to describe the experimental data.
Here we focus on determining the statistics of the signal alone; combining the signal with background is straightforward and described in more detail in App.~\ref{app:SigBkgDist}.
To make the discussion concrete, we frame the problem in the context of the recently proposed ABRACADABRA experiment~\cite{Kahn:2016aff}, operating in the broadband readout mode.
We emphasize, however, that the results we derive are much more general and are applicable to any experiment which seeks to measure time-series data based upon the local axion field.
An example of this generality is provided in the next section, where we extend the formalism to the resonant case.

Let us briefly review the operation of ABRACADABRA, a 10-cm version of which is currently under development~\cite{Battaglieri:2017aum}.
This experiment exploits the fact that the coupling between the axion and QED, given by the operator in~\eqref{eq:AxionEMCoupling}, induces the following modification to Amp\`ere's circuital law
\begin{equation}
\nabla \times \mathbf{B} = \frac{\partial \mathbf{E}}{\partial t} + \mathbf{J} - g_{a \gamma \gamma} \left( \mathbf{E} \times \nabla a - \mathbf{B} \frac{\partial a}{\partial t} \right)\,.
\end{equation}
The final term in this equation implies that in the presence of a magnetic field and axion DM, there is an effective current induced that follows the primary laboratory magnetic field lines and oscillates at the axion frequency.
ABRACADABRA sources this effective current via a toroidal magnet, which generates a large static magnetic field.
The axion then generates an oscillating current parallel to the magnetic field lines, which in turn sources an oscillating magnetic flux through the center of the torus.
By placing a pickup loop in the center of the torus, this oscillating magnetic field will induce an oscillating magnetic flux of the form
\begin{equation}
\Phi_{\rm pickup}(t) = g_{a \gamma \gamma} B_{\rm max} V_B m_a a(t)\,,
\label{eq:pickupflux}
\end{equation}
where $B_{\rm max}$ is the magnetic field at the inner radius of the torus, and $V_B$ is a factor that accounts for the geometry of the toroidal magnet and pickup loop and has units of m$^3$.
In the broadband configuration, the pickup loop, which is taken to have inductance $L_p$, is inductively coupled to a DC SQUID magnetometer of inductance $L$, which will then see a magnetic flux of
\begin{equation}
\Phi_{\rm SQUID} \approx \frac{\alpha}{2} \sqrt{\frac{L}{L_p}} \Phi_{\rm pickup}\,,
\label{eq:SQUIDflux}
\end{equation}
where $\alpha$ is an $\mathcal{O}(1)$ number characterizing how the SQUID geometry impacts the mutual inductance of the SQUID and pickup loop circuit.
A typical value we will use in calculations is $\alpha = 1/\sqrt{2}$.
The coupling will also induce a frequency independent phase difference between the pickup loop and magnetometer fluxes, but as we show below such an overall phase will not contribute to the measured PSD and so we do not keep track of it.

In this way, through repeated measurements of the magnetic flux detected by the SQUID, ABRACADABRA is able to build up a time series of data proportional to the local axion field.
If the experiment is sampling the magnetic flux at a frequency $f$ over a time period $T$, then it will collect a total of $N = f\,T$ data points separated by a time spacing $\Delta t = 1 / f$.
Storing all of the experimental data may pose a challenge.\footnote{To quantify this, if we take the realistic values of $f = 100$ MHz and $T = 1$ year, this amounts to almost 13 PB of data.}
In Sec.~\ref{sec:sensitivity} we will introduce a stacking procedure to cut down on the amount of stored data while maintaining the same level of sensitivity, but for now we will put this issue aside and assume that all the data is stored and analyzed.
Combining~\eqref{eq:alocal}, \eqref{eq:pickupflux}, and \eqref{eq:SQUIDflux}, we find that
\begin{equation}\begin{aligned}
\Phi_n =\,&\sqrt{A} \sum_j\, \alpha_j \sqrt{f(v_j) \Delta v} \\ \times &\cos \left[ m_a \left( 1 + \frac{v_j^2}{2} \right) n \Delta t + \phi_j \right]\,,
\label{eq:RealSpaceData}
\end{aligned}\end{equation}
where $n \in 0, 1, \ldots, N-1$ indexes the measurement at time $t = n \Delta t$, and for future convenience we have defined
\begin{equation}
A \equiv \frac{\alpha^2}{4} \frac{L}{L_p} g^2_{a \gamma \gamma} B^2_{\rm max} V^2_B \rho_{\rm DM}\,.
\label{eq:Adefn}
\end{equation}
$A$ is proportional to the terms that dictate the size of the axion signal in the experiment, and the specific form here is peculiar to ABRACADABRA.
We note that $A$ caries the SI units of ${\rm Wb}^2$, which conveniently makes it dimensionless in natural units.

To pick the axion signal out of this time-series data, given the signal is oscillating almost at a specific frequency $m_a$ plus small corrections coming from the velocity components, it is convenient to instead consider the discrete Fourier transform of the data:
\begin{equation}
\Phi_k = \sum_{n=0}^{N-1} \Phi_n e^{-i 2 \pi kn/N}\,,
\end{equation}
where now $k \in 0, 1, \ldots, N-1$.
In practice it is more useful to work with the PSD of the magnetic flux, given by
\begin{equation}\begin{aligned}
S_{\Phi\Phi}^k = &\frac{\left( \Delta t \right)^2}{T} \left| \Phi_k \right|^2 \\
= &A \frac{\left( \Delta t \right)^2}{T} \left| 
\sum_{n=0}^{N-1} \sum_j \alpha_j \sqrt{f(v_j) \Delta v} \right. \\
&\hspace{1cm}\times \left.\cos \left[ \omega_j n \Delta t + \phi_j \right] e^{-i 2 \pi kn/N}
\right|^2\,.
\end{aligned}\end{equation}
Note that in the second equality we defined $\omega_j \equiv m_a \left( 1 + v_j^2/2 \right)$.
For the moment, it is helpful to rewrite the PSD as a function of the angular frequency $\omega$, which we can do by noting that $k = \omega T/(2\pi) = \omega \Delta t N/(2\pi)$, giving
\begin{equation}\begin{aligned}
S_{\Phi\Phi}(\omega) = A &\left| 
\sum_j \alpha_j \sqrt{\frac{f(v_j) \Delta v}{T}} \right. \\
\times &\left. \Delta t \sum_{n=0}^{N-1}\cos \left[ \omega_j n \Delta t + \phi_j \right] e^{-i \omega n \Delta t}
\right|^2\,.
\end{aligned}\end{equation}
Our experimental resolution to frequency differences is dictated by the time the experiment is run for, specifically $\Delta f = 1/T$.
Then, given the definition of $\omega_j$, for large enough $T$ we have approximately $1/T \approx m_a v_j \Delta v / (2\pi)$, and so
\begin{equation}\begin{aligned}
S_{\Phi\Phi}(\omega) = A &\left| 
\sum_j \Delta v\,\alpha_j \sqrt{\frac{f(v_j) m_a v_j}{2\pi}} \right. \\
\times &\left. \Delta t \sum_{n=0}^{N-1}\cos \left[ \omega_j n \Delta t + \phi_j \right] e^{-i \omega n \Delta t}
\right|^2\,.
\end{aligned}\end{equation}

In a realistic experimental run, $T$ will usually be much larger than any other time scale in the problem considered so far.
Exceptions to this occur when there are ultra-coherent features in the dark matter distribution, which we discuss in detail in Sec.~\ref{sec:DMDistribution}.
Putting the exceptions aside for now, we can approximate $T \to \infty$, which means we can also treat $\Delta v \to dv$, $\Delta t \to dt$, and replace the sum over $j$ with an integral over $v$ as follows:
\begin{equation}\begin{aligned}
S_{\Phi\Phi}(\omega) \approx A &\left| 
\int dv\,\alpha_v \sqrt{\frac{f(v) m_a v}{2\pi}} \right. \\
\times &\left. dt \sum_{n=0}^{N-1}\cos \left[ \omega_v n dt + \phi_v \right] e^{-i \omega n dt}
\right|^2\,.
\label{eq:PSDvelint}
\end{aligned}\end{equation}
Note in the above result we have a subscript $v$ on $\alpha_v$ and $\phi_v$, indicating that for every value of $v$ in the integral we have a different random draw of these numbers.

\begin{figure*}[htb]
\includegraphics[scale=0.38]{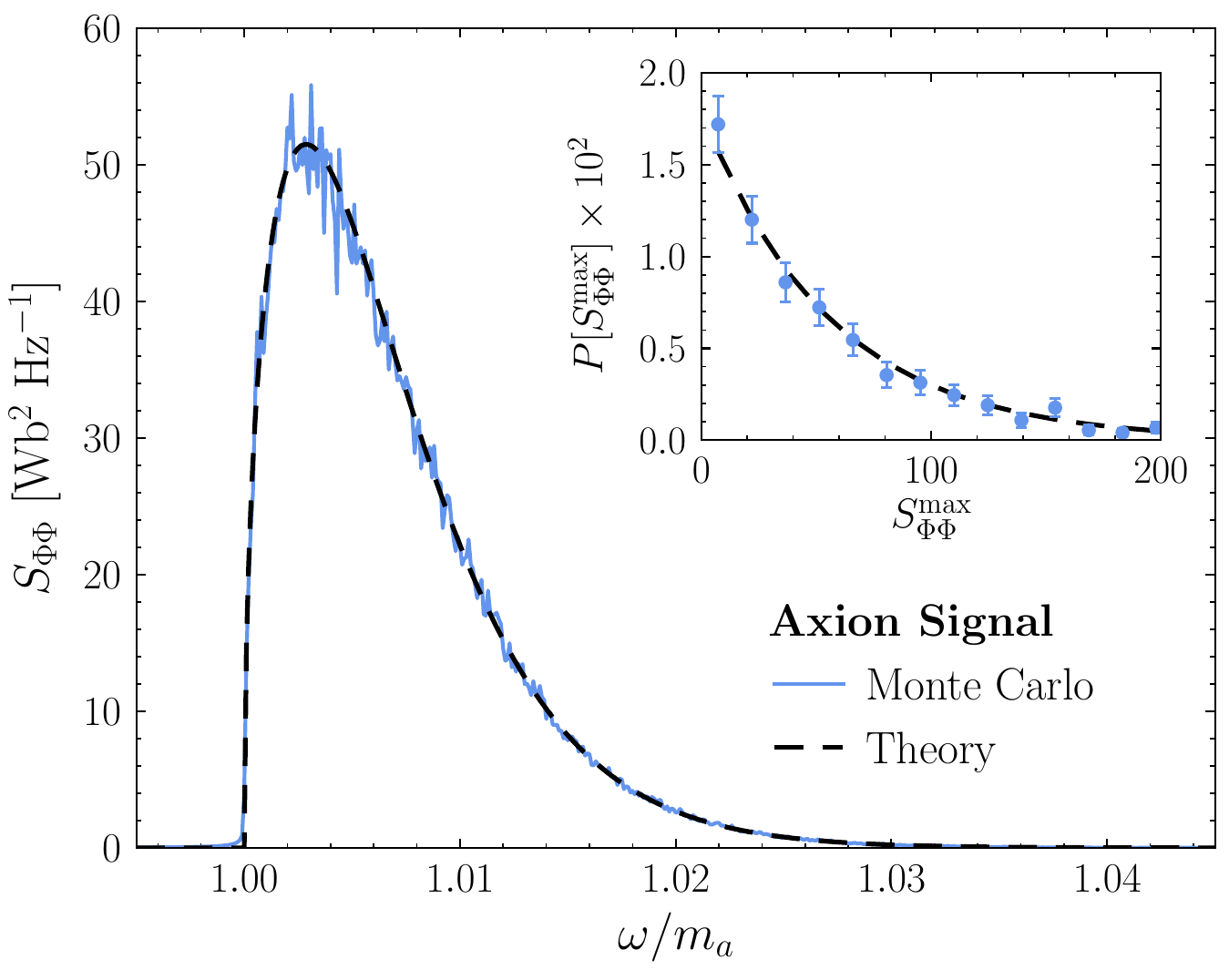} \hspace{0.2cm}
\includegraphics[scale=0.375]{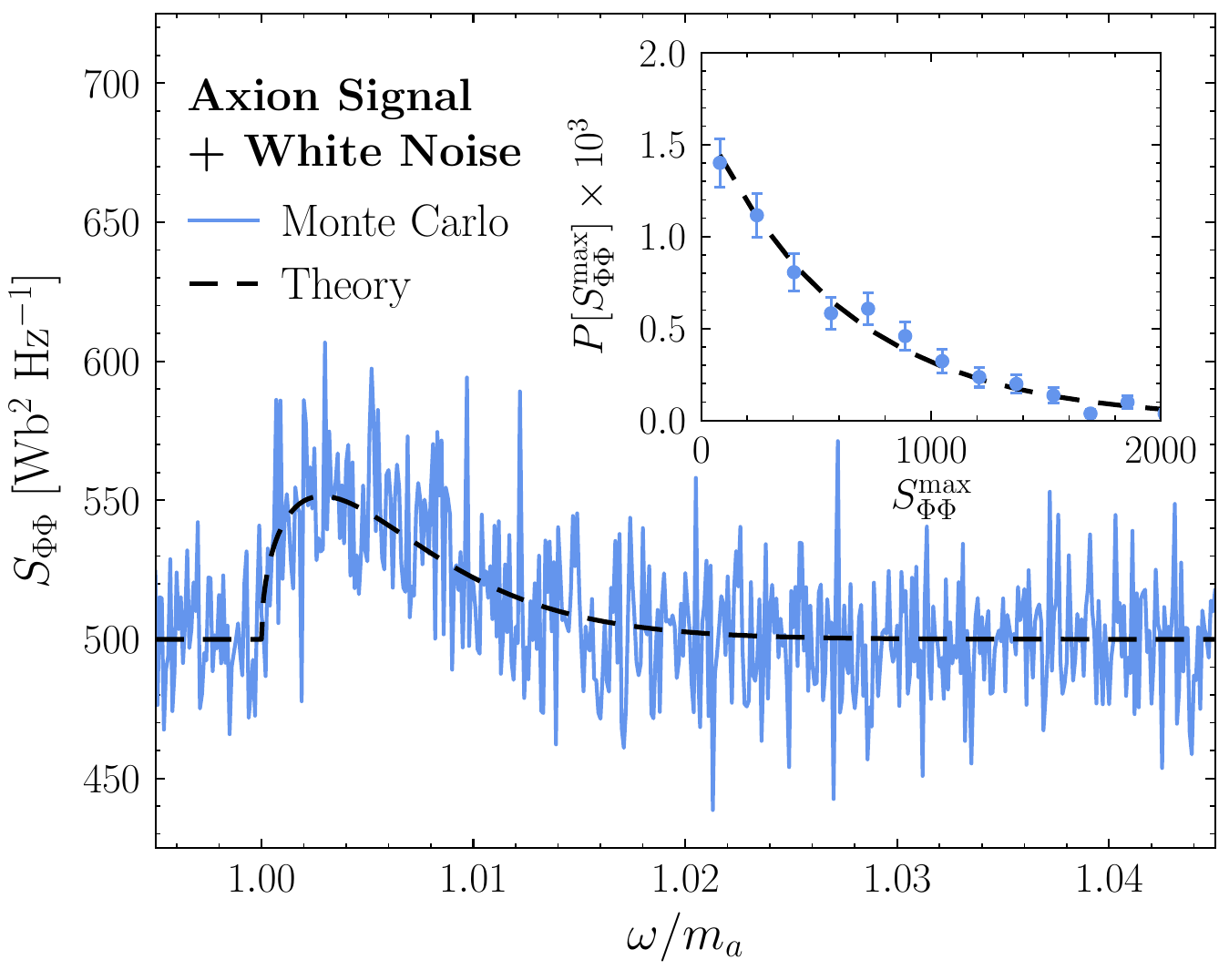}
\caption{
(Left) A comparison between the mean of 500 Monte Carlo simulations of a signal only PSD dataset (blue) and the analytic expectation given in~\eqref{eq:FullLambda} (black).
The inset shows the distribution of the 500 simulated $S_{\Phi \Phi}$ versus the predicted exponential distribution, as in~\eqref{eq:PSDasExponential}, at the frequency where the signal distribution is maximized, $\omega/m_a \approx 1.003$.
This example was generated assuming the unphysical but illustrative parameters $A=1$ Wb$^2$, $m_a = 2 \pi$ Hz, and $v_0 = v_{\rm obs} = $ 220,000 km/s.
Importantly the simulations were generated by constructing the full axion field starting from~\eqref{eq:individualaxion}, and so the agreement between theory and Monte Carlo is a non-trivial confirmation of the framework.
(Right) As on the left, but with Gaussian distributed white noise added into the time-series data with variance $\lambda_B/\Delta t$, and taking $\lambda_B = 500~{\rm Wb}^2~{\rm Hz}^{-1}$.
Again we see the theory prediction in good agreement with the average data, whilst at an individual frequency point the simulated data is exponentially distributed.
See text for details.
}
\label{fig:ConfirmDistributions}
\vspace{-0.4cm}
\end{figure*}

At this point, to make further progress we focus specifically on the sum over $n$ in the second line above. In detail,
\begin{equation}\begin{aligned}
&dt \sum_{n=0}^{N-1}\cos \left[ \omega_v n dt + \phi_v \right] e^{-i \omega n dt} \\
= &\frac{dt}{2} \left\{ e^{i\phi_v} \frac{1-\exp \left[ i \left( \omega_v - \omega \right) T \right]}{1-\exp \left[ i \left( \omega_v - \omega \right) dt \right]} \right. \\
&\left.+ e^{-i\phi_v} \frac{1-\exp \left[ -i \left( \omega_v + \omega \right) T \right]}{1-\exp \left[ -i \left( \omega_v + \omega \right) dt \right]} \right\} \\
\approx & \frac{e^{i (\phi_v + (\omega_v-\omega) T/2)}}{2} \left\{ \frac{\sin \left[ \frac{1}{2} (\omega_v - \omega) T \right]}{\frac{1}{2} (\omega_v - \omega)} \right. \\
&\left.+ e^{-i(2\phi_v + \omega_v T)} \frac{\sin \left[ \frac{1}{2} (\omega_v + \omega) T \right]}{\frac{1}{2} (\omega_v + \omega)} \right\}\,,
\end{aligned}\end{equation}
where in the final step we expanded using $(\omega_v \pm \omega) dt \ll 1$.
Then, taking the $(\omega_v \pm \omega) T \to \infty$ limit we can use the result that $\lim_{\epsilon \to 0} \sin(x/\epsilon)/x = \pi \delta(x)$ to rewrite the terms in angled brackets in terms of Dirac-$\delta$ functions which we can use to perform the integral over speeds.
There are terms associated with both positive and negative frequencies, but as we have $\omega_v > 0$ we only keep the positive result, and so conclude:
\begin{equation}\begin{aligned}
&dt \sum_{n=0}^{N-1}\cos \left[ \omega_v n dt + \phi_v \right] e^{-i \omega n dt} \\
&\approx \pi e^{i(\phi_v+(\omega_v-\omega) T/2)} \delta(\omega_v - \omega)\,.
\end{aligned}\end{equation}
With the above arguments we may perform the velocity integral in~\eqref{eq:PSDvelint}, obtaining 
\begin{equation}
S_{\Phi\Phi}(\omega) = A\left. \frac{\pi f(v)}{2 m_a v} \alpha^2 \right|_{v=\sqrt{2\omega/m_a-2}}\,.
\label{eq:PSDRayleighSq}
\end{equation}
Note that $\omega \approx m_a$, up to corrections that are $\mathcal{O}(v^2)$; where the distinction is not important, we write $m_a$ instead of $\omega$, as in the denominator above.
Further, in~\eqref{eq:PSDRayleighSq} we have dropped the subscript $v$ from $\alpha$, as it is just a single Rayleigh distributed number as given in~\eqref{eq:ajandalphajdist}.
Since $\alpha^2$ is exponentially distributed, this then implies that the PSD is also exponentially distributed: 
\begin{equation}\begin{aligned}
&P \left[ S_{\Phi\Phi}(\omega) \right] = \frac{1}{\lambda(\omega)} e^{-S_{\Phi\Phi}(\omega)/\lambda(\omega)}\,, \\
&\lambda(\omega) \equiv \langle S_{\Phi\Phi}(\omega) \rangle = A \left. \frac{\pi f(v)}{m_a v} \right|_{v=\sqrt{2\omega/m_a-2}}\,.
\label{eq:PSDasExponential}
\end{aligned}\end{equation}
Recall that $A$, which is effectively dictating the strength of the axion signal, has units of Wb$^2$, so $S_{\Phi\Phi}$ carries units ${\rm Wb}^2/{\rm Hz}$, or in natural units ${\rm eV}^{-1}$.

In any real experiment there will also be background sources of noise in the dataset.
For most sources we can think of this as mean zero Gaussian distributed noise in the time domain.\footnote{If the mean of the background distribution is non-zero, then this will only impact the $k=0$ mode of the PSD.
For reasons discussed in App.~\ref{app:SigBkgDist}, we will not include this mode in our likelihood, and as such we are only sensitive to the variance of the distributions, and so can choose them to have mean zero without loss of generality.}
For example, in ABRACADABRA the main background sources are expected to be noise within the SQUID for the broadband configuration or thermal noise in the resonant circuit~\cite{Kahn:2016aff}.
Both of these are well described by normally-distributed noise sources, and so they fall under this class of backgrounds.
In ADMX the dominant background is also thermal noise, and the Gaussian nature of this source has been discussed in Refs.~\cite{Daw:1998jm,Duffy:2006aa}; indeed, in~\cite{Duffy:2006aa} they noted the power due to thermal noise in the experiment should be exponentially distributed. 
It is likely that most other noise sources will also be normally distributed.
However, it may well be possible that certain axion direct detection experiments do suffer from background sources that are not well described by Gaussian noise.
In such a case the framework we present in this work will not go through directly, but the same logic can be used to derive a new likelihood that accounts for the specific background distribution.
Restricting ourselves to the Gaussian approximation, then, as demonstrated in App.~\ref{app:SigBkgDist}, if we have a series of Gaussian distributed backgrounds of variance $\lambda_B^i/\Delta t$, where $i$ indexes the various backgrounds, then the PSD formed from the combinations of all these will again be exponentially distributed with mean
\begin{equation}
\langle S_{\Phi\Phi}^{\rm bkg}(\omega) \rangle = \lambda_B \equiv \sum_i \lambda_B^i\,.
\label{eq:BkgExpMean}
\end{equation}
It is important to note that in general $\lambda_B$ will be a function of $\omega$, reflecting an underlying time variation in the backgrounds.

Given that the individual signal and background only cases are exponentially distributed, it is perhaps not surprising that the combined signal plus background is exponentially distributed also.
This fact is demonstrated in App.~\ref{app:SigBkgDist}, however we point out here that the correct way to think about this is that the two are combined at the level of the time-series data, not at the level of the PSD.
To highlight this, the sum of two exponential distributions is not another exponential.
Taking this fact, we arrive at the result that the full PSD will be exponentially distributed, with mean
\begin{equation}
\lambda(\omega) = A \left. \frac{\pi f(v)}{m_a v} \right|_{v=\sqrt{2\omega/m_a-2}} + \lambda_B\,.
\label{eq:FullLambda}
\end{equation}

As noted above, in the broadband mode noise within the SQUID magnetometer is expected to be the dominant source of background for ABRACADABRA, making it a useful example to keep in mind.
At high frequencies this noise source becomes frequency independent, with magnitude:
\begin{equation}
\sqrt{\lambda_B} \sim 10^{-6} \Phi_0 / \sqrt{\rm Hz}\,,
\label{eq:SQUIDnoiseHighf}
\end{equation}
which is written in terms of the flux quantum, $\Phi_0 = h/(2e) \approx 2.1 \times 10^{-15}$ Wb.
As such the typical value for the background is
\begin{equation}
\lambda_B \approx 4.4 \times 10^{-42}~{\rm Wb}^2~{\rm Hz}^{-1} = 1.6 \times 10^5~{\rm eV}^{-1}\,.
\label{eq:SquidNoise}
\end{equation}
With this example in mind, we will often assume we have a frequency independent background in our analysis to simplify results, but the formalism can in general account for an arbitrary dependence.
Despite this we note that in a real DC SQUID, there will also be a contribution to the noise scaling as $1/f$, that should dominate below $\sim 50$ Hz.
We refer to~\cite{Kahn:2016aff} for a more detailed discussion of these backgrounds.

To demonstrate how mock datasets compare to the theoretical expectations derived above, in Fig.~\ref{fig:ConfirmDistributions} we show the comparison directly, with (right) and without (left) background noise.
In both cases we show the PSD as a function of frequency averaged over 500 realization of the simulated data.
In the main figures we see that the frequency dependence of the mean of the signal only and signal plus background distributions, constructed from the simulations, are well described by the analytic relation in~\eqref{eq:FullLambda}.
The insets demonstrate that at a given frequency the simulated data is exponentially distributed in both cases, as predicted by~\eqref{eq:PSDasExponential}.
The agreement is a non-trivial check of the validity of the framework.
We emphasize that the Monte Carlo simulations are constructed in the time domain using~\eqref{eq:individualaxion} in the signal case and by drawing mean zero Gaussian noise with variance $\lambda_B/\Delta t$ for the background at each time step.
To generate these results we picked numerically convenient rather than physically realistic values.
Specifically we used $A=1$ Wb$^2$, $m_a = 2\pi$ Hz, $\lambda_B = 500$ Wb$^2$ Hz$^{-1}$, and we assumed the signal was drawn from an SHM as given in~\eqref{eq:SHM}, but with $v_0 = v_{\rm obs} =$ 220,000 km/s instead of the physical values.
However, we emphasize that these values were chosen for presentation purposes only and that we have explicitly verified that the formalism above is also valid for more realistic signal and background parameters.

Knowing how the data is distributed means we can now write down a likelihood function to constrain a signal and background model ${\mathcal M}$, with model parameters ${\boldsymbol{\theta}}$, for a given dataset $d$.
The dataset is given, in the case of ABRACADABRA, by $N$ measurements of the magnetic flux in the SQUID at time intervals $\Delta t$.
This data is then converted into a PSD distribution $S^k_{\Phi \Phi}$, measured at $N$ frequencies given by $\omega = 2 \pi k/T$, for $k \in 0,1,\ldots,N-1$.
The likelihood function for the model ${\mathcal M}$ then takes the form\footnote{The omission of $k=0$ from the likelihood is deliberate.
As described in App.~\ref{app:SigBkgDist}, the background is in fact not exponentially distributed for this value.
In addition the signal cannot contribute to the $k=0$ mode, as this would correspond to probing the velocity distribution at an imaginary value.
As such the $k=0$ or DC mode is only probing a constant contribution to the background, which we can always simply set to be zero and neglected, implying that we lose no sensitivity by simply excluding this case.
Moreover, in practice it is likely only necessary to include $k$ modes corresponding to frequencies in the vicinity of the mass under question.}
\begin{mdframed}[linewidth=1.2pt, roundcorner=5pt]
\vspace{-5pt}
\begin{equation}
\mathcal{L}(d \vert {\mathcal M}, \boldsymbol{\theta}) = \prod_{k=1}^{N-1} \frac{1}{\lambda_k(\boldsymbol{\theta})} e^{- S_{\Phi \Phi}^k / \lambda_k(\boldsymbol{\theta})}\,,
\label{eq:AxionLikelihood}
\end{equation}\vspace{-0.15cm}
\end{mdframed}
where we have used an index $k$ to denote quantities evaluated at a frequency $\omega = 2 \pi k/T$.  Note that the $\boldsymbol{\theta}$ completely specify the model expectation given in~\eqref{eq:FullLambda}.
Specifically, $\boldsymbol{\theta}$ includes parameters controlling the background contribution in $\lambda_B$, the DM halo velocity distribution $f(v)$, and the axion coupling $g_{a\gamma\gamma}$ that appears in $A$.
In the following section, we will show how to use this likelihood to set a limit on or claim a discovery of the axion, as well as constrain properties of the axion velocity distribution in the event of a detection.
First, however, we describe how the formalism above is modified for a resonant readout.

\subsection{Coupling to a Resonant Experiment}\label{sec:Resonant}

The discussion above was premised upon a broadband experimental set up.
The broadband circuit has the advantage of being able to search across a broad range of axion masses with the same dataset.
A common alternative is the resonant framework, where the resonant frequency is tuned to the axion mass under consideration before reading out the signal~\cite{Sikivie:1983ip}. 
Resonant experiments provide increased sensitivity at the frequencies under consideration.
The resonators may include physical resonators, such as that used by the ADMX experiment, or resonant circuits as used, for example, in Ref.~\cite{Sikivie:2013laa}.

In this section we demonstrate how the framework above is modified in these cases, and importantly will find that the same likelihood function applies, with a simple modification to the expected PSD given in~\eqref{eq:FullLambda}.
As a consequence, this will show that the various applications of the likelihood framework that we demonstrate throughout the rest of this work are applicable to resonant experiments, even though our examples will generally be couched in the language of a broadband framework for simplicity.

To avoid the discussion becoming too abstract, we will again work with the concrete set up of ABRACADABRA, this time in the resonant mode.
We assume, for simplicity, a simple resonant circuit, where the pickup loop is connected to an RLC circuit that is inductively coupled to the SQUID, though more complicated circuits, such as feedback damping circuits~\cite{seton1995use,seton1999gradiometer,Chaudhuri:2014dla}, may be preferable in practice~\cite{Kahn:2016aff}.
However, the analysis formalism described below should apply to any resonant circuit where thermal noise is the dominant noise source.

Our starting point is the magnetic flux due to the axion through the pickup loop, $\Phi_{\rm pickup}$, as given in~\eqref{eq:pickupflux}.
Instead of directly inductively coupling the pickup loop to the SQUID, this time we run the pickup loop through an RLC circuit with inductance $L_i$, resonant frequency $\omega_0$, and quality factor $Q_0$.
The strategy is to vary $\omega_0$ over time in order to probe a range of axion masses; we will discuss a strategy for how to choose the time variation later in this work.
Note that the quality factor also determines the bandwidth of the circuit, and so choosing a $Q_0$ corresponding to the width of the signal or better is preferable, though we leave a detailed optimization of the resonant strategy to future work.
If we inductively couple this circuit directly to the SQUID, then the flux received will be
\begin{equation}
\Phi_{\rm SQUID} = \alpha Q_0 \sqrt{\mathcal{T}(m_a)} \frac{\sqrt{L L_i}}{L_T} \Phi_{\rm pickup}\,,
\end{equation}
where we ignore constant phase shifts.
Note that we have defined the total inductance of the pickup loop and the RLC circuit as $L_T \equiv L_i + L_p$ and also a transfer function for the RLC circuit:
\begin{equation}
\mathcal{T}(\omega) \equiv \frac{1}{\left(1 - \omega_0^2/\omega^2 \right)^2 Q_0^2 + \omega_0^2/\omega^2} \,.
\label{eq:TransferFn}
\end{equation}

Following through the same steps as in the broadband case, we find that now our expected signal PSD is
\begin{equation}\begin{aligned}
\lambda^{\rm res}(\omega) =&\,A^{\rm res} Q_0^2 \mathcal{T}(\omega) \frac{\pi f(v)}{m_a v}\,, \\
A^{\rm res} \equiv&\,\alpha^2 \frac{L L_i}{L_T^2} g_{a \gamma \gamma}^2 B_{\rm max}^2 V_B^2 \rho_{\rm DM}\,, 
\end{aligned}\end{equation}
where again velocities are evaluated at $v=\sqrt{2\omega/m_a-2}$.
Comparing the expected resonant signal PSD, $\lambda^{\rm res}(\omega)$, with the expected broadband result, $\lambda(\omega)$ given in~\eqref{eq:PSDasExponential}, we see that other than the additional frequency dependence in $\mathcal{T}(\omega)$ the two only differ in experimental prefactors.

In the resonant case we also need to rethink what constitutes the dominant background source.
In particular, the addition of a resistor in the RLC circuit will generate a new source of background: Johnson–Nyquist or thermal noise.
This background is again expected to be normally distributed, with a variance $\lambda_B^\text{therm}/\Delta t$ and
\begin{equation}
\lambda_B^\text{therm}(\omega) = 2 \alpha^2 k_b T \frac{L L_i}{L_T}\frac{\omega_0}{\omega^2}Q_0 \mathcal{T}(\omega)\,,
\end{equation}
where $T$ is in this context the temperature of the circuit.
At the resonance frequency, for typical values of the parameters of interest, it may be verified that thermal noise dominates the intrinsic noise in the SQUID~\cite{Chaudhuri:2014dla,Kahn:2016aff}.
Accordingly, we neglect the background from the SQUID noise, and our full resonant model prediction is given by:\footnote{In practice, we can often approximate $L_T \approx L_i$ for a resonant configuration.}
\begin{align}
\lambda^{\rm res}(\omega) &=\left[A^{\rm res} Q_0 \frac{\pi f(v)}{m_a v}
+ \tilde \lambda_B^{\rm therm}(\omega) \right] Q_0 \mathcal{T}(\omega)\,, \nonumber\\
\tilde \lambda_B^{\rm therm}(\omega) &\equiv 2 \alpha^2 k_b T \frac{L L_i}{L_T} \frac{\omega_0}{\omega^2}\,.
\end{align}
As we will see below, the fact that the transfer function is common to both the signal and background will mean its dependence vanishes when computing our experimental sensitivity.
This point will be demonstrated in the next section.

Finally we note in passing several limitations with the simple configuration described above.
Firstly above we envisioned using a DC SQUID, which should be functional for the frequency range 100 Hz to $\sim$10 MHz.
At higher frequencies, the SQUID noise may begin to dominate over the thermal noise; moving to an AC SQUID can stave off this transition to 1 GHz~\cite{Chaudhuri:2014dla}.
Beyond this an entirely different set up would be required to read out the flux through the pickup loop, one example being provided by a parametric amplifier.
We refer to~\cite{Chaudhuri:2014dla} for a detailed discussion of each of these regimes.
Importantly, while more complicated circuits may lead to more complicated transfer functions in~\eqref{eq:TransferFn}, so long as the frequency-dependent factors are common to both the signal and the noise, the analysis formalism described below goes through unchanged.
Going forward, we assume that whenever discussing the resonant readout technique that we are in a thermal background dominated regime so the form of the transfer function is irrelevant.

\section{Experimental Sensitivity}\label{sec:sensitivity}

Armed with the likelihood given in~\eqref{eq:AxionLikelihood}, we will now determine the experimental sensitivity we can achieve.\footnote{In this and subsequent sections, we will predominantly use a frequentist statistical framework when applying the likelihood.
Nevertheless, we emphasize that our likelihood can be applied equally well within a Bayesian setting.
In particular, in Sec.~\ref{sec:DMDistribution}, we will use the Bayesian posterior as a tool for analyzing data in the presence of a putative signal.}
Below we will firstly define a series of useful statistics that will be the basic tools in our analysis.
After this we will then use an Asimov based analysis, following~\cite{Cowan:2010js}, to study the expected background and signal distributions.
We then introduce a procedure for stacking the data, which will reduce the computational demands associated with analyzing the enormous datasets axion direct detection experiments could potentially collect.
Following on from this, we will show how to use the Asimov framework to estimate our expected upper limits and discovery threshold, fully accounting for the look elsewhere effect.
Finally we will contrast our method to the simple $S/N=1$ approach commonly used in the literature.
An alternative analysis strategy to the one described in this section is to instead consider the average power in some frequency range near the expected signal location.
Such an approach is less sensitive to the one presented here, and so we have relegated its discussion to App.~\ref{app:BandAv}.

The starting point for our analysis is the likelihood $\mathcal{L}(d \vert \mathcal{M}, \boldsymbol{\theta})$.
To claim a discovery or set limits on the axion, we need to know properties of the likelihood as a function of the coupling strength, which is effectively given by $A$, and the axion mass $m_a$.
As such we separate out the parameters $\boldsymbol{\theta}$ into those of interest, $\{A, m_a\}$, and those describing the background, $\boldsymbol{\theta}_B$: $\boldsymbol{\theta} = \left\{ A, m_a, \boldsymbol{\theta}_B \right\}$.
Note that for now we fix the halo velocity distribution, though in the next two sections we generalize the model parameters to include ones that describe the DM velocity distribution.
With this distinction, we can now set up our basic frequentist tool for testing the axion model, based on the profile likelihood:
\begin{equation}\begin{aligned}
\Theta(m_a, A) = \,&2 [ \ln \mathcal{L}(d \vert {\mathcal M}, \{ A,m_a, \hat{\boldsymbol{\theta}}_B \} ) \\
&\hspace{-0.15cm}- \ln \mathcal{L}(d \vert {\mathcal M_B},  \hat{\boldsymbol{\theta}}_B ) ]\,,
\label{eq:Theta}
\end{aligned}\end{equation}
where in each of these terms $\hat{\boldsymbol{\theta}}_B$ denotes the values of the background parameters that maximize the likelihood for that dataset and model.
Note in the second line we have defined the background-only model ${\mathcal M_B}$ that has $A = 0$ and model parameters $\boldsymbol{\theta}_B$.

In terms of this basic object we can now define two useful quantities.
The first of these is a test statistic used for setting upper limits on $A$ and hence $g_{a\gamma\gamma}$:
\begin{equation}\begin{aligned}
q(m_a, A) =\,&\left\{ \begin{array}{lc} 
\Theta(m_a, A) - \Theta(m_a, \hat{A}) & A \geq \hat{A}\,, \\
0 & A < \hat{A}\,, \\
\end{array} \right.
\label{eq:TSlim}
\end{aligned}\end{equation}
where $\hat{A}$ is the value of $A$ that results in the maximum value of  $\Theta(m_a, A)$ at fixed $m_a$.
The rationale for setting this test statistic to zero for $A < \hat{A}$ is that when setting upper limits, the best we can hope to do is constrain a parameter corresponding to one stronger than the best fit value.
Observe that when $A \geq \hat{A}$, we have
\begin{equation}\begin{aligned}
q(m_a, A > \hat{A}) = \,&2 [ \ln \mathcal{L}(d \vert {\mathcal M}, \{ A,m_a, \hat{\boldsymbol{\theta}}_B \} \\
&\hspace{-0.15cm}- \ln \mathcal{L}(d \vert {\mathcal M}, \{ \hat{A},m_a, \hat{\boldsymbol{\theta}}_B \} ]\,,
\end{aligned}\end{equation}
and so this corresponds to the degradation in the likelihood as we increase $A$ beyond the best fit point.
According to Wilks' theorem, the statistic $q$, at fixed $m_a$, is asymptotically a half-chi-squared distributed with one degree of freedom.
It is a half and not full chi-squared distribution, as from the definition in~\eqref{eq:TSlim}, $q$ vanished by definition for $A < \hat{A}$.
This implies, in particular, that for a given $m_a$, the 95\% limit on $A$ will be set when $q(m_a,A_{95\%}) \approx -2.71$.
Note also that when setting limits we allow $A$ to float negative.

The second object of interest is a test statistic for discovery, denoted TS, which quantifies by how much a model with an axion of a given mass provides a better fit to the data than one without it.
This is defined as:
\begin{equation}
{\rm TS}(m_a) = \Theta(m_a, \hat{A})\,.
\label{eq:TSdisc}
\end{equation}
Below we will use the TS to quantify the 3 and 5$\sigma$ discovery thresholds, giving an accounting for the look elsewhere effect.
But the intuition is that the larger the ${\rm TS}$ the more preferred the axion.

Importantly both $q$ and TS are defined in terms of $\Theta$, implying that through an understanding of this object we can determine everything about our two test statistics.
As this will be the central object of interest, we will write out its form explicitly.
Combining~\eqref{eq:Theta} with our form of the likelihood in~\eqref{eq:AxionLikelihood}, we arrive at:
\begin{equation}\begin{aligned}
\Theta(m_a, A) = 2 \sum_{k=1}^{N-1} \left[ S_{\Phi \Phi}^k \left( \frac{1}{\lambda_B} - \frac{1}{\lambda_k} \right) - \ln \frac{\lambda_k}{\lambda_B} \right]\,.
\label{eq:Thetadef}
\end{aligned}\end{equation}
Recall that here $S_{\Phi \Phi}^k$ represents the data, whilst $\lambda_k$ and $\lambda_B$ represent the signal plus background and background only contributions respectively.
We also reiterate that only $\lambda_k$ is a function of $m_a$ and $A$, and further that $\lambda_B$ can also be $k$ dependent if the background varies with frequency.
Moreover, we stress that all $k$ modes need not be included in~\eqref{eq:Thetadef} in practice, but rather only the $k$ modes corresponding to frequencies in the vicinity of $m_a$.

\subsection{Asymptotic Distribution of the Test Statistics}

The object defined in~\eqref{eq:Theta} can be used immediately to quantify the preference for an axion signal in an experimental dataset, through the two test statistics defined above.
Before looking at any data, however, it is often useful to know what the expected sensitivity is of an experiment using these statistics.
Traditionally this is obtained via Monte Carlo simulations of the experiment, and through many realizations the expected distribution of $q$ and TS can be constructed.
The problem is also analytically tractable, however, using the method of the Asimov dataset~\cite{Cowan:2010js}, which allows us to determine the asymptotic properties of the test statistics over many realization of the data.
In this subsection we will exploit the Asimov approach to derive the asymptotic distribution of $\Theta$, and then in subsequent sections we use this formalism to determine the expected limit and discovery potential of a prospective experiment.

The key step in  the Asimov approach for our purposes is to take the dataset to be equal to the mean predictions of the model under question, neglecting statistical fluctuations.
Consider the case where we have a dataset that contains a signal of the axion with signal strength $A_t$, where the subscript $t$ indicates this is the true value.
In this case, the Asimov dataset is given by:
\begin{equation}
S_{\Phi\Phi}^{k,{\rm Asimov}} \equiv \lambda_k^t = A_t \frac{\pi f(v)}{m_a v} + \lambda_B\,,
\end{equation}
which is just~\eqref{eq:FullLambda} with $A \to A_t$.
Note that this expression should be evaluated at $v=\sqrt{4\pi k/(m_a T)-2}$, but here and below we leave the relation between $v$ and $k$ implicit.
Now using this Asimov dataset, $\Theta$ becomes (suppressing the dependence on $m_a$):
\begin{equation}
\tilde{\Theta}(A) = 2 \sum_{k=1}^{N-1} \left[ \lambda_k^t \left( \frac{1}{\lambda_B} - \frac{1}{\lambda_k} \right) - \ln \frac{\lambda_k}{\lambda_B} \right]\,,
\label{eq:AsimovThetaSum}
\end{equation}
where $\tilde{\Theta}$ denotes the asymptotic form of $\Theta$.
Importantly, one can check that this object is maximized exactly at $A=A_t$; in detail,
\begin{equation}
\max_A \tilde{\Theta}(A) = \tilde{\Theta}(A_t)\,.
\end{equation}

Now if we assume that the experiment has been run long enough that the width of frequency bins is much smaller than the range over which $\lambda_k$ or $\lambda_B$ varies,\footnote{Note that in general we would expect the signal to at least have a spread set by the velocity dispersion of the SHM, although in the presence of substructure the dispersion could be much smaller.} then we can approximate the sum over $k$ modes as an integral over velocity, just as we did in Sec.~\ref{sec:likelihood}:
\begin{equation}\begin{aligned}
\tilde{\Theta}(A) = &\frac{T m_a}{\pi} \int dv\,v \left[ \left( A_t \frac{\pi f(v)}{m_a v} + \lambda_B \right) \right. \\
\times &\left( \frac{1}{\lambda_B} - \frac{1}{A \pi f(v) / (m_a v) + \lambda_B} \right) \\
-&\left.\ln \left( 1 + A \frac{\pi f(v)}{m_a v \lambda_B} \right) \right]\,.
\label{eq:AsimovThetaInt}
\end{aligned}\end{equation}
To further simplify the expression above, we note a signal will likely be much smaller than the background in any individual bin, such that $A \pi f(v) / (m_a v),$ $A_t \pi f(v) / (m_a v) \ll \lambda_B$.
Expanding to leading order in $A$ and $A_t$, we then find
\begin{equation}\begin{aligned}
\tilde{\Theta}(A) \approx \frac{AT\pi}{m_a} \left (A_t-\frac{A}{2} \right) \int \frac{dv}{v} \frac{f(v)^2}{\lambda_B^2}\,,
\label{eq:Thetavelint}
\end{aligned}\end{equation}
where we have left $\lambda_B$ in the integral, as in general it will depend on frequency and hence velocity according to $\omega = m_a(1 + v^2/2)$.

The form of the integral over velocity as it appears in~\eqref{eq:Thetavelint} is worth commenting on, as it already implies interesting results for axion direct detection.
If we assume that the background is frequency independent, then this result tells us that the experimental sensitivity to the axion coupling $g_{a \gamma \gamma}^2$ scales as
\begin{equation}
g_{a \gamma \gamma}^2 \sim\, {1 \over  \sqrt{ \int_0^\infty dv\,\frac{f(v)^2}{v} } }\, \qquad {\rm (Field)}\,,\
\label{eq:AxionSpeedInt}
\end{equation}
with the DM velocity distribution.  This should be contrasted with the rate at WIMP\footnote{Here, we use weakly interacting massive particle (WIMP) direct detection to simply refer to the direct detection of massive DM particles at the $\sim$MeV scale and above, even if the particle models are not directly related to the WIMP paradigm.} direct detection experiments, which scales with the mean inverse speed (see, for example,~\cite{Freese:2012xd}).
In particular, the limit on the DM cross-section $\sigma_{\rm DD}$ to scatter off ordinary matter, which generically scales with the coupling $g$ to ordinary matter as $g^2$, scales with the velocity distribution as
\begin{equation}
\sigma_{\rm DD} \sim \, {1 \over \int_{v_{\rm min}}^\infty dv\,\frac{f(v)}{v} } \, \qquad {\rm (Particle)}\,,
\end{equation}
where $v_{\rm min}$ is the minimum speed required to cause the target nucleus in the detector to recoil at a given recoil energy.
This cut off scales with the inverse reduced mass of the WIMP nucleon system, $v_{\rm min} \propto 1/\mu$, so that for lighter DM particles the rate is particularly sensitive to the upper end of the speed profile.
In the axion case, the significance of an axion signal depends on an integral over the full speed profile.
Importantly, the quadratic scaling of the integrand with the speed distribution implies that axion direct detection experiments are particularly sensitive to small scale structures in the speed profile, such as those that can be induced by local DM substructure.
This stands in contrast to WIMP direct detection, where substructure is generally thought to only have a minimal impact, see, {\it e.g.},~\cite{Vogelsberger:2008qb}.

We will explore the sensitivity of axion direct detection experiments to DM substructure in Sec.~\ref{sec:DMDistribution}, but for now we illustrate the difference between axion and WIMP experiments noted above with a simple example.
Suppose that there is a contribution to the local DM velocity distribution that can be modeled as a cold stream, with $f_\text{str}(v) = {1 /\delta v}$ for $v_\text{str} < v < v_\text{str} + \delta v$ and zero otherwise.
We assume that the stream width $\delta v \ll v_\text{str}$, where $v_\text{str}$ is the stream boost speed in the Earth frame.  Then, then in the WIMP case we find $\sigma_{\rm DD} \sim v_\text{str}$, where we have assumed $v_\text{str} > v_\text{min}$.
However, in the axion case there is an extra enhancement for small stream widths such that $g_{a \gamma \gamma}^2 \sim \sqrt{v_\text{str} \delta v}$.
Note that this implies that as $\delta v$ decreases we can probe smaller values of $g_{a \gamma \gamma}$ in the axion case, while conversely decreasing $\delta v$ does not improve our sensitivity to $\sigma_{\rm DD}$ in the WIMP case.

Finally we note that if we repeated the analysis leading to~\eqref{eq:Thetavelint} for the resonant case, we would instead have arrived at
\begin{equation}\begin{aligned}
\tilde{\Theta}^{\rm res}(A^{\rm res}) = &\frac{Q_0^2 A^{\rm res}T\pi}{m_a} \left (A^{\rm res}_t-\frac{A^{\rm res}}{2} \right) \\
\times &\int \frac{dv}{v} \frac{f(v)^2}{(\tilde \lambda_B^{\rm therm})^2}\,,
\end{aligned}\end{equation}
which is essentially the same result but with the broadband quantities replaced with their appropriate resonant counterparts.
Importantly, note that the transfer function and its associated frequency dependence has dropped out of this result because it involved a ratio of the signal to the background, both of which are linear in $\mathcal{T}(\omega)$.
This justifies the claim that going forward our estimates for the resonant case can be obtained straightforwardly from the broadband results provided we make the substitutions:
\begin{equation}\begin{aligned}
A &\to {Q_0} A^{\rm res}\,, \\
\lambda_B &\to \tilde \lambda_B^{\rm therm}\,.
\end{aligned}\end{equation}

\subsection{A Procedure for Stacking the Data}\label{sec:Stack}

We would like a method to reduce the number of PSD components that need to be stored, without sacrificing sensitivity, given that if we are sampling at a high rate, for example $\sim$100 MHz or higher, over an extended time, the amount of data to be stored and analyzed could become substantial.
As we will now show, stacking the PSD data provides exactly such a method.\footnote{We thank Jon Ouellet for conversations related to this point.}

The central idea is to break the data up into $N_T$ subintervals of duration $\Delta T = T/N_T$, each with $\Delta N = N/N_T$ PSD components.\footnote{The choice of notation here is used to emphasize that for $N_T \gg 1$ we have $\Delta T \ll T$ and $\Delta N \ll N$, but of course neither quantity should ever be thought of as infinitesimal.}
In each of these subintervals we calculate the PSD $S_{\Phi\Phi}^{k,\ell}$, where now $k$ only indexes the integers from $0$ to $\Delta N-1$, and we have the new index $\ell = 0, 1, \ldots, N_T-1$ that identifies the relevant subinterval.
Using this data, our likelihood takes the form
\begin{equation}
\mathcal{L}(d \vert \boldsymbol{\theta}) = \prod_{\ell=0}^{N_T-1} \prod_{k=1}^{\Delta N-1} \frac{1}{\lambda_k(\boldsymbol{\theta})} e^{- S_{\Phi \Phi}^{k,\ell} / \lambda_k(\boldsymbol{\theta})}\,.
\label{eq:StackLikelihoodAllSum}
\end{equation}
Importantly, we assume that the model prediction in each subinterval is identical, which we comment on more below.
With this assumption, it is natural to define a stacked PSD
\begin{equation}
\overbar{S}_{\Phi\Phi}^k \equiv \frac{1}{N_T} \sum_{\ell=1}^{N_T-1} S_{\Phi \Phi}^{k,\ell}\,.
\label{eq:StackedPSD}
\end{equation}
The averaged PSD components will be distributed as the average of a sum of exponentially distributed random variables with mean $\lambda_k$, which is given by the Erlang distribution:
\begin{equation}
P[\overbar{S}_{\Phi\Phi}^k] = \frac{N_T^{N_T}}{(N_T-1)!} \frac{\left( \overbar{S}_{\Phi\Phi}^k \right)^{N_T-1}}{\lambda_k^{N_T}} e^{-N_T \overbar{S}_{\Phi\Phi}^k/\lambda_k}\,.
\end{equation}

Using this stacked data, we can simplify~\eqref{eq:StackLikelihoodAllSum} by removing the sum over $\ell$:
\begin{equation}
\mathcal{L}(d \vert \boldsymbol{\theta}) = \prod_{k=1}^{\Delta N-1} \frac{1}{\lambda_k(\boldsymbol{\theta})^{N_T}} e^{- N_T \overbar{S}_{\Phi\Phi}^k / \lambda_k(\boldsymbol{\theta})}\,,
\end{equation}
where in this result we can already see the reduction in computational requirements as it only involves a product over $\Delta N \ll N$ numbers, since the $\overbar{S}_{\Phi\Phi}^k$ can be precomputed and updated as more data comes in.

Our next task is to determine how this choice will impact our sensitivity, using the test statistics defined in the previous subsections.
It is sufficient to consider $\Theta(m_a,A)$, defined in~\eqref{eq:Theta} and from which the other statistics of interest can be derived.
Doing so, we can repeat the Asimov analysis from the previous subsection to determine the asymptotic form of the stacked $\Theta$, given by
\begin{equation}\begin{aligned}
\tilde{\Theta}_{\rm stacked}(A) = \frac{A N_T \Delta T \pi}{m_a} \left (A_t-\frac{A}{2} \right) \int \frac{dv}{v} \frac{f(v)^2}{\lambda_B^2}\,.
\end{aligned}\end{equation}
Yet as $N_T \Delta T = T$, the stacked and unstacked form of $\tilde{\Theta}$ are identical.
This implies that our stacking procedure, which for $N_T \gg 1$ dramatically reduces the required computation, has no impact on our sensitivity to an axion signal.

There is, however, a catch.
Stacking implies that we are only sensitive to frequency shifts of size $\Delta f = 1 / \Delta T$, which can be much larger than the shifts we were sensitive to in the full dataset, where $\Delta f = 1 / T \ll 1 / \Delta T$.
This could mean, depending on the size of the frequency spacings, that ultra-cold local DM substructure is no longer resolved, and therefore the enhancement it would have given to the integral over velocity discussed above is lost.
In this sense stacking can lead to a degradation in sensitivity, and so choosing a stacking strategy should be done with careful consideration of the features being searched for.
To provide some intuition, if we are searching for an axion at a mass corresponding to a frequency $f$ and drawn from a velocity distribution with dispersion $v_0$, then the coherence time is $\sim 1/(fv_0 v_\mathrm{obs})$.
To be able to fully resolve the axion signal we would then want $\Delta T \gg 1/(fv_0 v_\mathrm{obs})$.
For the SHM, and scanning in frequencies from 100 MHz down to 100 Hz, the coherence time varies from 20 ms up to 5 hours.
In such a scenario, if data were collected for a year, many stacking procedures would be feasible.
On the other hand if searching for the signal from a cold stream with a dispersion of $v_0 = 1$ km/s, then over the same frequency range the coherence time varies from 4 seconds up to 45 days.
For the lowest frequencies in this case, any stacking procedure would be sacrificing sensitivity to such cold substructure.
On the other hand, at the lowest frequencies high sampling rates are not necessary.
Thus, a hybrid approach may be preferable in practice, where the data is stacked in Fourier space at high frequencies while at low frequencies the data is stacked in time ({\it i.e.} down-sampled) in order to reduce the data size without sacrificing the sensitivity to cold substructure at any possible axion mass.

Another relevant consideration is that due to the Earth's acceleration, lab-frame frequencies may shift throughout the day and year, which would invalidate our assumption that the model predictions are identical between subintervals.
The rotational speed of the Earth's surface about its axis is roughly $0.46 \cos(\delta)$ km/s, where $\delta$ is the latitude.
This value is small enough that it can safely neglected for any cold flow with a velocity dispersion greater than this.
The rotation of the Earth about the Sun, however, occurs at roughly 30 km/s and is thus harder to ignore when searching for cold substructure, as we discuss later in this work.
Annual and daily modulation can lead to striking additional signatures, which we explore in detail in Sec.~\ref{sec:DMDistribution}.

\subsection{Expected Upper Limit}\label{sec:95limits}

We are now in a position to write down the expected 95\% limit on $A$.
In the case of a limit, the appropriate Asimov dataset to use is a background only distribution, so that $A_t = 0$.
Then by combining our definition of the likelihood profile in~\eqref{eq:TSlim} with our Asimov result in~\eqref{eq:Thetavelint}, we arrive at the 95\% limit where $q(m_a,A_{95\%}) = -2.71$, given by
\begin{equation}
\tilde{A}_{95\%} = \sqrt{2.71 \left[ \frac{T\pi}{2 m_a} \int \frac{dv}{v} \frac{f(v)^2}{\lambda_B^2} \right]^{-1}}\,.
\label{eq:A95}
\end{equation}
Note that again the tilde indicates this is an Asimov, or median, quantity.
Of course what we actually want, however, is a limit on $g_{a \gamma \gamma}$, and so for the particular example of ABRACADABRA we can insert the form of $A$ given in~\eqref{eq:Adefn}, yielding
\begin{equation}\begin{aligned}
\tilde{g}_{a\gamma\gamma}^{95\%} = &\frac{2.71^{1/4} \sqrt{L_p/L}}{\alpha B_{\rm max} V_B \sqrt{\rho_{\rm DM}}} \\
\times &\left[ \frac{T\pi}{32 m_a} \int \frac{dv}{v} \frac{f(v)^2}{\lambda_B^2} \right]^{-1/4}\,.
\label{eq:g95}
\end{aligned}\end{equation}

One of the real powers of the Asimov analysis is that not only can we determine the median expected limit, we can also derive analytically the expected size of fluctuations away from the central value, without having to revert to Monte Carlo simulations.
The details of this statistical procedure are discussed in~\cite{Cowan:2010js}.
As we are constructing power-constrained 95\% one-sided limits, we obtain confidence intervals via
\begin{equation}
q(m_a,A_{95\% \pm N\sigma}) = - \left( \Phi^{-1} \left[ 0.95 \right] \pm N \right)^2\,,
\label{eq:CLSB}
\end{equation}
where $\Phi$ is the cumulative distribution function of the standard normal distribution (zero mean and unit variance), and $\Phi^{-1}$ is the inverse of this  (so $\Phi^{-1} \left[ 0.95 \right] \approx 1.64$).  
Note that if we take $N=0$, then the above just reduces to $q(m_a,A_{95\%}) = -2.71$, but this more general result contains the information about the error bands in the expected limit.
In this way, by replacing the 2.71 that appears in~\eqref{eq:g95} with the appropriate value for the $N \sigma$ uncertainty band on the 95\% limit, we can construct the median and uncertainty bands on $\tilde{g}_{a\gamma\gamma}^{95\%}$ analytically.
For completeness, in App.~\ref{app:MC_Expectations} we verify that the bands constructed in this manner agree with those generated using Monte Carlo simulations.
Finally, to be conservative we use power-constrained limits~\cite{Cowan:2011an}, which in practice means we do not allow ourself to set a limit below our 1$\sigma$ uncertainty band on the upper limit.

\subsection{Expected Discovery Reach}\label{sec:discoveryreach}

In order to find evidence for a signal, we need to understand the expected distribution of the TS under the null hypothesis.
The reason is that this distribution determines how likely the background is to produce a given TS value, and hence what threshold ${\rm TS}_{\rm thresh}$ we should set to establish the existence of a signal at a given confidence level.
Once we have such a threshold test statistic, applying our Asimov results above to the case of discovery, we find we would be sensitive to discover a signal with the following strength
\begin{equation}\begin{aligned}
\tilde{g}_{a\gamma\gamma}^{\rm thresh} = &\frac{{\rm TS}^{1/4}_{\rm thresh} \sqrt{L_p/L}}{\alpha B_{\rm max} V_B \sqrt{\rho_{\rm DM}}} \\
\times &\left[ \frac{T\pi}{32 m_a} \int \frac{dv}{v} \frac{f(v)^2}{\lambda_B^2} \right]^{-1/4}\,.
\label{eq:gthresh}
\end{aligned}\end{equation}

Locally, the significance in favor of the axion model is expected to be approximated by $\sqrt{\text{TS}}$~\cite{Cowan:2010js}; that is, a value $\text{TS} = 25$ corresponds to approximately 5$\sigma$ local significance.
However, when scanning over multiple independent mass points, the look elsewhere effect must be accounted for in quoting values for the global rather than local significance.  The look elsewhere effect may be determined through Monte Carlo simulations.
However, in this section we will derive an analytic approximation to ${\rm TS}_{\rm thresh}$, which accounts for the look elsewhere effect, and as we will show provides an accurate representation to the output from such Monte Carlo studies.
The result will be a mapping between the desired global significance threshold and the value of ${\rm TS}_{\rm thresh}$ that should be taken, depending on the mass range scanned.
We note that there are also other proposals in the literature for approaching this problem; for a recent one see, {\it e.g.},~\cite{Beaujean:2017eyq}.

Our starting point is to note that the asymptotic form of the survival function for the local TS under the null hypothesis is given by
\begin{equation}
S[{\rm TS}_{\rm thresh}] 
= 1 - \Phi\left( \sqrt{ {\rm TS}_{\rm thresh} } \right) \,,
\label{eq:survival}
\end{equation}
where $S[{\rm TS}_{\rm thresh}] $ is the probability that the TS, under the null hypothesis, takes a value greater than ${\rm TS}_{\rm thresh}$.
This is derived explicitly in App.~\ref{app:asymTS} starting from the likelihood function, and it is equivalent to the statement that the asymptotic local significance is given by $\sqrt{\text{TS}}$.
However in any realistic experiment, we will look in a number of independent frequency windows corresponding to different axion masses.
To account for this we need to note that in any of these windows there could be an upward fluctuation. 
To do so let us say that we look at $N_{m_a}$ independent mass points, and we want to set the threshold test statistic, ${\rm TS}_{\rm thresh}$, such that the probability that the background will not fake the signal in any bin is $1-p$.
To relate these two quantities, if we assume that $p$ is small enough, we can write the probability that at least one of the TSs, from the set over all mass points, is greater than ${\rm TS}_{\rm thresh}$ as
\begin{equation}\begin{aligned}
p &= 1 - \left( 1 - S[{\rm TS}_{\rm thresh}] \right)^{N_{m_a}} \\
&\approx N_{m_a} S[{\rm TS}_{\rm thresh}]\,.
\end{aligned}\end{equation}
From here we can then substitute the survival function from~\eqref{eq:survival}, and expanding this out gives
\begin{equation}
{\rm TS}_{\rm thresh} = \left[ \Phi^{-1} \left( 1 - {p \over N_{m_a}} \right) \right]^2 \,.
\label{eq:TSthreshLE}
\end{equation}

Using this result, as soon as we know $N_{m_a}$ we can determine ${\rm TS}_{\rm thresh}$ as it should be used in our formula for $\tilde{g}_{a\gamma\gamma}^{\rm thresh}$ in~\eqref{eq:gthresh}.
To give some intuition, in the case where we ignore the look elsewhere effect and set $N_{m_a} = 1$, then the 3$\sigma$ requirement is that $p \approx 1.35 \times 10^{-3}$, yielding ${\rm TS}_{\rm thresh} = 9$, as expected.
Importantly, note that the $p$ values here correspond to that for 1-sided fluctuations~\cite{Cowan:2010js}.

In any realistic experiment, we expect $N_{m_a} \gg 1$.
However, estimating the correct value for $N_{m_a}$ is complicated by the fact that we may scan over a continuum of different possible mass points in practice, though not all of the mass points have independent data.
We expect a mass point as frequency $m_a$ to extend over a frequency bandwidth $\sim$$m_a v_0^2$, for the SHM.  Thus, we expect to be able to characterize a set of independent mass point by the relation
\begin{equation}
m_a^{(i)} = m_a^{(0)} \left( 1 + \alpha v_0 v_\mathrm{obs} \right)^i\,, 
\label{eq:mai}
\end{equation}
where $m_a^{(0)}$ is the first mass point, $i=0,\ldots,N_{m_a}-1$, and $\alpha$ is a number order unity that should be tuned to Monte Carlo simulations.
Given the parameterization in~\eqref{eq:mai}, we may estimate the number of mass points by relating $m_a^0$ with the minimum frequency $f_\text{min}$ and $m_a^{(N_{m_a} - 1)}$ with the highest frequency $f_\text{max}$; solving for $N_{m_a}$ in the limit $N_{m_a} \gg 1$ then gives
\begin{equation}
N_{m_a} \approx \frac{1}{\alpha \, v_0 v_\mathrm{obs} } \ln \frac{f_{\rm max}}{f_{\rm min}}\,.
\label{eq:Nmaapprox}
\end{equation}

\begin{figure}[t]
\includegraphics[scale=0.38]{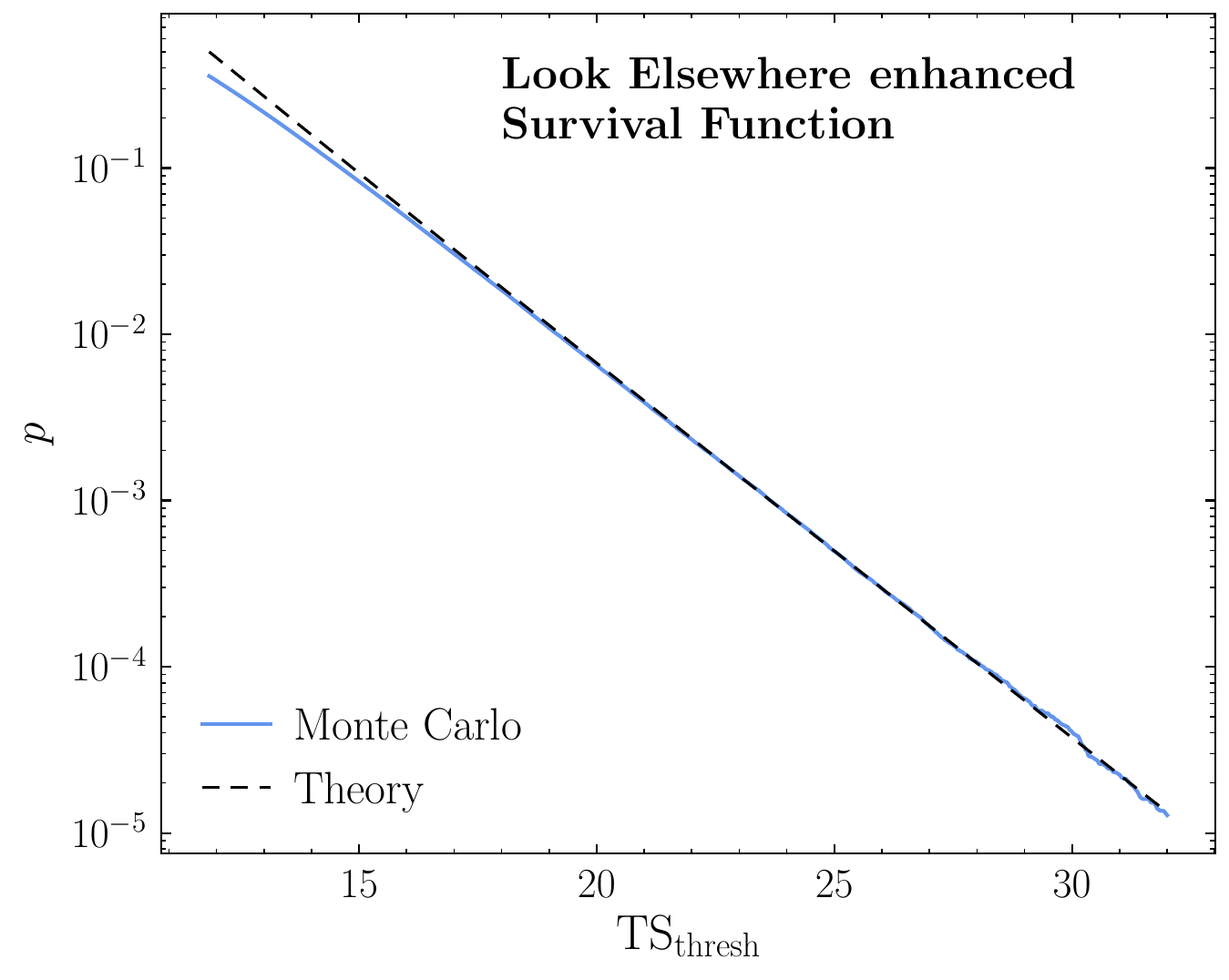}
\caption{
A comparison between the look elsewhere effect improved survival function approximate result derived between~\eqref{eq:TSthreshLE} and~\eqref{eq:Nmaapprox}, and the equivalent values derived directly from Monte Carlo simulations.
The good agreement between the two, especially at large ${\rm TS}_{\rm thresh}$ demonstrates that our approximate result is useful for estimating how often the background can fluctuate to fake the signal at a given confidence level.
Note the values plotted here correspond to signals varying from 0 to 4$\sigma$, for derived values of $\lambda_B$ given in~\eqref{eq:SquidNoise} and 2.5 million Monte Carlo simulations.
We do not extend the plot up to the 5$\sigma$ value relevant for discovery, as this would require roughly 100 times as many simulations.
This statement in itself already demonstrated the utility of our approximate analytic result.
}
\label{fig:LESurvival}
\vspace{-0.4cm}
\end{figure}

In Fig.~\ref{fig:LESurvival} we compare the analytic prediction in~\eqref{eq:TSthreshLE}, combined with~\eqref{eq:Nmaapprox}, with the result of 2.5 million Monte Carlo simulations.
From the ensemble of simulations, we are able to compute the value of $p$ for each value of TS$_\text{thresh}$.
Note that in each simulation we scan for axion DM over a frequency range $f_\text{max} / f_\text{min} \approx 1.0007$; setting $v_0 = 220$ km/s and $v_\mathrm{obs} = 232$ km/s then gives, through~\eqref{eq:Nmaapprox}, $N_{m_a} \approx 1.23 \times 10^3 / \alpha$.
The analytic results are found to agree well with the simulations for $\alpha \approx 3/4$; this value may also be understood by thinking more carefully about the extent of the SHM.
Note that the real power of the analytic formalism is that once we have tuned the relations in~\eqref{eq:TSthreshLE} and~\eqref{eq:Nmaapprox} to Monte Carlo, in order to find the appropriate value of $\alpha$, we may extrapolate to smaller values of $p$, where the number of Monte Carlo simulations required to directly determine TS$_\text{thresh}$ would be intractable.

To give some more realistic examples, if we assume the experiments scans from 100 Hz to 100 MHz, using the SHM values we obtain $N_{m_a} \sim 3 \times 10^7$.
This then increases the 3$\sigma$ (5$\sigma$) threshold TS to 40.9 (57.5).
To contrast if instead our significance was dominated by a stream with dispersion roughly 20 km/s, then instead we would find $N_{m_a} \sim 4 \times 10^9$, and the 3$\sigma$ (5$\sigma$) threshold TS becomes 50.3 (67.0).

\subsection{Comparison with $S/N=1$}\label{sec:SN1}

In the absence of a full likelihood framework, a common method employed for estimating sensitivity is obtained by setting the signal equal to the expected background, or $S/N=1$.
For example, this approach was used in the original ABRACADABRA proposal~\cite{Kahn:2016aff} and also for the proposed CASPEr experiment~\cite{Budker:2013hfa}.
In this section we want to contrast this simple estimate to the output from our full likelihood machinery.

Now, following these earlier references, in our notation the signal-to-noise ratio can be written as
\begin{equation}
S/N = \left| \Phi_{\rm SQUID} \right| (T \tau)^{1/4} / \sqrt{| \lambda_B |}\,,
\label{eq:SN1}
\end{equation}
where $\tau$ is the signal coherence time.
This $S/N \propto T^{1/4}$ scaling occurs when the collection time is longer than the coherence time.
If $T < \tau$, instead the significance grows as $S/N \propto T^{1/2}$, as demonstrated in \cite{Budker:2013hfa}. 
In App.~\ref{app:RootTScaling} we demonstrate that this same scaling can also be seen directly from our likelihood.

In order to make a concrete comparison, we consider ABRACADABRA with the axion following only the bulk velocity distribution.
In this case, the coherence of the bulk halo, as discussed above, will effectively ensure we always have $T \gg \tau$, implying the signal grows as $T^{1/4}$.
To simplify~\eqref{eq:SN1}, firstly consider $\left| \Phi_{\rm SQUID} \right|$.
Combining~\eqref{eq:SQUIDflux} and \eqref{eq:pickupflux}, we have:
\begin{equation}
\left| \Phi_{\rm SQUID} \right| = \frac{\alpha}{2} \sqrt{\frac{L}{L_p}} g_{a\gamma \gamma} B_{\rm max} V_B m_a \left| a(t) \right|\,.
\end{equation}
For the purposes of determining the average axion field over a time $T \gg \tau$, we can simply consider the axion field in the zero velocity limit, where
\begin{equation}
\left| a(t) \right| = \frac{\sqrt{2 \rho_{\rm DM}}}{m_a} \left| \cos (m_a t) \right|  = \frac{\sqrt{\rho_{\rm DM}}}{m_a}\,.
\end{equation}
Note that since it is the PSD that is measured in practice, we calculate the average as $\sqrt{\left| \cos^2 (m_a t) \right|} = 1/\sqrt{2}$. 
The coherence time is determined by the kinetic energy $\frac{1}{2} m_a v^2$, which perturbs the axion frequency.
Once the phase shift from this correction equals $\pi$, the field will be fully out of phase, so we take
\begin{equation}
\tau = \frac{2\pi}{m_a v_0 v_\mathrm{obs} }\,,
\end{equation}
where again with the bulk halo in mind, we took values appropriate for the SHM.
Finally, we assume that we have a frequency independent background PSD $\lambda_B$.
Combining these results with the threshold $S/N=1$, we obtain a sensitivity estimate of
\begin{equation}
g_{a\gamma \gamma} = \frac{2 \sqrt{\lambda_B} \sqrt{L_p/L}}{\alpha B_{\rm max} V_B \sqrt{\rho_{\rm DM}}} \left( \frac{m_a v_0 v_\mathrm{obs} }{2\pi T} \right)^{1/4}\,.
\label{eq:ABRASN1}
\end{equation}

We want to contrast this estimate with the exact value we obtain from the analysis method outlined in this section.
For this purpose we take our result, but evaluated at some ${\rm TS}_{\rm req}$ which is schematic---it can be 2.71 for the case of a 95\% limit, or $\sim$58 for a $5 \sigma$ discovery accounting for the look elsewhere effect.
If we assume $f(v)$ follows the SHM and further take $v_{\rm obs} = v_0$, then the equivalent result is:
\begin{equation}\begin{aligned}
g_{a\gamma \gamma} = &\left( \frac{64\,{\rm TS}_{\rm req}\,\sqrt{2\pi}}{{\rm erf} \left[ \sqrt{2} \right]} \right)^{1/4} \\
\times &\frac{\sqrt{\lambda_B} \sqrt{L_p/L}}{\alpha B_{\rm max} V_B \sqrt{\rho_{\rm DM}}} \left( \frac{m_a v_0^2}{2\pi T} \right)^{1/4}\,.
\end{aligned}\end{equation}
Note that the formula above is equivalent to the statement that
\begin{equation}
S/N \approx 1.8\,{\rm TS}_{\rm req}^{1/4}\,.
\end{equation}
For example, the 95\% expected upper limit would require $S/N = 2.31$, whilst a 5$\sigma$ discovery accounting for the look elsewhere effect assuming the SHM, requires $S/N = 4.97$.
We will see in the next section that the comparisons are similar for a resonant experiment also. 
In general the various thresholds are achieved with a larger signal than the naive $S/N=1$ suggests.
Nonetheless, the standard estimate is not a terrible approximation to the true results, especially considering that $S \sim g_{a \gamma \gamma}^2$.
We emphasize, however, that there is a lot more that can be extracted from having the full likelihood framework, which we turn to in the subsequent sections.

\section{Application to the Bulk Halo}\label{sec:BulkHalo}

In this section we apply the formalism developed so far to ABRACADABRA and ADMX. 
For this purpose we take a simple concrete example, where $f(v)$ describes only the bulk halo, which we further assume follows the SHM as defined in~\eqref{eq:SHM}.
Additionally we assume that over the frequency band of the signal,\footnote{By the frequency band we simply mean the range of frequencies over which the signal will be significant, which for the SHM is approximately $[m_a, m_a(1+v_0 v_\mathrm{obs} )]$.} the mean of the background distribution in frequency space is approximately frequency independent.
These assumptions imply that the integral appearing in~\eqref{eq:g95} and~\eqref{eq:gthresh} can be evaluated exactly:
\begin{equation}\begin{aligned}
\int \frac{dv}{v} \frac{f(v)^2}{\lambda_B^2} = \frac{{\rm erf} \left[ \sqrt{2} v_{\rm obs}/v_0 \right]}{\sqrt{2\pi} v_0 v_{\rm obs} \lambda_B^2}\,,
\end{aligned}\end{equation}
with $\sqrt{\lambda_B} \approx 10^{-6} \Phi_0/\sqrt{\rm Hz}$ as given in~\eqref{eq:SQUIDnoiseHighf}.
In the following subsections, we will demonstrate explicitly how to construct projected limits and detection sensitivities, under the assumption of the SHM velocity distribution, and we will show in the event of a detection the parameters of the SHM may be determined using the likelihood framework.
We will extend this framework to more realistic $f(v)$, including DM substructure, in the next section.

\begin{figure*}[htb]
\includegraphics[scale=0.38]{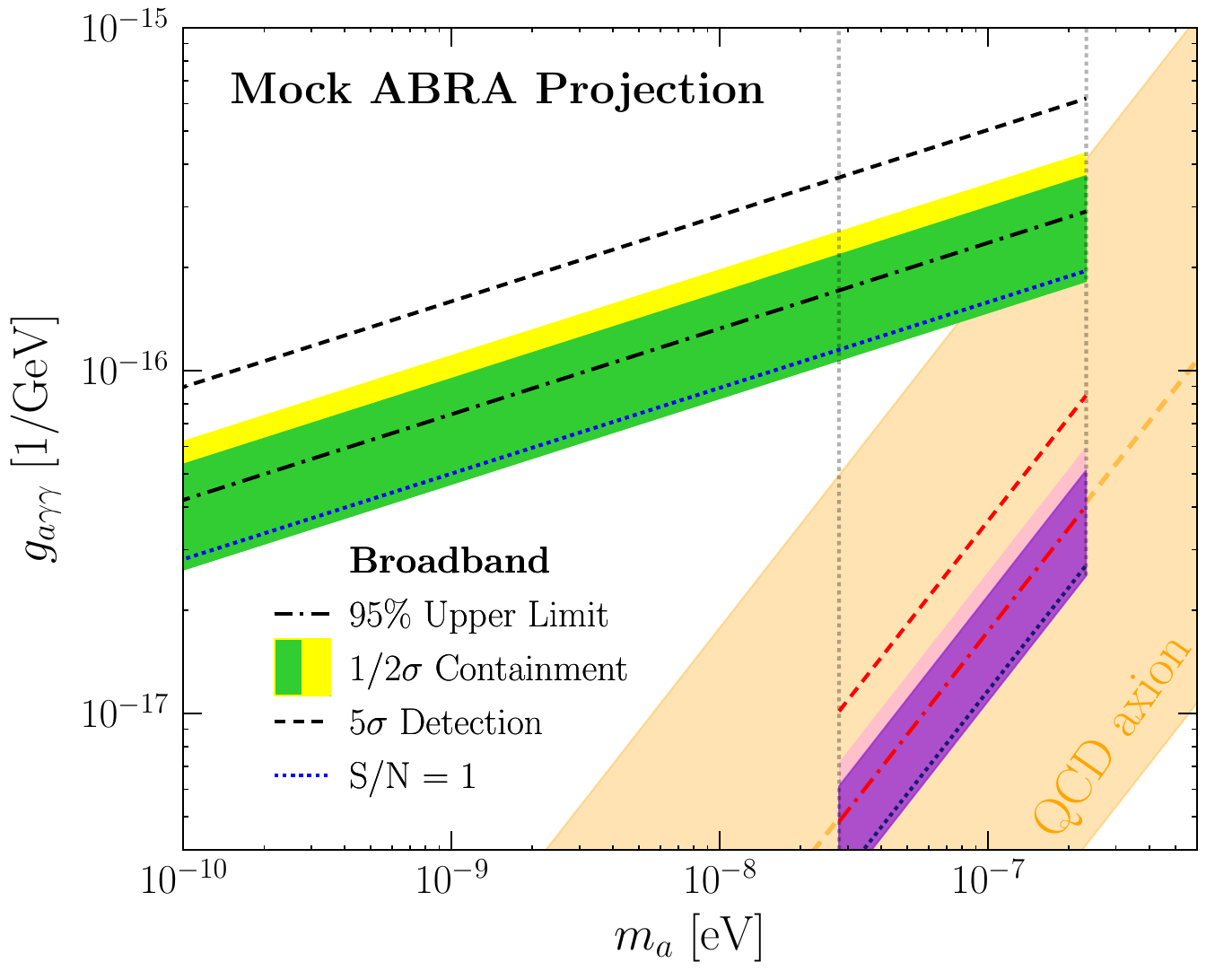} \hspace{0.2cm}
\includegraphics[scale=0.38]{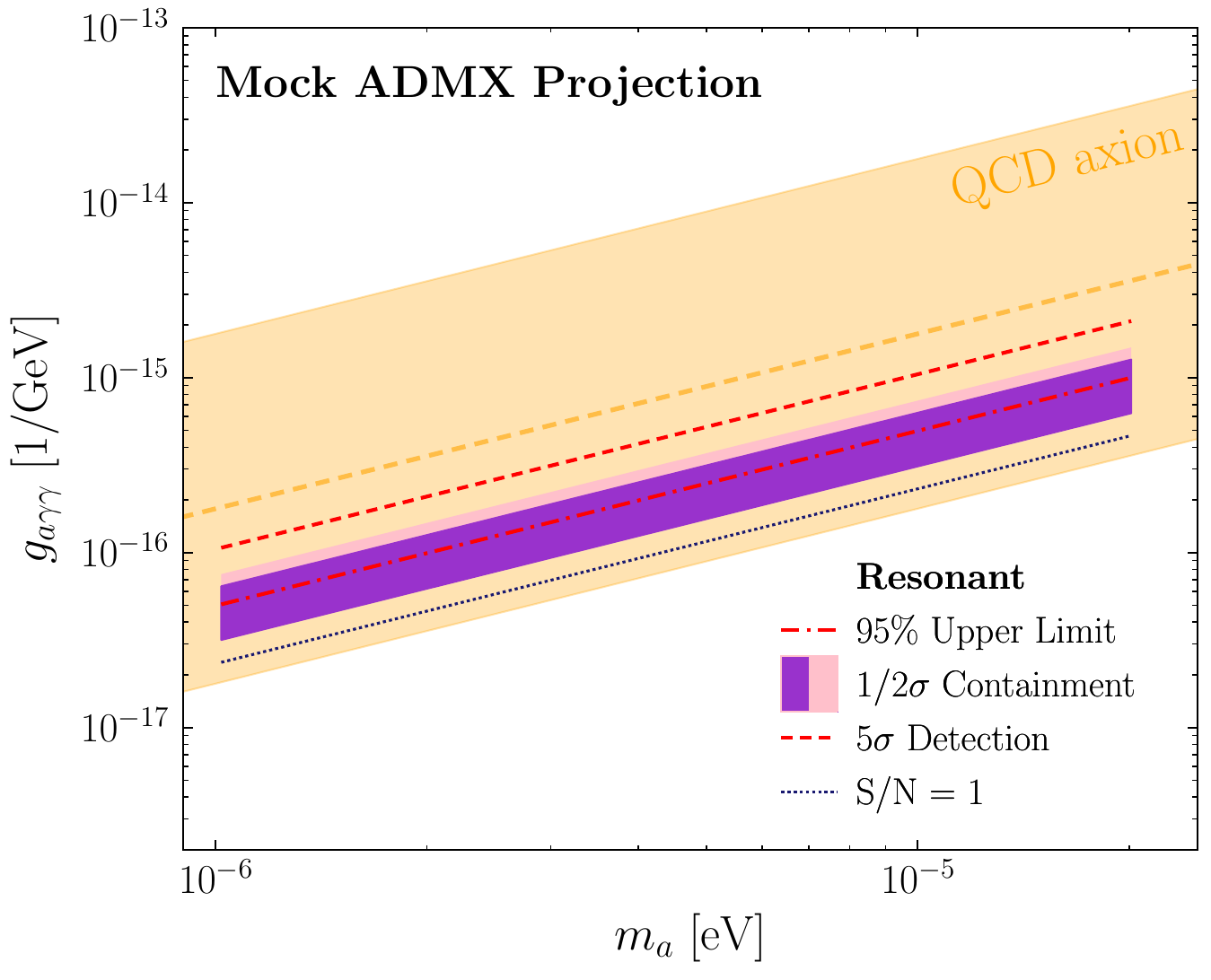}
\caption{
(Left)
A comparison of the projected sensitivities for a hypothetical version of the ABRACADABRA (ABRA) experiment~\cite{Kahn:2016aff}, with inner toroidal radius $R = 0.85$ m, an outer toroidal radius double this value, and a height $h = 4 \, R$.
A maximum magnetic field of 10 T is assumed, along with an interrogation time of 1 year.
(Right) An equivalent comparison of projections for a future ADMX experiment.
Here we take a total run time of 5 years, a volume of 500 L, quality factor of $10^5$, magnetic field of 7 T, and a system temperature of 148 mK.
In both panels the exact sensitivities are contrasted with an estimate obtained from the signal-to-noise ratio, $S/N=1$.
}
\label{fig:Sens}
\vspace{-0.4cm}
\end{figure*}

\subsection{Sensitivity}

In Fig.~\ref{fig:Sens} we illustrate the formalism introduced in Sec.~\ref{sec:sensitivity} for hypothetical future versions of the ABRACADABRA and ADMX experiments. 
To be specific, for ABRACADABRA we assumed that the radius of the pickup loop is identical to the inner radius of the torus, $R$, and also equal to the width of the torus, so that the total radius out to the outer edge of the toroid is $2 R$.
For concreteness, we took $R = 0.85$ m and then set the height of the torus to be $h= 4 \, R$.
For the remaining parameters we generally follow~\cite{Kahn:2016aff}, taking $\alpha^2 = 0.5$, pickup-loop inductance $L_p = \pi R^2/h$, SQUID inductance $L = 1$ nH, and local DM density $\rho_{\rm DM} = 0.4~{\rm GeV}/{\rm cm}^3$.
In the broadband mode we assume a flat spectrum of SQUID noise of $\sqrt{\lambda_B} = 10^{-6} \Phi_0/\sqrt{\rm Hz}$.  In the resonant mode, we take a temperature of $100$ mK and $Q_0 = 10^6$ for the RLC circuit.
Note that we cut off our projections when the Compton wavelength of the axion is equal to the inner radius of the detector.
The reason for this is that at high frequencies the magnetoquasistatic approximation used in the original analysis of~\cite{Kahn:2016aff}, which we follow, breaks down.
ABRACADABRA  is still expected to set limits in this regime, albeit weaker, however in the absence of a detailed treatment we leave this region out.\footnote{Preliminary simulations indicate that good sensitivity is likely maintained to somewhat higher frequency values.
We thank Kevin Zhou for these preliminary results.}

For ADMX, we use the projected values recently presented in~\cite{ADMXtalk}, which updated the earlier projections from~\cite{Shokair:2014rna,Stern:2016bbw}.
We take the volume $V = 500$ L, quality factor $Q = 10^5$, magnetic field $B = 7$ T, and system temperature $T_s = 148$ mK.
So far, we have not described how our analysis framework is modified for the case of ADMX.
Nevertheless, it is again a simple modification of the framework presented in Sec.~\ref{sec:likelihood}.
Starting from the power the axion field and thermal noise sources generate in the ADMX cavity, which is described in detail in a number of references, see {\it e.g.},~\cite{Daw:1998jm,Hotz:2013xaa,Hong:2014vua,McAllister:2015zcz,Lee:2016jfi,OHare:2017yze}, we find
\begin{equation}\begin{aligned}
A^{\rm ADMX} = &g_{a \gamma \gamma}^2 \frac{\rho_{\rm DM}}{m_a} Q B^2 V C_{010}\,, \\
\lambda_B^{\rm ADMX} = &k_B T_s\,,
\end{aligned}\end{equation}
where $C_{010} \approx 0.692$ is the cavity form factor for the ${\rm TM}_{010}$ mode, which dominates for the ADMX configuration.
In terms of these quantities, the mean PSD is given by 
\begin{align}
\lambda^{\rm ADMX}(\omega) = &\left( A^{\rm ADMX} \left. \frac{\pi f(v)}{m_a v} \right|_{v=\sqrt{2\omega/m_a-2}} + \lambda_B^{\rm ADMX} \right) \nonumber \\
\times &\mathcal{T}^{\rm ADMX}(\omega)\,,
\end{align}
where $\mathcal{T}^{\rm ADMX}(\omega)$ is the transfer function for the ADMX resonant cavity.
The transfer function has support over a frequency interval of width $\sim$$\omega_0 Q^{-1}$, where $\omega_0$ is the resonant frequency, in analogy to~\eqref{eq:TransferFn}.
However, the exact form of this transfer function is not important for our purposes, since it is common to the noise and signal contributions.  
In addition to computing the sensitivity of ADMX using our likelihood framework, we also derive an $S/N=1$ estimate for the sensitivity from the Dicke radiometer equation~\cite{Dicke:1946}.

In Fig.~\ref{fig:Sens}, the dashed curves represents the sensitivity for a 5$\sigma$ discovery, using the formalism derived in Sec.~\ref{sec:discoveryreach}, including the look elsewhere effect.\footnote{We caution that in the resonant case, looking for upwards fluctuations in excess of the 5$\sigma$ look elsewhere effect enhanced detection thresholds is unlikely to be the optimal discovery strategy.
Instead, one could, for example, further interrogate masses where a 2$\sigma$ upward fluctuation is observed.
For example, ADMX implements exactly such a strategy, as described in~\cite{Asztalos:2001tf}.
We make no attempt to determine the ideal resonant discovery strategy in this work.
}
We also show the median expected 95\% limit, as well as the 1 and 2$\sigma$ bands on the expectations for these quantities, derived using the procedure described in Sec.~\ref{sec:95limits}.
We reiterate that we present power-constrained limits~\cite{Cowan:2011an}, so that we do not allow ourselves to set limits stronger than the expected 1$\sigma$ downward fluctuation.
In addition we have also added the naive $S/N=1$ estimated sensitivity line for the broadband mode, as given in~\eqref{eq:ABRASN1}.
As shown in Sec.~\ref{sec:SN1}, the 95\% limit and detection threshold differ only from the naive estimate by factors of order unity.
The figure also includes the theoretically motivated region for the QCD axion in orange.

For the resonant results shown in Fig.~\ref{fig:Sens}, we adjusted the scanning strategy such that the mean limit under the null hypothesis is parallel to the QCD line in the $g_{a\gamma \gamma} - m_a$ plane.
For ABRACADABRA, we chose a minimum mass $m_a = 2.8 \times 10^{-8}$ eV and a maximum mass $m_a = 2.3 \times 10^{-7}$ eV, and the total number of bins scanned in the resonant search was $1.3 \times 10^6$.
A total scanning time of 1 year was used.
The lowest-frequency bin was scanned for $T = 704$ s, while the highest-frequency bin was scanned for $T = 0.0175$ s; the amount of time spent at the $i^\text{th}$ mass scales as $T \propto (m_a^i)^{-5}$.
Note that we have not considered the possibility of incorporating an additional broadband readout in the resonant scan to increase the sensitivity, though such an approach may be feasible.
For ADMX, we instead scanned between masses of $1.0 \times 10^{-6}$ and $20.1 \times 10^{-6}$ eV, using a total of $1.8 \times 10^6$ mass bins.
Here a total scanning time of 5 years was broken up as follows: the smallest and largest masses were scanned for 268 and 13.5 s, and now the time spent at the $i^\text{th}$ mass scales as $T \propto (m_a^i)^{-1}$.

\begin{figure}[t]
\includegraphics[scale=0.38]{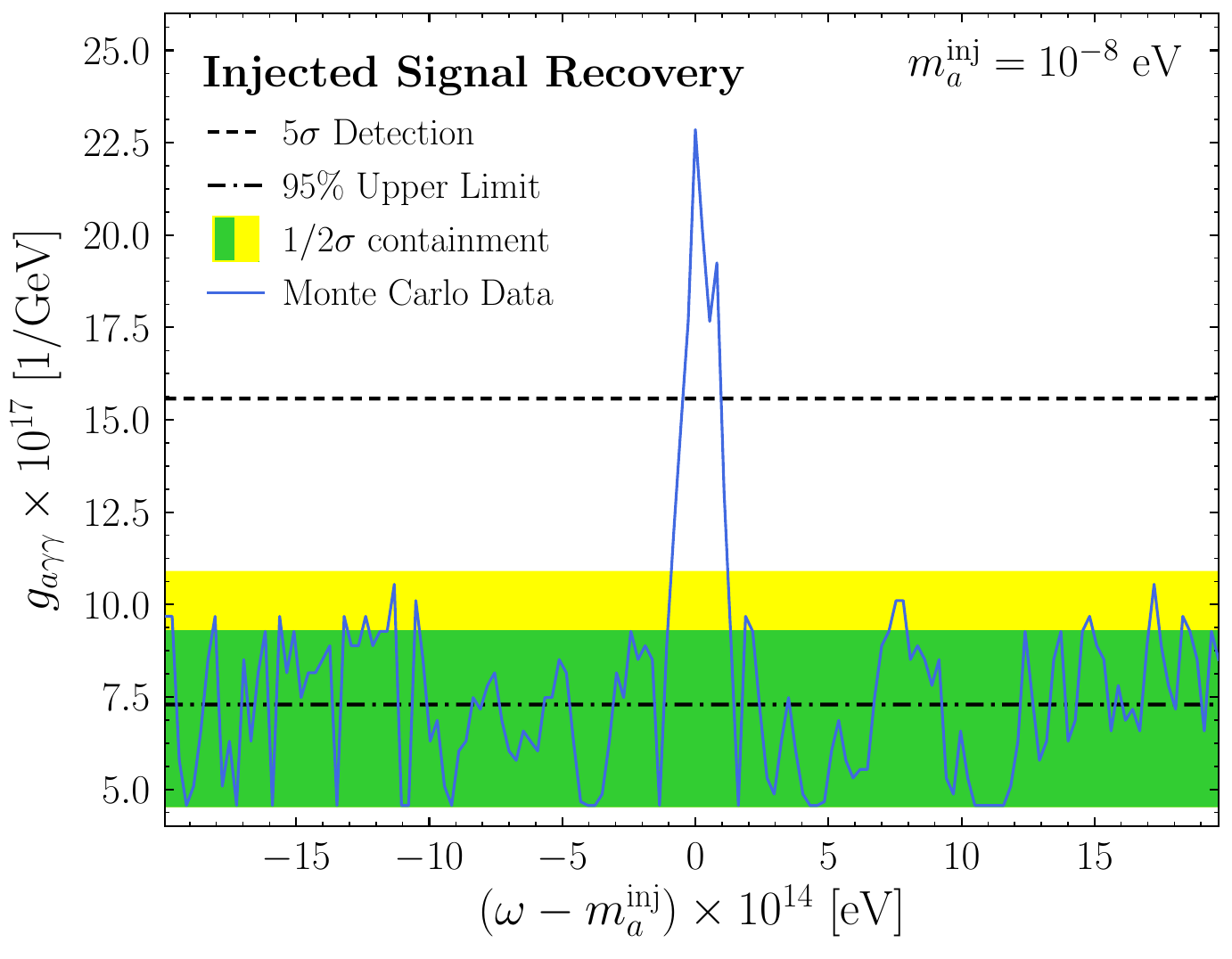}
\caption{An actual limit obtained from a single Monte Carlo simulation, with the broadband readout, compared to the various expectations for the broadband ABRACADABRA framework used in Fig.~\ref{fig:Sens}.
The data was simulated with an injected signal corresponding to $m_a = 10^{-8}$ eV and $g_{a\gamma \gamma}  = 2.21 \times 10^{-16}$ GeV$^{-1}$, and indeed we can see that right near the frequency corresponding to the axion mass, we are unable to exclude the corresponding signal strength.
}
\label{fig:Inj}
\vspace{-0.4cm}
\end{figure}

To simulate what an actual limit would look like as derived from real data, we generate Monte Carlo data for the mock broadband ABRACADABRA experimental setup under the assumption that the axion explains all of DM with $m_a = 10^{-8}$ eV and $g_{a \gamma \gamma} =  2.21 \times 10^{-16}\, \text{GeV}^{-1}$. 
Fig.~\ref{fig:Inj} shows the resulting limit in the vicinity of the true mass; the region has been magnified so that the bin to bin fluctuations can be seen.
The figure shows that in general the limit moves around between the expected bands, however right at the center, at the location of the true mass, the limit weakens considerably.

\subsection{Parameter Estimation}

In this section, we show how to estimate the DM coupling to photons and aspects of the DM phase-space distribution in the event of a detection or a detection candidate.
This is done in practice by scanning over the likelihood function with the relevant degrees of freedom given to the parameters of interest.
In this section, we show how to anticipate the uncertainties on the parameter estimates using the Asimov framework.
We proceed in an analogous fashion to previous sections, where we studied the asymptotic form of the background only distribution; in this section, we study the asymptotic form of the likelihood in the presence of a signal.

As a starting point, consider estimating the signal strength $A$ from a dataset drawn from a distribution where the true value is $A_t$.
Note that we use $A$ rather than $g_{a \gamma\gamma}$ only to simplify the expressions; the extension to the actual parameter of interest is straightforward.
Recall we have actually already shown in the previous section that the asymptotic form of $\Theta$ given in~\eqref{eq:Thetavelint} has the key property that it is maximized at the correct value of the signal strength, $A_t$.\footnote{Recall we assumed $A_{(t)} \pi f(v) / (m_a v) \ll \lambda_B$ in deriving that expression.}
We can determine the uncertainty on the estimated $A$ from the curvature around the maximum.
In detail,
\begin{equation}
\sigma_A^{-2} = - \frac{1}{2} \partial_A^2 \Tilde{\Theta}(A) \vert_{A=A_t} = \frac{T\pi}{2m_a} \int \frac{dv}{v} \frac{f(v)^2}{\lambda_B^2}\,,
\label{eq:SigmaA}
\end{equation}
where $\sigma_A$ is the expected uncertainty on the measurement.
Using the SHM velocity distribution, this simplifies to
\begin{equation}
\sigma_A = \sqrt{\frac{2\sqrt{2}\,m_a\,\lambda_B^2\,v_0 v_{\rm obs}}{T \sqrt{\pi}\,{\rm erf} \left[ \sqrt{2} v_{\rm obs}/v_0 \right]}} = \frac{A_t}{\sqrt{\rm TS}} \,.
\end{equation}
From this we can see that, as expected, the uncertainty on the signal strength increases with the background, decreases with a longer experimental run time, and scales inversely proportional to the square root of the TS for detection.
The last point is important because it says that the central value $A_t$ is $\sqrt{\text{TS}}$ standard deviations away from zero, which matches our interpretation of $\sqrt{\text{TS}}$ as the significance.

We can readily extend this strategy to the estimation of other signal parameters.
For example, we can use this to estimate the best fit SHM parameters, $v_0$ and $v_{\rm obs}$, and their associated uncertainties.
Let us denote by $f_t(v) = f_{\rm SHM}(v \vert v_0^t,v_{\rm obs}^t)$ the speed distribution given by the true SHM parameters, and then $f(v) = f_{\rm SHM}(v \vert v_0,v_{\rm obs}^t)$ represents the distribution for some arbitrary value of $v_0$.
To repeat the Asimov analysis, we now use the dataset and model predictions given by
\begin{equation}\begin{aligned}
S_{\Phi\Phi}^{k,{\rm Asimov}} \equiv \lambda_k^t =& A_t \frac{\pi f_t(v)}{m_a v} + \lambda_B\,, \\
\lambda_k =& A_t \frac{\pi f(v)}{m_a v} + \lambda_B\,,
\end{aligned}\end{equation}
respectively.
Then, through the same process as above we arrive at
\begin{equation}\begin{aligned}
\tilde{\Theta}(v_0) = \frac{A_t^2T\pi}{m_a} \int \frac{dv}{v} \frac{f(v)}{\lambda_B^2} \left( f_t(v) - \frac{f(v)}{2} \right)\,.
\end{aligned}\end{equation}
Again this asymptotic expression satisfies the central Asimov requirement that
\begin{equation}
\max_{v_0} \tilde{\Theta}(v_0) = \tilde{\Theta}(v_0^t)\,.
\end{equation}
Beyond this, however, we can again estimate the uncertainty on the best fit velocity dispersion:
\begin{equation}\begin{aligned}
\sigma_{v_0}^{-2} = &- \frac{1}{2} \partial_{v_0}^2 \Tilde{\Theta}(v_0) \vert_{v_0=v_0^t} \\
= &\frac{A_t^2 T\pi}{2m_a} \int \frac{dv}{v} \frac{\left( \partial_{v_0} f(v) \vert_{v_0=v_0^t} \right)^2}{\lambda_B^2}\,,
\end{aligned}\end{equation}
so that if we assume $\lambda_B$ is independent of frequency, we have
\begin{align}
\sigma_{v_0} = &\frac{v_0^t}{\sqrt{\rm TS}} \left( \frac{3}{4} + \frac{v_{\rm obs}^t \left( 9 v_0^{t2} - 4 v_{\rm obs}^{t2} \right) e^{-2v_{\rm obs}^{t2}/v_0^{t2}}}{\sqrt{2\pi} v_0^{t3}\,{\rm erf} \left[ \sqrt{2} v_{\rm obs}^t/v_0^t \right]} \right)^{-1/2} \nonumber \\
\approx &1.02 \frac{v_0^t}{\sqrt{\rm TS}}\,.
\end{align}
Above, we have taken the SHM values for the approximate result.
Applying the same strategy for $v_{\rm obs}$, we would find the maximum is again obtained at the true value, with the uncertainty now given by
\begin{align}
\sigma_{v_{\rm obs}} = &\frac{v_0^t}{\sqrt{\rm TS}} \left( 1 - \frac{4 v_{\rm obs}^t e^{-2v_{\rm obs}^{t2}/v_0^{t2}}}{\sqrt{2\pi} v_0^t\,{\rm erf} \left[ \sqrt{2} v_{\rm obs}^t/v_0^t \right]} \right)^{-1/2} \nonumber \\
\approx & 1.11 \frac{v_0^t}{\sqrt{\rm TS}}\,.
\end{align}

From these three results for parameter estimation using our likelihood we can see that in general if we are estimating a parameter $\alpha_t$, the estimated mean value will be $\mu_{\alpha} = \alpha_t$, and the uncertainty tends to scale as $\sigma_{\alpha} \sim {\rm TS}^{-1/2}$.
Thus exactly as expected, the more significant the detection of axion, or specifically the larger the TS, the greater precision with which we can estimate parameters.

\section{Impact of a Realistic and Time-Varying DM Distribution}\label{sec:DMDistribution}

In the previous sections, we have developed a framework for the analysis of a signal sourced by axion DM drawn from the SHM distribution $f_{\rm SHM}(v \vert v_0, v_{\rm obs})$. 
However, this neglects a number of effects that modify the DM speed distribution; in particular: annual modulation, gravitational focusing, and the possible presence of local velocity substructure. 
As we have verified by Monte Carlo simulations, the exclusion of these features from our analysis has a negligible effect on our ability to successfully constrain or discover an axion signal in our data, even when features excluded from the analysis are included in the data sets.
Consequently, the framework of Sec.~\ref{sec:BulkHalo} is sufficient for the first stage of the data analysis.
Nonetheless, since we do expect these effects to be manifest in a hypothetically discovered signal, they present opportunities to gain sharper insight on the local DM distribution.
Moreover, because annual modulation and gravitational focusing result in distinct signatures expected to be present only in the presence of a genuine axion signal, the identification of these features would further strengthen any candidate detection.
In addition, if we are within a cold stream or debris flow, a significant enhancement to the signal is possible.
In this section, we specify the details of annual modulation, gravitational focusing, and velocity substructure and their inclusion in the DM speed distribution.

Because the signatures of annual modulation, gravitational focusing, and velocity substructure are necessarily time-dependent, we are forced to promote our likelihood to incorporate variation in time.\footnote{Cold velocity substructure is more subject to annual and daily modulation, which is why these effects are time-dependent in the Earth frame even if they are not in the Solar frame.}
To do so, we will make use of the stacking procedure described in Sec.~\ref{sec:sensitivity}.
We assume that the full dataset is broken into $N_T$ subintervals of duration $\Delta T = T/N_T$ containing $\Delta N = N/N_T$ PSD measurements.
Now, however, we will assume that $\Delta T$ is sufficiently small that the speed distribution does not change appreciably within a given interval.
As the distribution will change over the full collection time $T$, we have a different model prediction in each time interval given by:
\begin{equation}
\lambda_{k,\ell} = A \frac{\pi f(v, t_{\ell})}{m_a v} + \lambda_B\,,
\end{equation}
which leads to the following modified likelihood
\begin{equation}
\mathcal{L}(d \vert \boldsymbol{\theta}) = \prod_{\ell=0}^{N_T-1} \prod_{k=1}^{\Delta N-1} \frac{1}{\lambda_{k,\ell}(\boldsymbol{\theta})} e^{-S_{\Phi\Phi}^{k,\ell}/\lambda_{k,\ell}(\boldsymbol{\theta})}\,.
\end{equation}
This is the form of the likelihood we will use throughout this section.
Note that the $\ell$ dependence on the model prediction invalidates the stacking analysis performed in Sec.~\ref{sec:sensitivity}, though the data may still be stacked over time intervals that are sufficiently smaller than a year (day) for annual (daily) modulation.

\subsection{Halo Annual Modulation}

\begin{figure*}[htb]
\includegraphics[scale=0.45]{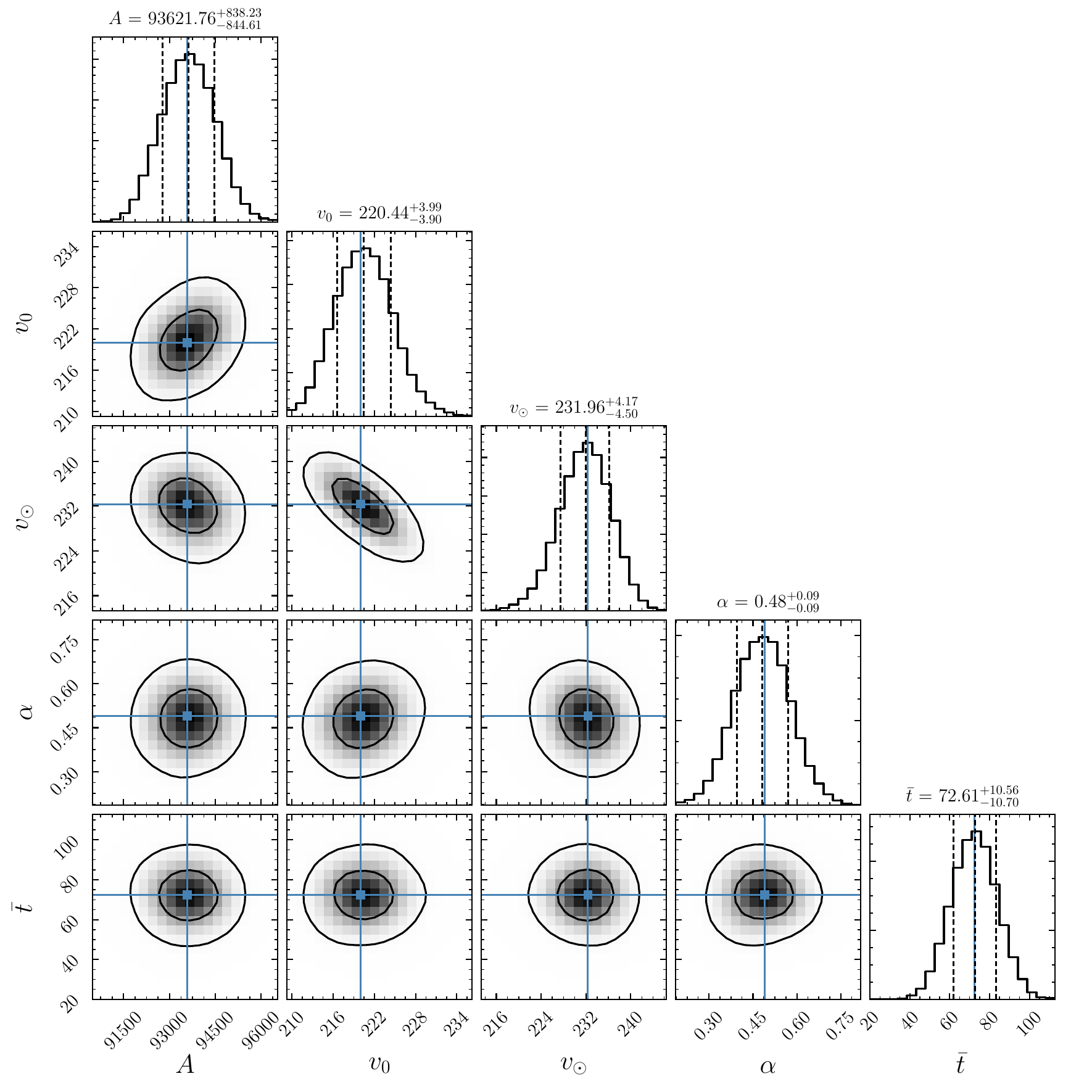} 
\caption{
The posterior distribution for a model with annual modulation where the signal strength is at the threshold of annual modulation detection at 5$\sigma$.
The true parameter values are indicated in blue, with the $1\sigma$ confidence intervals on the parameter estimations indicated by the dashed black lines in the one parameter posteriors.
The two parameter posteriors show the 1 and 2$\sigma$ contours.
The axion mass, $m_a$, was also scanned over, and is recovered accurately but not shown here.
Note that this example uses the Asimov dataset.
All times are measured in days and velocities in km$/$s, while the units of $A$ are arbitrary. 
}
\label{fig:AnnualMod}
\vspace{-0.4cm}
\end{figure*}

Before studying how annual modulation impacts the expected axion signal, we first review how it modifies the DM speed distribution.\footnote{We refer to~\cite{Lee:2013xxa} for a comprehensive review of these details.}
Our starting point for this is the SHM distribution given in~\eqref{eq:SHM}.
Throughout the year the detector's speed in the Galactic halo frame, $v_{\rm obs}$, is expected to oscillate as the Earth orbits the Sun.
In the lab frame, this results in an effectively time-dependent halo distribution $f_{\rm SHM}(v, t)$.
All of the time dependence, neglecting that from gravitational focusing, which will be dealt with separately, can be accounted for by upgrading the relative detector-halo speed to a time-dependent parameter $v_{\rm obs}(t)$.
To determine this speed, first note that $\mathbf{v}_{\rm obs}(t) = \mathbf{v}_{\odot} + \mathbf{v}_{\oplus}(t)$, where $\mathbf{v}_{\odot}$ and $\mathbf{v}_{\oplus}(t)$ are the velocity of the Sun with respect to the Galactic frame and the velocity of the Earth with respect to the Sun, respectively.
These are specified by\footnote{Corrections to $\mathbf{v}_\oplus(t)$ are suppressed by the eccentricity of the Earth's orbit, given by $e \approx 0.016722$, and so can safely be neglected.}
\begin{equation}\begin{aligned}
\mathbf{v}_{\odot} &= v_{\odot} (0.0473, 0.9984, 0.0301)\,, \\
\mathbf{v}_\oplus(t) &\approx v_{\oplus} \left( \cos \left[\omega(t-t_1)\right] \hat{\boldsymbol{\epsilon}}_1 + \sin \left[\omega(t-t_1)\right] \hat{\boldsymbol{\epsilon}}_2 \right)\,,
\end{aligned}\end{equation}
where the magnitudes are given by $v_{\odot} \approx 232.37 $ km/s and $v_{\oplus} \approx 29.79$ km/s.
We have further introduced \mbox{$\omega \approx 2 \pi / (365 \, \, \text{days})$} as the period of the Earth's revolution, $t_1$ as the time of the vernal equinox (which occurred on March 20 in 2017), and the unit vectors $\hat{\boldsymbol{\epsilon}}_1$ and $\hat{\boldsymbol{\epsilon}}_2$ specifying the ecliptic plane.
These vectors are given in Galactic coordinates by
\begin{equation}\begin{aligned}
\hat{\boldsymbol{\epsilon}}_1 &\approx (0.9940, 0.1095, 0.0031)\,, \\
\hat{\boldsymbol{\epsilon}}_2 &\approx (-0.0517, 0.4945, -0.8677)\,.
\end{aligned}\end{equation}
We may then find the time-varying Galactic-frame speed
\begin{equation}
v_{\rm obs}(t) = \sqrt{v_{\odot}^2 + v_{\oplus}^2 + 2 v_{\odot} v_{\oplus} \alpha \cos \left[ \omega(t-\bar{t}) \right]}\,,
\end{equation}
given in terms of the parameters
\begin{equation}\begin{aligned} \label{alpha_t}
\alpha &\equiv \sqrt{(\hat{\mathbf{v}}_{\odot} \cdot \hat{\boldsymbol{\epsilon}}_1)^2 + (\hat{\mathbf{v}}_{\odot} \cdot \hat{\boldsymbol{\epsilon}}_2)^2} \approx 0.491\,, \\
\bar{t} &\equiv t_1 + \frac{1}{\omega} \arctan\left(\frac{\hat{\mathbf{v}}_{\odot} \cdot \hat{\boldsymbol{\epsilon}}_2}{\hat{\mathbf{v}}_{\odot} \cdot \hat{\boldsymbol{\epsilon}}_1}\right)\approx t_1 + 72.5~{\rm days}\,.
\end{aligned}\end{equation}
Whilst we have given the accepted values for the various parameters above, if a definitive axion signal was detected we could then take for example $v_{\odot}$, $\alpha$, and $\bar{t}$ as unknown parameters to be estimated from the likelihood.
Their agreement with the accepted values would be a highly non-trivial test of the signal.
We will show an example of this below, but before doing so we use the Asimov formalism to estimate how significant a signal we would need to detect annual modulation from the bulk halo.

\begin{figure*}[htb]
\includegraphics[scale=0.45]{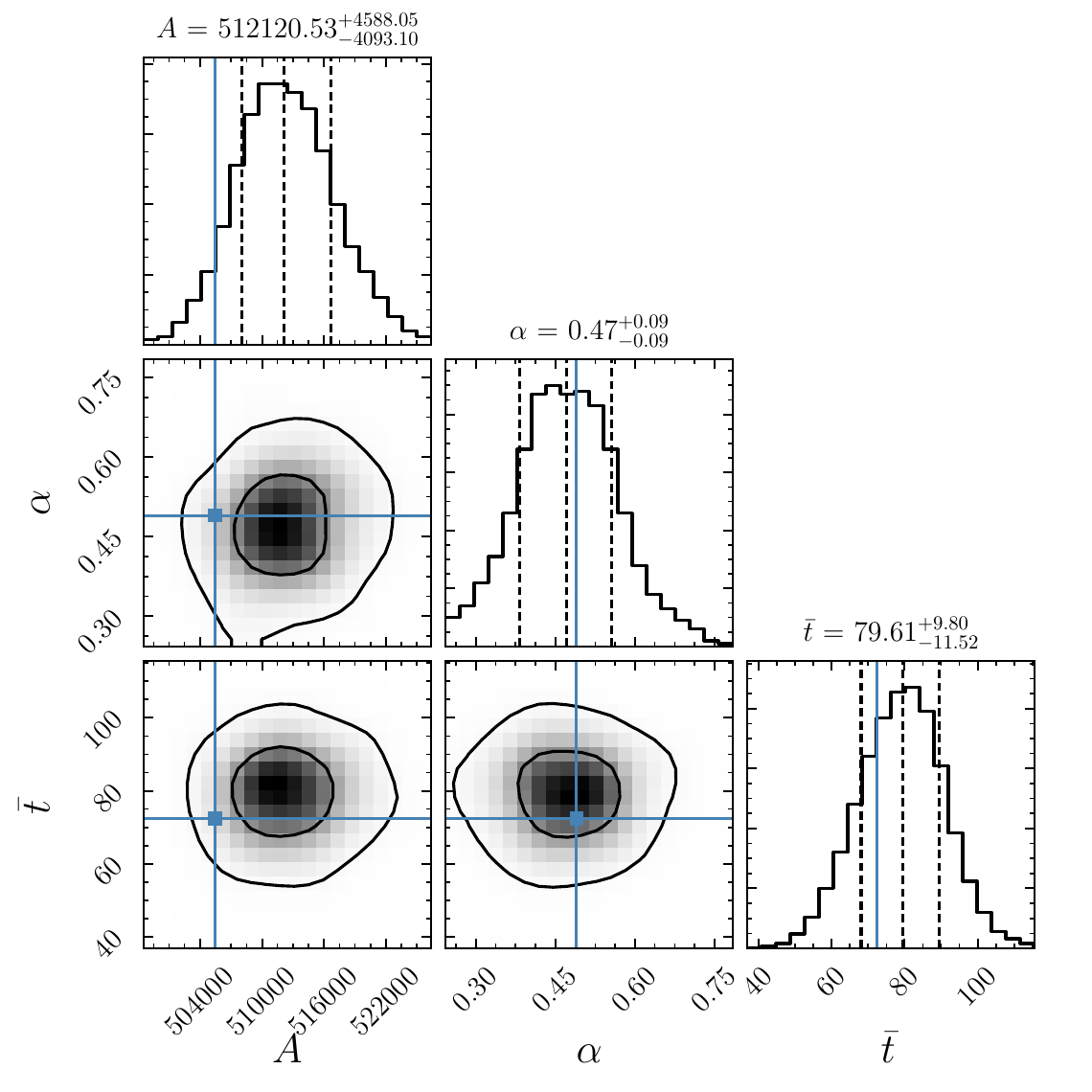}  
\includegraphics[scale=0.45]{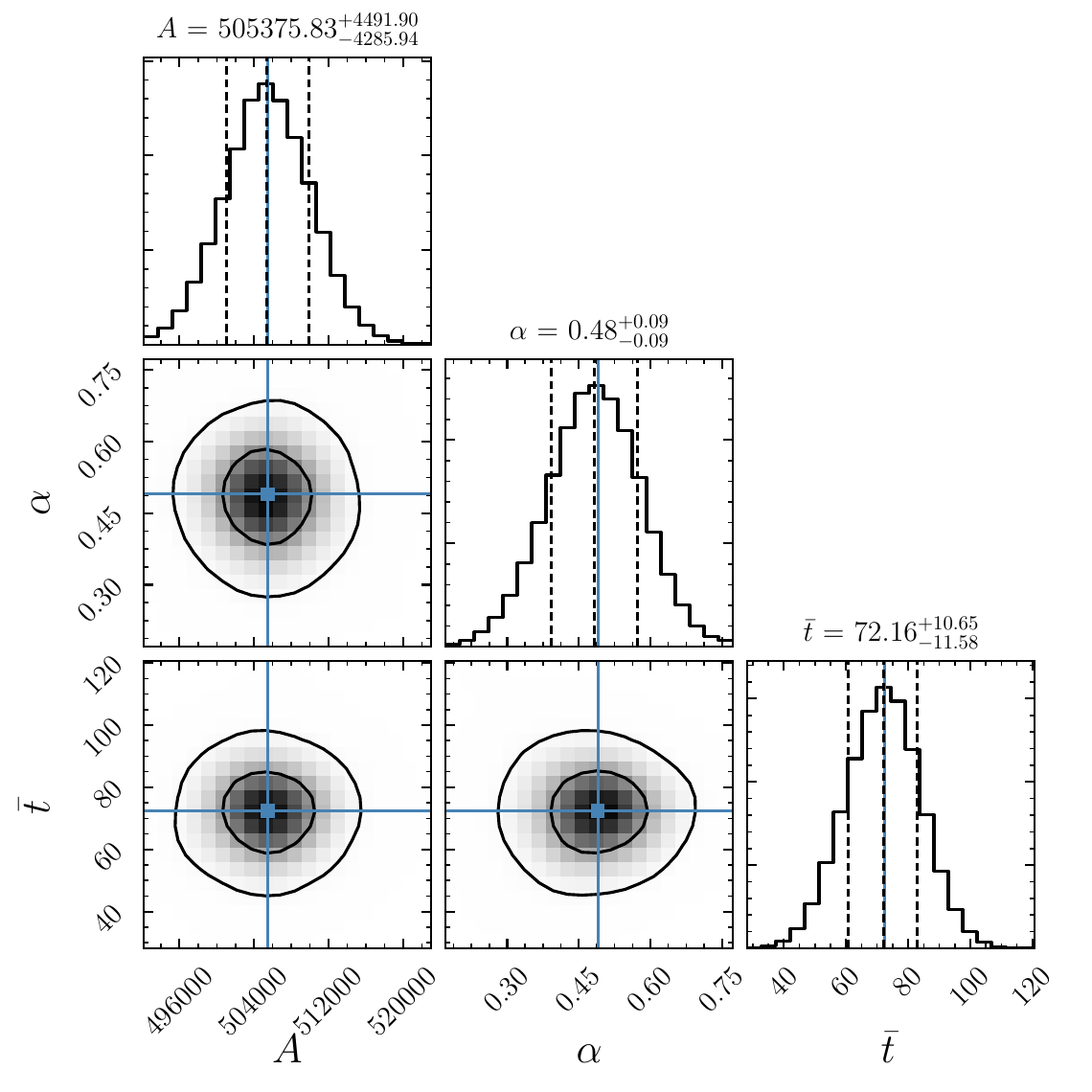}
\caption{
As in Fig.~\ref{fig:AnnualMod}, but this time the data includes gravitational focusing and the model only includes the parameters $A$, $\alpha$ and $\bar t$.
(Left) Gravitational focusing, while present in the Asimov data, is excluded from the model template.
The estimations of $A$ and $\bar t$ are off at the $\sim$$2 \sigma$ and $\sim$$1 \sigma$ levels, respectively.
(Right) As in the left panel but including gravitational focusing in the model template.
As expected, the parameter estimation is quite accurate in this case.
}
\label{fig:GFCornerPlots}
\vspace{-0.4cm}
\end{figure*}

Ignoring annual modulation, the detection significance of a SHM signal scales with the parameters of interest as
\begin{equation}\begin{aligned}
{\rm TS} &= \frac{A^2T\pi}{2 m_a \lambda_B^2} \frac{{\rm erf} \left[ \sqrt{2} v_{\rm obs}/v_0\right]}{\sqrt{2\pi}v_0 v_{\rm obs}}\,,
\end{aligned}\end{equation}
where here and throughout this section we assume the background is frequency independent over the width of the signal.
The relevant question is, on average, at what value of TS do we detect annual modulation at a given significance?
To estimate this, we calculate the test statistics between models with and without annual modulation included; in order to discover annual modulation we can think of the model without it included as the null hypothesis.
We denote this test statistic by TS$_\text{a.m.}$.  
We can estimate the median value for TS$_\text{a.m.}$ as a function of the model parameters using the asymptotic form of $\Theta$ and the Asimov formalism; in this case, the Asimov dataset includes annual modulation.  
Specifically, we find
\begin{equation}\begin{aligned}
{\rm TS}_\text{a.m.} = \frac{A^2T\pi}{m_a \lambda_B^2} &\int {dv \over v} \left[ f_t(v)^2    \vphantom{- f(v) \left( f_t(v) - \frac{f(v)}{2} \right)}\right.  \\
& \left. - f(v) \left( f_t(v) - \frac{f(v)}{2} \right)  \right] \,.
\end{aligned}\end{equation}
Above, $f_t$ features annual modulation while $f$ does not.
In order to simplify the calculation, we define an expansion parameter:
\begin{equation}
\epsilon \equiv \frac{v_{\odot} v_{\oplus}}{v_{\odot}^2 + v_{\oplus}^2} \approx 0.126\,,
\end{equation}
in terms of which we can write:
\begin{equation}
v_{\rm obs}(t) \approx v_{\rm obs} \left( 1 + \epsilon \alpha \cos \left[ \omega(t-\bar{t}) \right] \right)\,,
\end{equation}
with $v_{\rm obs} \approx 232$ km/s.
Using this and averaging all time dependence over one period in the final result, we calculate the ratio of TS$_\text{a.m.}$ to TS in the SHM as
\begin{align}
{ \text{TS}_\text{a.m.} \over \text{TS} } = \,& \frac{\alpha^2 \epsilon^2 v_{\rm obs}^2}{2v_0^2} \left( 1 - \frac{4 v_{\rm obs} e^{-2v_{\rm obs}^2/v_0^2}}{\sqrt{2\pi} v_0 {\rm erf} \left[ \sqrt{2} v_{\rm obs}/v_0\right]} \right) \nonumber\\
\approx\,& 0.00173\,.
\end{align}

From the discussion above, we see that if it took a time $T$ to detect the axion at a given significance, it would take a time $580 T$ to detect annual modulation at the same significance.
Alternatively, as the test statistic scales like $g_{\alpha \gamma \gamma}^4$, the coupling for the threshold of discovery for annual modulation will be $\sim$$5$ times larger, on average, than the coupling for the threshold of discovery of a signal.
On the other hand, in the resonant setup large increases in the TS are readily obtainable since after the axion mass is known we can stay at the correct frequency for an extended period instead of scanning over multiple frequencies.

In Fig.~\ref{fig:AnnualMod} we show the posterior distribution generated in a Bayesian framework from an analysis of the Asimov dataset with $g_{a\gamma \gamma}$ at the threshold for detection of annual modulation at 5$\sigma$.
Note that we float $A$, $m_a$, $v_0$, $v_\odot$, $\alpha$, and $\bar t$ as model parameters with linear-flat priors in the fit.
All model parameters are seen to be well converged, including $m_a$ which is not shown in the figure.
This analysis was performed using \texttt{Multinest}~\cite{Feroz:2008xx,Buchner:2014nha} with 500 live points.
The Asimov results are consistent with those found from an ensemble of simulated datasets, as expected.

\subsection{Halo Gravitational Focusing}

An additional source of annual modulation in the axion signal is sourced by the focusing of the axion flux by the Sun's gravitational potential.
This effect is already known to have a significant impact on annual modulation in the context of WIMP direct detection, as pointed out in~\cite{Lee:2013wza}.
The intuition behind gravitational focusing is that in the frame of the Sun the DM velocity distribution appears as a wind.
The gravitational field of the Sun focuses the DM ``down-wind" of the Sun, leading to an enhanced rate when the Earth is ``down-wind" relative to when the Earth is ``up-wind."  
Here we investigate the impact of gravitational focusing on the corresponding axion signal.

In~\cite{Lee:2013wza} an exact closed-form expression was used to model the perturbation to the DM phase-space distribution from the Sun's potential.
The perturbed phase-space distribution is derived using Liouville's theorem and exactly solving for the trajectories of the DM particles in the gravitational field.
However, in this work we take advantage of a perturbative result (to leading order in Newton's constant), valid when the DM speeds are much larger than the Solar escape velocity, that allows us to write~\cite{Buschmann:2017ams}
\begin{equation}
f(v, t) = f_{\rm halo}(v, t) + f_{\rm GF}(v, t)\,,
\end{equation}
where $f_{\rm halo}(v, t)$ is the unperturbed velocity distribution in the Earth frame, and where the perturbation by gravitational focusing $f_\text{GF}$ is given by 
\begin{align}
f_\text{GF}(v, t) &\equiv -\frac{2 G M_\odot}{x_\oplus(t)} \int  \frac{v^2 d\Omega}{\pi ^{\frac{3}{2}} v_0 ^ 5} \frac{e^{-(\mathbf{v} + \mathbf{v}_\oplus(t) + \mathbf{v}_\odot)^2/v_0^2}}{v} \label{eq:GFDef}
\\ & \times \frac{(\mathbf v + \mathbf v_\oplus (t) + \mathbf v_\odot) \cdot \left(\hat{\mathbf{x}}_\oplus (t) - \frac{\mathbf{v} + \mathbf{v}_\oplus(t)}{|  \mathbf{v} + \mathbf{v}_\oplus(t)|}\right)}{1 -\hat{\mathbf{x}}_\oplus (t) \cdot \left(\frac{\mathbf{v} + \mathbf{v}_\oplus(t)}{|  \mathbf{v} + \mathbf{v}_\oplus(t)|}\right)}\,. \nonumber
\end{align}
Note that in this equation, $v^2 d\Omega$ is written out explicitly to account for the measure.
Here, $\mathbf{x}_\oplus(t)$ denotes the position of the Earth in the Solar frame; an explicit form for this in Galactic coordinates can be found in~\cite{Lee:2013xxa}.
Note that $f(v,t)$ is no longer normalized to integrate to unity, but rather the change in $\int dv f(v,t)$ throughout the year indicates the fractional change in the DM density do to gravitational focusing.
We have explicitly verified that the perturbative formalism for gravitational focusing is a good approximation to the exact formalism used in~\cite{Lee:2013wza} for the SHM.

To determine the impact of gravitational focusing, we perform two analyses using the Asimov dataset at the 5$\sigma$ detection threshold for annual modulation but this time including gravitational focusing.
We analyze the Asimov data in the Bayesian framework including with two models; the first model does not account for gravitational focusing, while the second one does.
The results of these analyses are shown in Fig.~\ref{fig:GFCornerPlots}.
The use of a limited number of live points is the most likely source of the residual disagreement between the injected and median value of $\bar{t}$ in the right panel.
Note that in these analyses we only float $A$, $\alpha$, and $\bar t$ for simplicity.
Neglecting gravitational focusing in the model (left panel) only leads to a approximately $2\sigma$ overestimate in the value of the $A$ parameter, while the central value of $\bar t$ is on average off by $\sim$10 days.
On the other hand, when gravitational focusing is included in the model (right panel), the halo parameters and the normalization are correctly inferred.

\subsection{Local DM Substructure}

So far, we have only considered an axion signal sourced by dark matter contained within the bulk halo, but there additionally exist a number of well-motivated classes of velocity substructure that have the potential to leave dramatic signatures in the direct detection data.
One large class of substructure relates to the DM subhalos that are expected to be present in the Milky Way~\cite{Maciejewski:2010gz}.
DM subhalos are believe to persist down to very small mass scales, potentially $\sim$$10^{-6}$ $M_\odot$ and below, due to the nearly scale-invariant spectrum of density perturbations generated during inflation.
Low-mass DM subhalos have low velocity dispersions, and so if we happen to be sitting in a DM subhalo, even if it only makes up a small fraction of the local DM density, it could show up as a narrow spike in velocity space over the bulk SHM contribution.
Even if we are not directly in a bound DM subhalo, we could still be affected by the tidally stripped debris that in-falling subhalos leave throughout the Galaxy.
There are two types of tidally-stripped substructure, in velocity space, that are important for direct detection (for a review of the importance of tidal debris at WIMP experiments, see~\cite{Freese:2012xd}): DM streams and debris flows.

As an in-falling subhalo descends through the potential of the Milky Way, the outer regions of the DM subhalo are expected to become tidally stripped and form an ultra-cold trailing stream~\cite{Vogelsberger:2008qb,Maciejewski:2010gz}.
Such streams should trail from DM subhalos of all sizes, with smaller subhalos having colder streams.
Eventually, the tidal debris dragged away from in-falling subhalos will become fully virialized.
However, before that occurs the debris becomes homogeneously distributed in position space but remains coherent in velocity space, forming the substructure known as debris flow~\cite{Lisanti:2011as}.
While it is unlikely that a DM substructure from in-falling subhalos dominates the local DM density~\cite{Vogelsberger:2008qb,Maciejewski:2010gz}, as we show in this subsection, even if the substructure only makes up a small fraction of the local DM density, due to the coherence in velocity space the signature of substructure at axion experiments can be substantial and even dominate over the SHM contribution.
This can be contrasted to the case in WIMP direct detection experiments, where substructure is expected to play an important role in annual modulation studies but not necessarily have a significant impact on the total rate~\cite{Kamionkowski:2008vw,Freese:2012xd,Lee:2013xxa}.
DM streams were recently considered in the context of axion direct detection in~\cite{OHare:2017yze}.

\begin{figure}[t]
\includegraphics[scale=0.38]{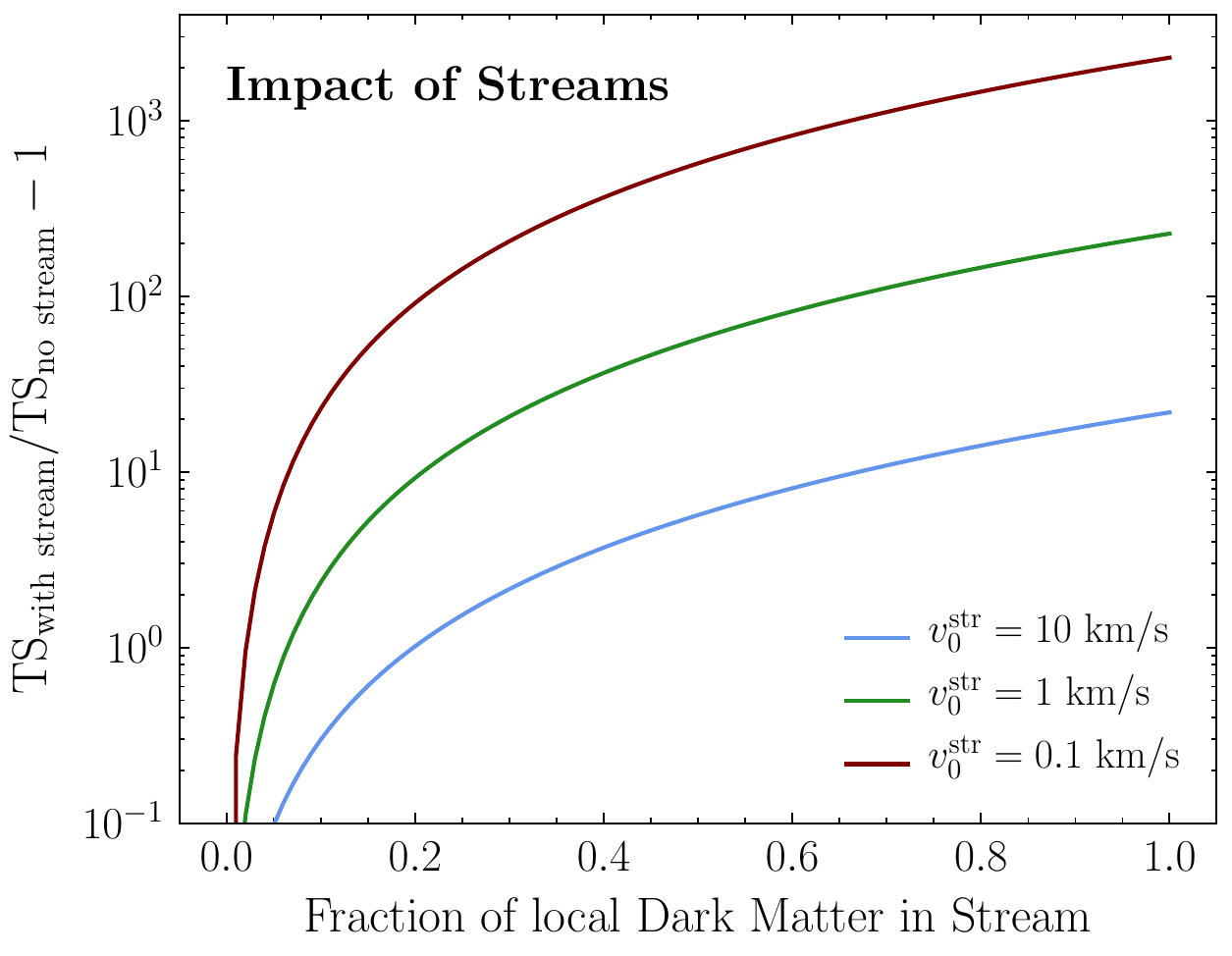}
\caption{
The enhancement expected in the TS in the presence of a coherent DM stream, as given in~\eqref{eq:TSstream}.
The TS is shown as a ratio with respect to the case where only the bulk halo is present and as a function of the fraction of the local DM within the substructure.
}
\label{fig:StreamImpact}
\vspace{-0.4cm}
\end{figure}

\begin{figure*}[htb]
\includegraphics[width=0.47\textwidth]{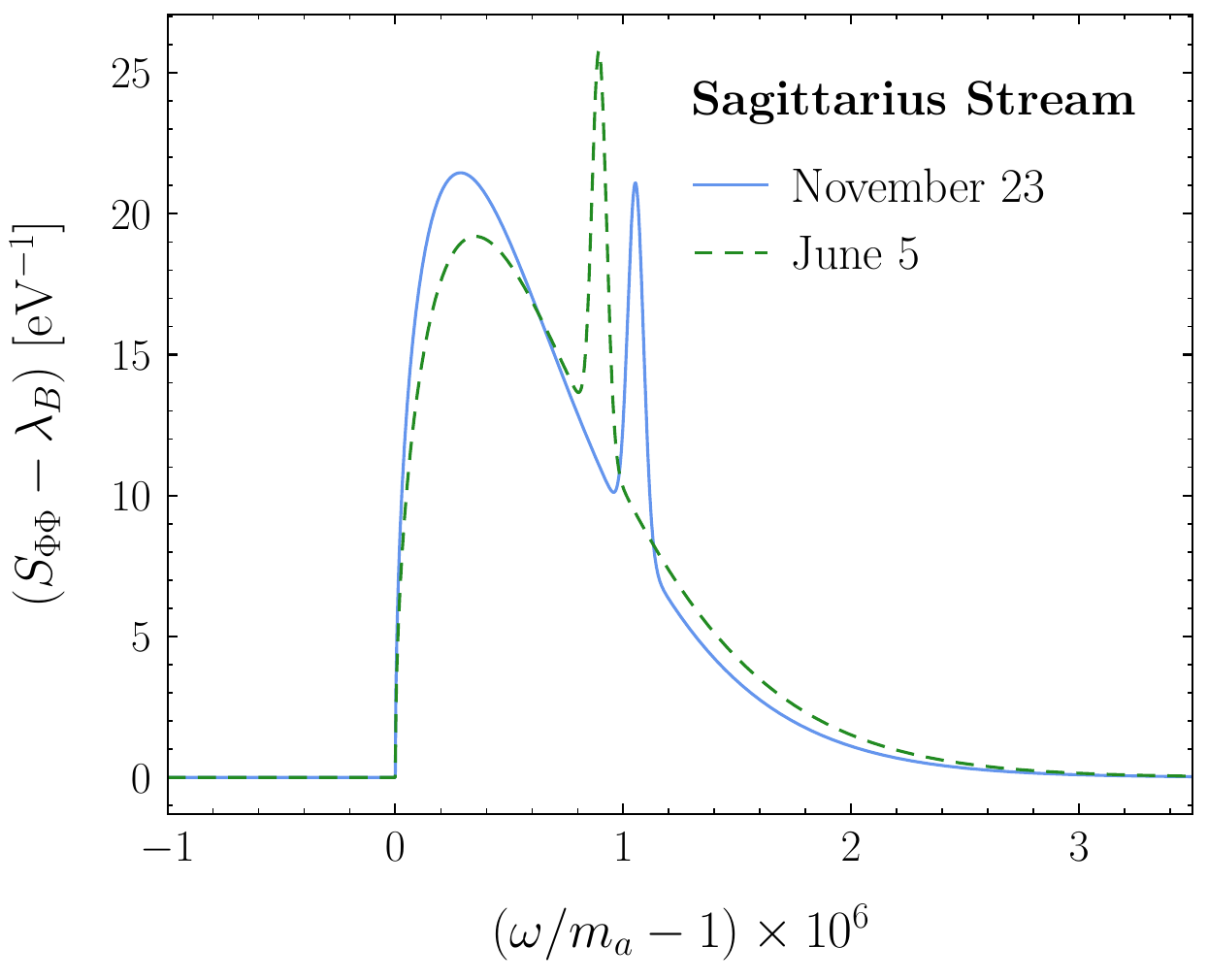} \hspace{0.2cm}
\includegraphics[width=0.482\textwidth]{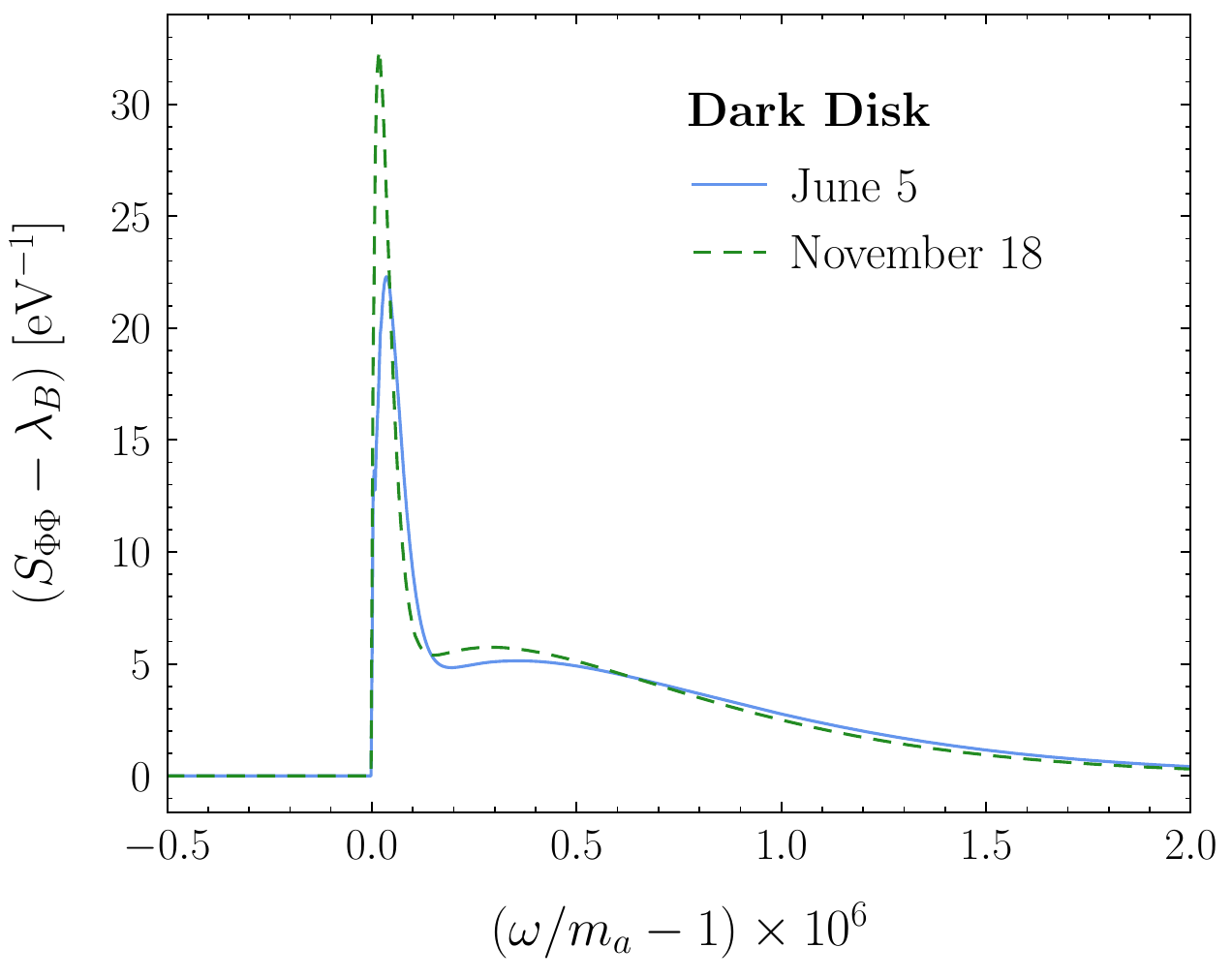} 
\caption{
The axion contribution to the PSD as a function of frequency in the presence of DM substructure.
(Left) We show the effect of a Sagittarius-like stream that makes up $\sim$5\% of the local DM density at two different times of year, corresponding to the dates of maximum TS (June 5) and minimum TS (November 23), where all dates are for 2017.
Annual modulation plays an important role for cold substructure because the Earth's orbital velocity may be larger than the substructure velocity dispersion.
(Right)  As in the left panel, but for a dark disk that makes up $\sim$20\% of the local DM density.
The dark-disk is co-rotating with the baryonic disk, with a lag speed $\sim$50 km$/$s, and so the contribution to the PSD is at lower speeds compared to the stream case.
Gravitational focusing also plays an important role for the disk since the solar-frame velocities are relatively low.
In this case the maximum and minimum TS occur on November 18 and June 5 respectively.
For both of these panels, the signal is generated using $m_a = 1$ MHz, $A$ set to the value for the threshold for detection of the SHM, and $\lambda_B$ set to the minimum SQUID noise.
}
\label{fig:templates}
\vspace{-0.4cm}
\end{figure*}

\begin{figure*}[htb]
\includegraphics[scale=0.4]{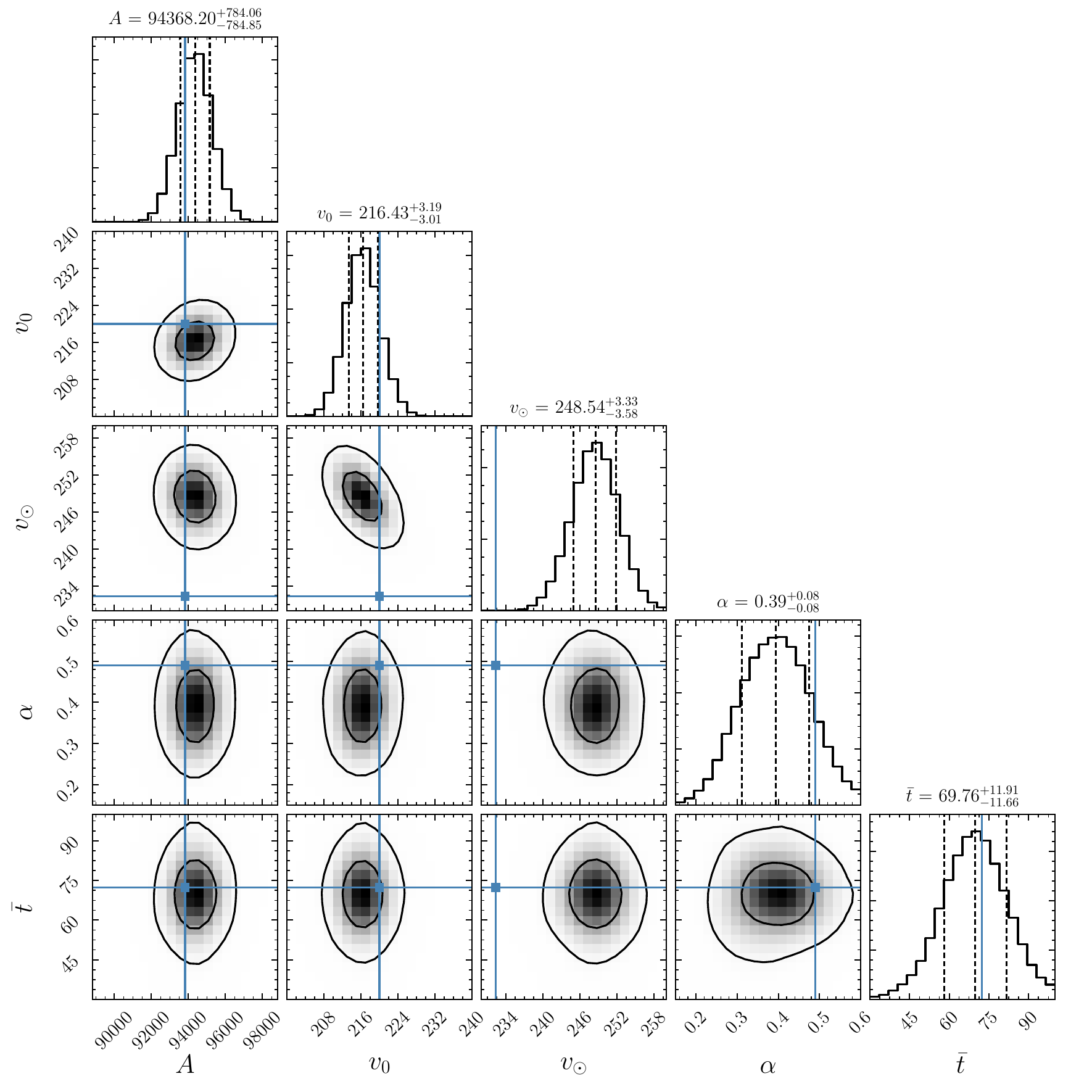}
\caption{
A Monte Carlo parameter estimation for the bulk halo parameters at the threshold of detection for annual modulation in the presence of a Sagittarius-like stream containing 5\% of the DM and with a narrow velocity dispersion of 10 km/s.
The accuracy of the parameter scan is worsened by the failure to account for the substructure in the analysis.
}
\label{fig:Parameter_Estimation_Substructure_Wrong}
\vspace{-0.4cm}
\end{figure*}

\begin{figure*}[htb]
\includegraphics[scale=0.295]{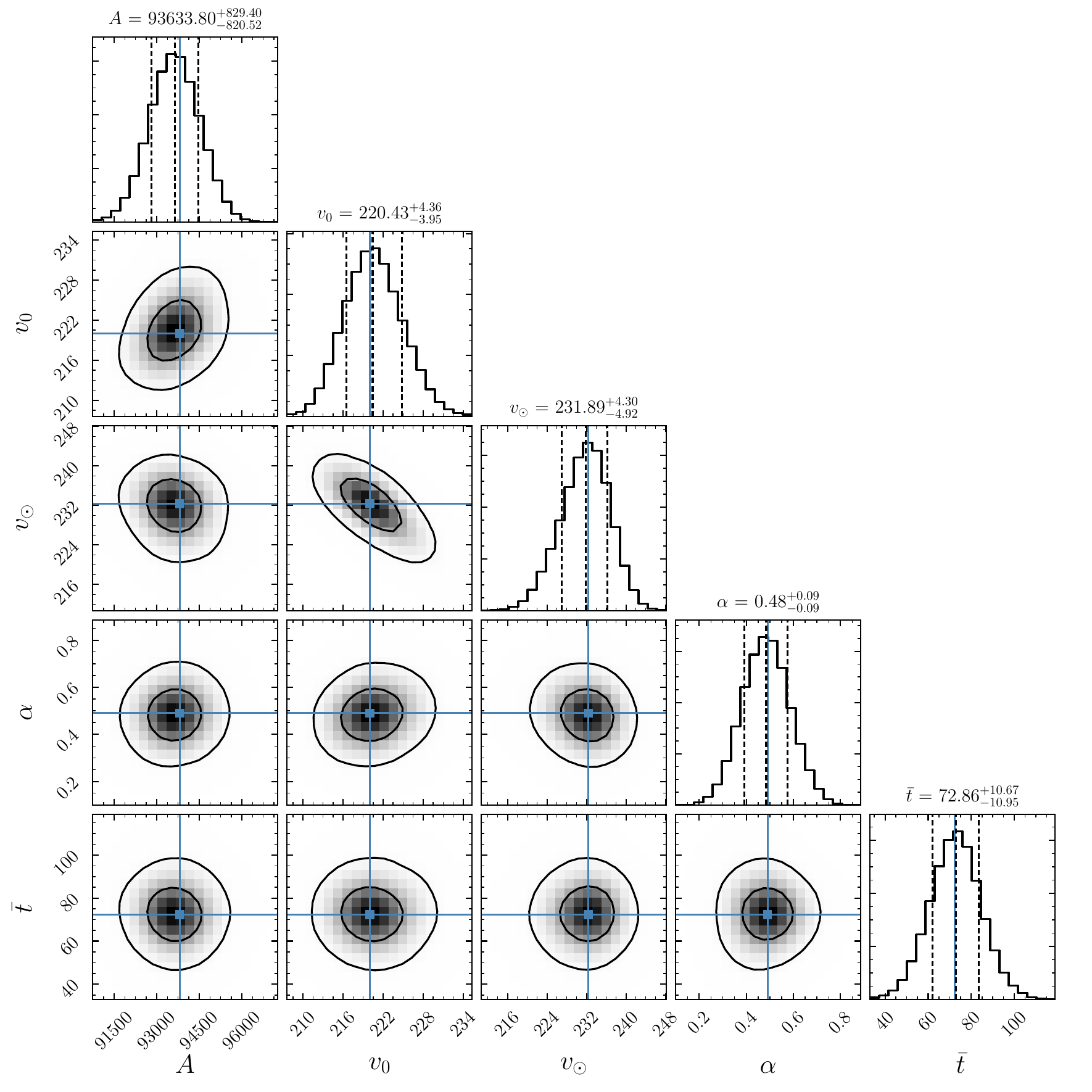}
\includegraphics[scale=0.295]{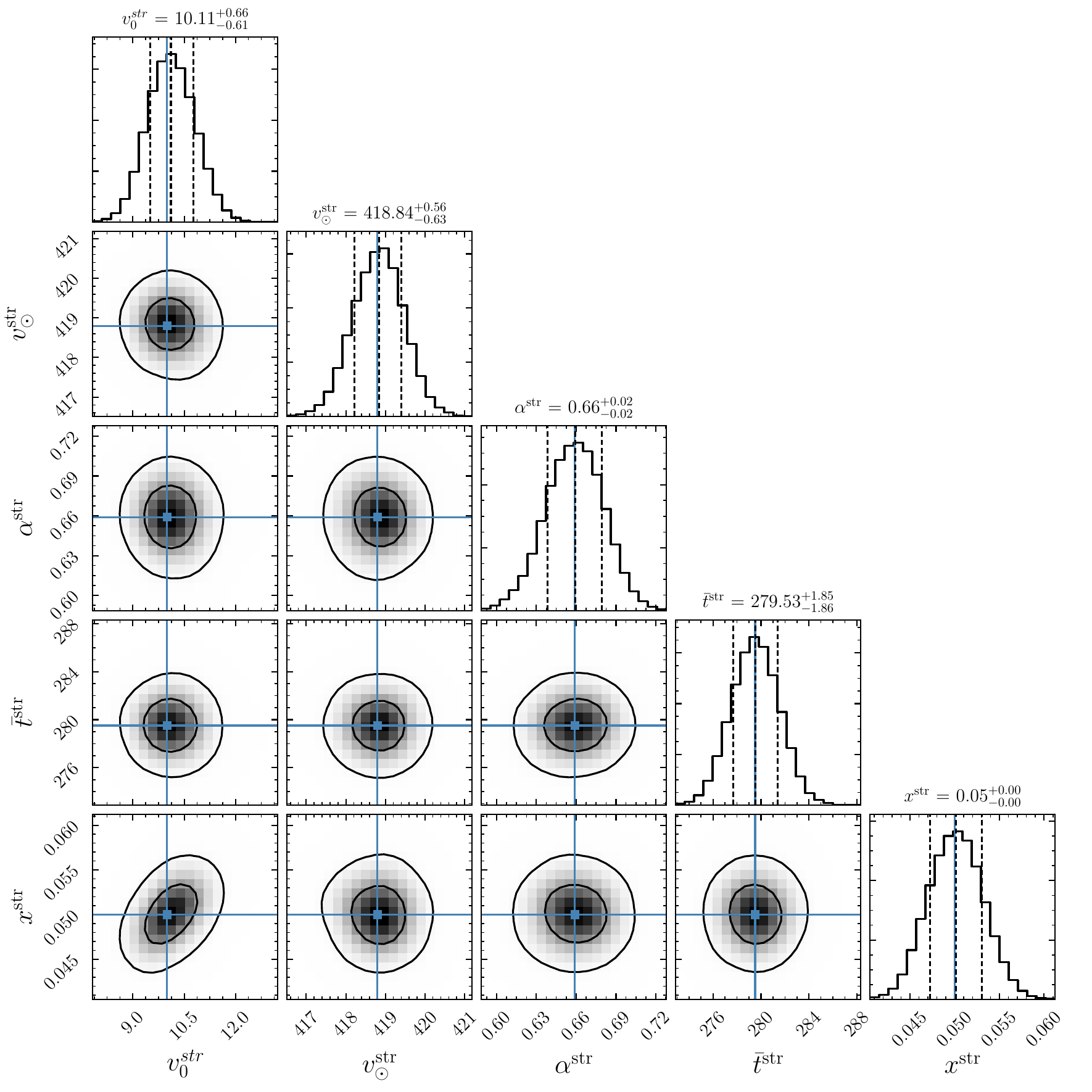}
\caption{
A simultaneous Monte Carlo parameter estimation for a signal containing a bulk halo and a Sagittarius-like stream with 5\% of the DM using identical seed parameters as Fig.~\ref{fig:Parameter_Estimation_Substructure_Wrong}. 
Scanning for the bulk halo and substructure simultaneously allows us to accurately recover the signal parameters. Left, the bulk parameter scan results, right, the stream parameter scan results.
}
\label{fig:Substructure_Correct}
\vspace{-0.4cm}
\end{figure*}

One DM stream in particular has received a significant amount of attention with regards to WIMP direct detection and that is the potential DM component  of the Sagittarius stream.
The Sagittarius stream consists of a winding stream of stars wrapping through the Milky Way that is thought to have formed from tidal stripping of the Sagittarius dwarf galaxy.
It is possible that the DM component of the Sagittarius stream contributes at the few percent level to the local DM density (see, {\it e.g.},~\cite{Vogelsberger:2008qb,Maciejewski:2010gz}).
We follow~\cite{Freese:2003tt,Savage:2006qr,OHare:2014nxd} and model the stream as a boosted Maxwellian distribution with a narrow velocity dispersion of $v_0 = 10$ km/s and a stream velocity of $v_{\rm str} = (0, 93.2, -388)$ km/s, in Galactic coordinates.
Further we assume that the Sagittarius stream constitutes 5\% of the local DM.
We will show that even though the stream may only be a small component of the local DM density, it can still leave an important signature in axion direct detection experiments, due to its small velocity dispersion.

Another possible source of DM substructure that has low velocity dispersion is a dark disk.
Co-rotating thick dark disks are found to form in certain $N$-body simulations with baryons~\cite{Read:2008fh,Bruch:2008rx,2009MNRAS.397...44R,2009ApJ...703.2275P} due to the disruption of merging satellites galaxies that are pulled into the disk.
In the simulations, the dark disks are found to be co-rotating with lag speeds and velocity dispersions both $\sim$$50$ km$/$s.
They may even dominate the local DM density~\cite{Read:2008fh,2009MNRAS.397...44R}; however, as we will see, even if the dark disk is only a small fraction of the local DM density, it can still leave a significant signature in the direct detection data due to the small velocity dispersion and lag speed.

To develop some intuition for how important substructure could be, let us take the oversimplified scenario in which the substructure of interest makes up a fraction $x$ of the local DM distribution and also follows the Maxwellian distribution with the same $v_{\rm obs}$ as in the SHM, but with a much smaller dispersion parameter $v_0^{\rm str}$.
Then we can write
\begin{equation}\begin{aligned}
f(v) = &(1-x) f_{\rm SHM}(v \vert v_0, v_{\rm obs}) \\
+ &x f_{\rm SHM}(v \vert v_0^{\rm str}, v_{\rm obs})\,.
\end{aligned}\end{equation}
Using this we can explicitly calculate the expected test statistic (in favor of the model of the SHM plus the stream over the null hypothesis of no DM) of a signal with a frequency independent background as:
\begin{widetext}
\begin{equation}\begin{aligned}
{\rm TS} = &\frac{A^2T\pi}{2m_a \lambda_B^2} \left[ (1-x)^2 \frac{{\rm erf} \left[ \sqrt{2} v_{\rm obs}/v_0 \right]}{\sqrt{2\pi} v_0 v_{\rm obs}} 
+ x^2 \frac{{\rm erf} \left[ \sqrt{2} v_{\rm obs}/{v_0^{\rm str}} \right]}{\sqrt{2\pi} {v_0^{\rm str}} v_{\rm obs}} + \frac{2 x (1-x)}{\sqrt{\pi} \left( v_0^2 + {v_0^{\rm str}}^2 \right)^{3/2} v_{\rm obs}} \right.\\
\times&\left. \left( ({v_0^{\rm str}}^2 + v_0^2)\,{\rm erf} \left[ \frac{v_{\rm obs}\sqrt{v_0^2 + {v_0^{\rm str}}^2}}{v_0 {v_0^{\rm str}}} \right]
 + ({v_0^{\rm str}}^2 - v_0^2)\,{\rm erf} \left[ 
\frac{v_{\rm obs}(v_0^2 - {v_0^{\rm str}}^2)}{v_0 {v_0^{\rm str}} \sqrt{v_0^2 + {v_0^{\rm str}}^2}}
\right] \exp \left[ - \frac{4v_{\rm obs}^2}{v_0^2 + {v_0^{\rm str}}^2} \right]
\right)
\right]\,.
\label{eq:TSstream}
\end{aligned}\end{equation}
\end{widetext}

In Fig.~\ref{fig:StreamImpact} we show this TS plotted as a function of the fraction of the DM in the stream $x$ for various values of ${v_0^{\rm str}}$, normalized to the TS when no stream is present.
The figure makes it clear that if the detector is within an ultra-cold DM stream the impact on the expected axion signal can be significant, even if the stream only makes up a small fraction of the DM.
For example, if 5\% of the local DM is in a stream with $v_0^{\rm str} \approx 0.1 $ km$/$s, then the TS in favor of the model with DM is nearly 10 times larger when the stream is modeled versus when it is not.
This emphasizes the importance of searching for cold DM substructure in addition to the SHM component.

Even though velocity substructures are not intrinsically time-dependent features, annual modulation is considerably more important for the detection of substructure, which is typically characterized by a speed dispersion less than the peak-to-peak variation of the Earth's velocity with respect to a given substructure frame.
The result is an observational signature of a given substructure feature poorly localized in frequency data collected over a year.
Therefore we need a more careful treatment than the one above, as we can only search for these features in a model framework which accounts for time-varying signals.

Under the assumption that velocity substructure can still be reasonably modeled by a boosted Maxwellian distribution, it is easily accommodated within our time-dependent model template.\footnote{Even if the velocity distribution is not Maxwellian, the relevant signal template is a straightforward generalization of that presented here for a Maxwellian.}
The direction of the stream in the ecliptic plane is specified through the parameters $\alpha^\text{sub}$ and $\bar{t}^\text{sub}$, which are defined in analogy to~\eqref{alpha_t} but where $\mathbf{v}_\odot^\text{sub} = v_\odot^\text{sub} \hat{\mathbf{v}}_\odot^\text{sub}$ is the stream boost velocity in the Solar frame.
The generalized velocity distribution, including gravitational focusing, for both the SHM and the substructure components is then given by
\begin{equation}\begin{aligned}
f = &(1-x)  f^\text{SHM} (v | v_{\odot}, \alpha, \bar{t}, v_0) \\
+ &x f^\text{sub}(v | v_{\odot}^{\rm sub}, \alpha^{\rm sub}, \bar{t}^{\rm sub}, v_0^{\rm sub})\,,
\end{aligned}\end{equation}
where the superscripts ``sub" and ``SHM" denote the generalized substructure and SHM velocity distributions, respectively, after gravitational focusing has been accounted for.
The generalization to multiple substructure components is straightforward.

The importance of annual modulation for cold substructure is illustrated in Fig.~\ref{fig:templates}, where we show, in the left panel, the mean PSD assuming the Sagittarius stream parameters taken at two different times throughout the year.
We have chosen the dates where the TS in favor of the stream is maximized, June 5, and minimized, November 23, both for 2017.
Since the stream is narrow in frequency space, the sharp peaks at these two different times of year are almost completely non-overlapping.
On the contrary, at frequencies where the stream does not contribute appreciably, annual modulation does not significantly affect the contribution from the SHM.

Just as we performed parameter estimations for the bulk halo component, we can also estimate the parameters defining the contribution of velocity substructure to the speed distribution.
It should be noted that the parameter estimation for the bulk halo component can be substantially affected by the presence of velocity substructure if the substructure is not properly accounted for.
An example of this can be seen in Fig.~\ref{fig:Parameter_Estimation_Substructure_Wrong}, where we have included a stream with Sagittarius-like parameters in the data, as given earlier, and used the Asimov dataset.
However, we have not accounted for the stream in the model that is fit to the data.  Note that the TS in favor of DM in this case is $\sim$$10^4$.
Our estimates for the SHM parameters $v_0$ and $v_\odot$ are significantly affected by the presence of the stream and disagree with the true values by multiple standard deviations.
In contrast, in Fig.~\ref{fig:Substructure_Correct} we display the posterior distribution for a fit including a Maxwellian stream.
Note that while both the SHM and the stream parameters are floated at the same time, we display the posteriors for the SHM and stream model parameters separately.
In this case both the stream and the SHM model parameters are accurately estimated.
Comparing the model that included the stream to that without, we find a TS value $\sim$400 in favor of the model with the stream over that without.\footnote{To simplify the analysis, we have neglected gravitational focusing in considering this Sagittarius-like stream.
Gravitational focusing is more important at lower speeds, and therefore is generally less relevant for such a stream than it would be in considering, for example, a dark disk.
We note, however, that if the stream is well-aligned with the ecliptic plane, it is possible to get large enhancements to the rate over short periods of time during the year~\cite{Patla:2013vza,Bertolucci:2017vgz,Zioutas:2017klh}, although such a configuration is not present for the Sagittarius stream.
}

Note that for our fiducial set of model parameters for the Sagittarius stream, we find that when the SHM is detected at $5\sigma$ significance ($\text{TS} \sim 58$), including the look elsewhere effect, the stream may barely start to become visible at $\sim$1.6$\sigma$ significance.  
We stress, however, that is possible that other, colder DM streams would contribute more substantially even if they are a smaller fraction of the local DM density.
While we illustrated the stream example for simplicity, the effects of the other types of velocity substructure may be worked out similarly.
For example, we find that with our fiducial choice of parameters for the dark disk lag speed and velocity dispersion, the dark disk would be detectable at the same significance as the SHM even if the dark disk only makes up $\sim$20\% of the local DM density. 
Moreover, the dark disk should be more affected by annual modulation and gravitational focusing than the SHM component, since the DM in the dark disk is on average slower moving in the Solar frame.
The PSD template is illustrated, assuming the dark disk makes up 20\% of the local DM density, in the right panel of Fig.~\ref{fig:templates}.
The dark disk leads to a significant increase in the PSD at low velocities, corresponding to frequencies near the axion mass.
As in the stream case, we show the PSD at two different times of year, corresponding to the date of maximal TS, November 18, and minimal TS, June 5.

\section{Conclusion}

The QCD axion, and axion like particles more generally, is a well motivated class of DM candidates, and if it constitutes the DM of our universe, then the burgeoning experimental program searching for such DM could be on the verge of a discovery.
With such possibilities it is important to be able to clearly and accurately quantify any emerging signal and set limits in their absence.
The likelihood framework we have introduced allows for exactly this.
In addition, through the use of the Asimov dataset, we have derived a number of analytic results that make quantifying these thresholds possible without recourse to Monte Carlo simulations.

In the event of an emerging signal, one would always worry about the possibility of unanticipated backgrounds.
Nevertheless DM provides its own way of addressing this concern through unique fingerprints in the frequency and time domains.
For example, we showed the form the local DM velocity distribution uniquely determines the frequency dependence of the PSD data, and that by exploiting this knowledge one is able to, through the likelihood framework, constrain properties of the local velocity distribution. 
Since the bulk of the DM halo is expected, locally, to follow a Maxwellian distribution with velocity dispersion set by the local rotation speed, correctly measuring the Maxwellian parameters will provide a non-trivial check of the nature of the signal.
In the time domain, any true signal should undergo annual modulation, including the subtle effect of gravitational focusing, and we quantified how this may be verified using the likelihood formalism.
Further, the likelihood is sensitive to the presence of local DM substructure such as cold streams, which can enhance the expected signal through an associated increase in the axion coherence time.
For example, we showed that the Sagittarius stream could leave a unique signature in the PSD data.
Nevertheless there are a great many possible types of DM substructure, beyond those considered here, that could be present at the position of the Earth, and we leave a careful study of these to future work.

Taken together the results of this work provide a set of tools that will prove useful in moving towards a possible DM axion detection, and, if we should be so lucky, into the era of axion astronomy that would follow.
Towards that end, we have provided an open-source code package at \url{https://github.com/bsafdi/AxiScan} for performing all the likelihood analyses discussed in this work and also simulating data at axion direct detection experiments for different background and signal models.

\section*{Acknowledgments}

First and foremost we would like to thank the ABRA-10cm collaboration for a number of discussions and insights about the operation of the pilot ABRACADABRA experiment.
The implementation of many of the techniques introduced in this paper will be presented within the full context of the ABRA-10cm experiment in an upcoming work~\cite{ABRA10cmReach}.
We are grateful to Yonatan Kahn and Jon Ouellet for contributions during the early stages of this project, and for a number of helpful conversations thereafter.
We would like to additionally thank Reyco Henning, Kent Irwin, Alex Millar, Siddharth Mishra-Sharma, Ciaran O'Hare, Lyman Page, Aaron Pierce, Jesse Thaler, Mark Vogelsberger, Lindley Winslow, Kevin Zhou, and Konstantin Zioutas for useful discussions related to this work.
NLR is supported by the U.S. Department of Energy under grant Contract Numbers DE-SC00012567 and DE-SC0013999.

\appendix

\section{Distribution of the Combined Signal and Background Model}\label{app:SigBkgDist}

In Sec.~\ref{sec:likelihood} of the main text we demonstrated that the signal only distribution is exponentially distributed, as given in~\eqref{eq:PSDasExponential}.
However, we simply asserted that the background only and signal plus background distributions were also exponentially distributed.
In this appendix we demonstrate both of these results.
We reiterate at the outset that in all cases the correct starting point for determining these distributions is the time-series data, which is where the different contributions are combined.
We cannot straightforwardly think about combining distributions at the level of the PSD.
To emphasize this, even though the PSD in the background and signal only cases are individually exponentially distributed, the sum of two exponentially distributed numbers is not itself exponentially distributed, and yet the PSD formed from the sum of the background and signal is.

Consider firstly the background only distribution.
Imagine we have time-series data collected in the presence of $n_B$ independent background sources, each Gaussian distributed random variables with mean zero and variance $\lambda_B^i/\Delta t$, where $i$ indexes the different backgrounds and the inclusion of $\Delta t$ in the variance is for later convenience.
Note that we can choose the backgrounds to have zero mean without loss of generality, because the mean will only impact the $k=0$ mode of the PSD, which for reasons described below we will not include in our likelihood.
In the presence of this noise, the time-series data will take the form
\begin{equation}
\Phi_n = \sum_{j=1}^{n_B} x_n^j\,,
\end{equation}
where $n = 0, 1, \ldots, N-1$ indexes the times at which the measurements were taken and the $x_n^i$ satisfy
\begin{equation}
\langle x_n^i \rangle = 0\,,\;\;\; \langle x_n^j x_m^l \rangle = \delta_{nm} \delta_{jl} \frac{\lambda_B^j}{\Delta t}\,.
\label{eq:BkgMeanVar}
\end{equation}
The second relation here follows as we assume our backgrounds are independent, and for a given background the values measured at different times are independent and identically distributed.
Moving towards the PSD, consider the discrete Fourier transform of this data:
\begin{equation}
\Phi_k = \sum_{n=0}^{N-1} \Phi_n e^{-i 2\pi k n/N} = \sum_{n=0}^{N-1} \sum_{j=1}^{n_B} x_n^j e^{-i 2\pi k n/N}\,.
\end{equation}
It is convenient to expand the exponential and analyze the real and imaginary parts of this separately.
In detail:
\begin{equation}\begin{aligned}
\Phi_k = &\sum_{n=0}^{N-1} \sum_{j=1}^{n_B} x_n^j \cos \left( \frac{2\pi k n}{N} \right) \\
- i &\sum_{n=0}^{N-1} \sum_{j=1}^{n_B} x_n^j \sin \left( \frac{2\pi k n}{N} \right) \\
\equiv & R_k + i I_k\,.
\end{aligned}\end{equation}

The real and imaginary parts, $R_k$ and $I_k$ respectively, are both Gaussian distributed since they are sums of Gaussian distributed random variables.
Accordingly they are completely specified by their means and variances, which we can determine using~\eqref{eq:BkgMeanVar}.
Consider the real part first, as the argument for the imaginary part proceeds in exactly the same fashion.
For the mean we have
\begin{equation}\begin{aligned}
\langle R_k \rangle = &\left\langle\sum_{n=0}^{N-1} \sum_{j=1}^{n_B} x_n^j \cos \left( \frac{2\pi k n}{N} \right) \right\rangle \\
=&\sum_{n=0}^{N-1} \sum_{j=1}^{n_B} \langle x_n^j \rangle \cos \left( \frac{2\pi k n}{N} \right) \\
= &0\,.
\end{aligned}\end{equation}
Similarly
\begin{equation}\begin{aligned}
\langle R_k^2 \rangle &= \sum_{j=1}^{n_B} \frac{\lambda_B^j}{\Delta t} \sum_{n=0}^{N-1} \cos^2 \left( \frac{2\pi k n}{N} \right) \\
&= \frac{\lambda_B}{\Delta t} \sum_{n=0}^{N-1} \cos^2 \left( \frac{2\pi k n}{N} \right) \,.
\end{aligned}\end{equation}
where we used $\lambda_B \equiv \sum_j \lambda_B^j$ following~\eqref{eq:BkgExpMean}.
We can evaluate the remaining sum using\footnote{Note that if $N$ is even, then for the $k=N/2$ mode the sum evaluates to the $k=0$ result.
This extends to~\eqref{eq:A8} and~\eqref{eq:A9}, and indeed when propagated through to the likelihood, implies that this mode will also be gamma and not exponentially distributed.
}
\begin{equation}
\sum_{n=0}^{N-1} \cos^2 \left( \frac{2\pi k n}{N} \right) = \left\{ \begin{array}{ll} N & k = 0 \\
N/2 & 0<k<N \end{array} \right. \,.
\end{equation}
Putting these together, we conclude the real part has a variance given by
\begin{equation}
\langle R_k^2 \rangle = \left\{ \begin{array}{ll} \frac{\lambda_B N}{\Delta t} & k = 0 \\
\frac{\lambda_B N}{2\Delta t} & 0<k<N \end{array} \right. \,.
\label{eq:A8}
\end{equation}
The argument for the imaginary part is almost identical, and we find again that $\langle I_k \rangle = 0$, whilst
\begin{equation}
\langle I_k^2 \rangle = \left\{ \begin{array}{ll} 0 & k = 0 \\
\frac{\lambda_B N}{2\Delta t} & 0<k<N \end{array} \right. \,.
\label{eq:A9}
\end{equation}

Knowing how contributions to the Fourier transform are distributed, we now move to the PSD, which will again be a random variable given by:
\begin{equation}
S_{\Phi \Phi}^k = \frac{\left( \Delta t \right)^2}{T} \left| \Phi_k \right|^2
= \frac{\Delta t}{N} \left( R_k^2 + I_k^2 \right)\,.
\end{equation}
There are many ways to determine the probability density function (pdf) obeyed by $S_{\Phi \Phi}^k$.
A particularly straightforward one in this case is to start by determining the cumulative distribution function (cdf), $F[S_{\Phi \Phi}^k]$.
We will do this for $N>k>0$ first, and return to the $k=0$ case afterwards.
To obtain the cdf, we simply integrate the distributions for $R_k$ and $I_k$ over all values up to some $S_{\Phi \Phi}^k$.  In detail,
\begin{equation}\begin{aligned}
F[S_{\Phi \Phi}^k] = &\int^{S_{\Phi \Phi}^k} dR_k dI_k\,\frac{\Delta t}{\pi \lambda_B N} \\
\times &\exp \left[ - \frac{\Delta t}{\lambda_B N} \left( R_k^2 + I_k^2 \right) \right]\,.
\end{aligned}\end{equation}
To perform this integral it is convenient to change to polar coordinates, $u^2 = R_k^2 + I_k^2$ and $\theta$, so that
\begin{equation}\begin{aligned}
F[S_{\Phi \Phi}^k] = &\int_0^{\sqrt{N S_{\Phi \Phi}^k/\Delta t}} du\,\frac{2\Delta t u}{\lambda_B N} \exp \left[ - \frac{\Delta t u^2}{\lambda_B N} \right] \\
= & 1 - e^{-S_{\Phi \Phi}^k/\lambda_B} \,.
\end{aligned}\end{equation}
The pdf is just the derivative of this, so we find
\begin{equation}\begin{aligned}
P[S_{\Phi \Phi}^k] = \frac{1}{\lambda_B} e^{-S_{\Phi \Phi}^k/\lambda_B}\,,
\end{aligned}\end{equation}
demonstrating that for $0 < k < N$ the background is exponentially distributed as claimed in the main body.

Consider next the case for $k=0$.
Utilizing an identical approach, we find firstly that
\begin{align}
F[S_{\Phi \Phi}^0] =\,&\int_{-\sqrt{N S_{\Phi \Phi}^k/\Delta t}}^{\sqrt{N S_{\Phi \Phi}^k/\Delta t}} \sqrt{\frac{\Delta t}{\pi \lambda_B N}} \exp \left[ - \frac{\Delta t}{\lambda_B N} R_0^2 \right] \nonumber \\
=\,& {\rm erf} \left[ \sqrt{S_{\Phi \Phi}^0 /\lambda_B} \right]\,,
\end{align}
implying
\begin{equation}
P[S_{\Phi \Phi}^0] = \frac{1}{\sqrt{\pi \lambda_B S_{\Phi \Phi}^0}} e^{-S_{\Phi \Phi}^0/\lambda_B}\,.
\end{equation}
Clearly the $k=0$ mode is not exponentially distributed: it is in fact gamma distributed with shape parameter $1/2$ and scale parameter $\lambda_B$.
In practice, however, this mode does not contribute to the likelihood function in~\eqref{eq:AxionLikelihood} since all of the axions we search for have finite mass and thus finite oscillation frequency.
Moreover, the $k=0$ mode is degenerate with the mean background values that we have chosen to neglect.

Finally we want to show that the combined signal and background dataset is also exponentially distributed for $0 < k < N-1$.
We will show this in a somewhat indirect manner.
Firstly, given that the signal is exponentially distributed, as shown in the main text, we will show that the real and imaginary parts of the discrete Fourier transform of such a dataset must be normally distributed.
Then we can combine the signal in as if it was just another background in the argument presented above, and it will follow immediately that the full distribution must be exponential.
Our starting point is~\eqref{eq:PSDasExponential}, where we showed the signal only PSD is exponentially distributed.
We repeat this result here for convenience:
\begin{equation}\begin{aligned}
P[S_{\Phi \Phi}^k] &= \frac{1}{\lambda_k} e^{-S_{\Phi \Phi}^k/\lambda_k}\,,\\
\lambda &\equiv A \left. \frac{\pi f(v)}{m_a v} \right|_{v=\sqrt{4\pi k/(m_a T)-2}}\,.
\end{aligned}\end{equation}
As an intermediate step, consider $S_{\Phi \Phi}^k = x + y$, where $x = (\Delta t/N) R_k^2$ and $y = (\Delta t/N) I_k^2$.
As the real and imaginary parts are independent and identically distributed for the signal dataset, then so too are $x$ and $y$, and we denote their pdf by $g$.
Given that $x,y \geq 0$, we can relate their distributions to that of the signal PSD via
\begin{equation}\begin{aligned}
P[S_{\Phi \Phi}^k] = &\int_0^{\infty} dx dy\,g[x] g[y] \delta(S_{\Phi \Phi}^k-x-y) \\
= &\int_0^{S_{\Phi \Phi}^k} dx\,g[x] g[S_{\Phi \Phi}^k-x]\,.
\end{aligned}\end{equation}
To solve this equation for $g$ we take the Laplace transform, denoting transformed quantities with a tilde.  This yields
\begin{equation}\begin{aligned}
\tilde{g}[\tilde{x}] = \frac{1}{\sqrt{1+\tilde{x}\lambda_k}}\,,
\end{aligned}\end{equation}
which when inverted becomes
\begin{equation}\begin{aligned}
g[x] = \frac{1}{\sqrt{\pi \lambda_k x}} e^{-x/\lambda_k}\,.
\end{aligned}\end{equation}
From here, to derive the pdf for $R_k$ we can change variables using $x = (\Delta t/N) R_k^2$.
In doing so we need to account for the Jacobian and also the fact that whilst $x \in [0,\infty)$, this is only half the domain of possible $R_k$ values.
Doing so we find
\begin{equation}\begin{aligned}
P[R_k] = \frac{1}{\sqrt{\pi N\lambda_k/\Delta t}} \exp \left[ - \frac{R_k^2}{N \lambda_k / \Delta t} \right]\,,
\end{aligned}\end{equation}
which is exactly a normal distribution with mean zero and variance $N \lambda_k / (2 \Delta t)$.
The distribution for $I_k$ will be identical, and thus we find the signal is distributed just like a single background but with $\lambda_B^j \to \lambda_k$.
If we then repeat the background only argument shown at the start of this appendix with the signal contribution added, we will find the full PSD is again exponentially distributed with mean $\lambda_k + \lambda_B$, completing the required derivation.

\section{Comparison to a Bandwidth Average}\label{app:BandAv}

An alternative analysis strategy to that presented in the main text is to take the average PSD (or power) measured across a given bandwidth range and compare that directly to the average model prediction.
This should be contrasted with taking the product of exponential likelihoods across $k$ modes as we introduced in~\eqref{eq:AxionLikelihood}, and at face value it should have less discriminating power as the information regarding how the axion signal is distributed within the bandwidth has been lost.
In this section we quantify this statement by deriving the expected sensitivity of such an approach.
As a side point we will also demonstrate how to derive the optimum bandwidth range in performing a bandwidth averaged search.

To begin with, we note that in each frequency bin the PSD formed from the data will still be exponentially distributed.
Then, if we are searching in some bandwidth range $\Omega_{\omega}$, which contains $n_{\omega}$ frequency bins, the mean PSD can be formed from a sum of these exponentials and will thus be Erlang distributed.
In detail, the likelihood will have the form
\begin{equation}
\mathcal{L}(d \vert \boldsymbol{\theta}) = \frac{n_{\omega}^{n_{\omega}}}{(n_{\omega}-1)!} \frac{\left( \bar{S}_{\Phi\Phi} \right)^{n_{\omega}-1}}{\bar{\lambda}^{n_{\omega}}} e^{- n_{\omega} \bar{S}_{\Phi\Phi}/\bar{\lambda}}\,,
\label{eq:AvPSDLikelihood}
\end{equation}
where we have defined:
\begin{equation}
\bar{S}_{\Phi\Phi} = \frac{1}{n_f} \sum_{k \in \Omega_{\omega}} S_{\Phi\Phi}^k\,,
\end{equation}
similarly to what we had when discussing the stacked data procedure in Sec.~\ref{sec:Stack}.
In the above equation we also introduced the mean model prediction, which assuming we have a frequency independent background will be given by
\begin{equation}\begin{aligned}
\bar{\lambda} = &\bar{\lambda}_S + \lambda_B\,, \\
\bar{\lambda}_S \equiv & \frac{1}{n_{\omega}} \sum_{k \in \Omega_{\omega}} A \frac{\pi f(v)}{m_a v}\,.
\label{eq:avsigPSD}
\end{aligned}\end{equation}

Consider the average signal prediction.
This average is taken over some frequency range, or bandwidth, which we denote by $\Delta \omega$, and is equivalent to a range in velocities, $v \in [0,v_{\rm max}]$.\footnote{In principle the lower velocity could be $v_{\rm min}$ rather than 0, and this value can also be optimized for.
Nevertheless as the signal distribution rises sharply from $v=0$, approximating $v_{\rm min}=0$ is sufficient for the argument in this appendix.}
Consequently we have
\begin{equation}
\Delta \omega = \frac{1}{2} m_a v^2_{\rm max}\,.
\end{equation}
The bandwidth can also be written as $\Delta \omega = n_{\omega} d\omega$, where $d\omega$ is the width of an individual frequency bin.
Assuming sufficient run time, as $d\omega = 2\pi/T$, then we can also write
\begin{equation}
\Delta \omega = n_{\omega} m_a v dv\,.
\end{equation}
Taken together, these show that
\begin{equation}
\frac{\Delta \omega}{\Delta \omega} = \frac{2 n_{\omega}}{v_{\rm max}^2} v dv\,.
\end{equation}
Substituting this into~\eqref{eq:avsigPSD}, we can rewrite the signal prediction as
\begin{equation}
\bar{\lambda}_S = \frac{2A \pi}{m_a v_{\rm max}^2} \int_0^{v_{\rm max}} dv\, f(v)\,.
\label{eq:avsigPSDsimp}
\end{equation}

To estimate the sensitivity it is most convenient to return to $\Theta$ as introduced in~\eqref{eq:Thetadef}.
This is modified for the averaged PSD likelihood given in~\eqref{eq:AvPSDLikelihood} to
\begin{equation}
\Theta(A) = 2 n_{\omega} \bar{S}_{\Phi\Phi} \left[ \frac{1}{\lambda_B} - \frac{1}{\bar{\lambda}} \right] - 2 n_{\omega}\ln \frac{\bar{\lambda}}{\lambda_B}\,,
\end{equation}
where as in Sec.~\ref{sec:sensitivity}, we suppress the axion mass dependence.
As in the main body, to analytically estimate the sensitivity we can use the Asimov dataset.
Here we denote this by $\bar{\lambda}_S^t + \lambda_B$, where $\bar{\lambda}_S^t$ is identical to~\eqref{eq:avsigPSDsimp}, but with the signal strength replaced by its true value: $A \to A_t$.
To simplify the resulting form of $\tilde{\Theta}$, we again assume that we are in the limit where the true and modeled average signal strength are subdominant to the average background, such that we obtain
\begin{equation}
\tilde{\Theta}(A) = \frac{2A T \pi}{m_a \lambda_B^2} \left( A_t - \frac{A}{2} \right) \left( \int_0^{v_{\rm max}} dv\, \frac{f(v)}{v_{\rm max}} \right)^2\,.
\end{equation}

To compare this directly to results obtained from the analysis in the main body, we need to determine a value for $v_{\rm max}$.
A procedure for doing so is to choose the $v_{\rm max}$ that maximizes the significance of any emerging signal, or in detail one that maximizes the test statistic of discovery.
Using TS as defined in~\eqref{eq:TSdisc}, for the present case we have
\begin{equation}
\widetilde{\rm TS} = \frac{A^2_t T \pi}{m_a \lambda_B^2} \left( \int_0^{v_{\rm max}} dv\, \frac{f(v)}{v_{\rm max}} \right)^2\,,
\end{equation}
which we want to maximize as a function of $v_{\rm max}$.
The value that does so depends critically on the form of $f(v)$, and so needs to be re-evaluated for each assumption.
For example, if we take the simple SHM ansatz as per~\eqref{eq:SHM}, then we find $v_{\rm max} \approx 453$ km/s.
Using this value we can then construct the ratio between the TS using our default bin-by-bin approach, denoted ${\rm TS}^{\rm full}$, to that obtained here, denoted ${\rm TS}^{\rm av.}$, which is explicitly:
\begin{align}
\frac{\widetilde{\rm TS}^{\rm full}}{\widetilde{\rm TS}^{\rm av.}} = &\left( \frac{1}{2} \int dv \frac{f(v)^2}{v} \right) \left( \int_0^{v_{\rm max}} dv\, \frac{f(v)}{v_{\rm max}} \right)^{-2} \nonumber\\
\approx &1.14\,,
\end{align}
where in the final step we again assumed a default SHM form for the speed distribution.
Thus as claimed at the outset, even when optimized, this averaging procedure is not as sensitive as our full construction.
The optimization is important; if we had instead taken $v_{\rm max} = 300$ (600) km/s, we would have obtained a ratio of 1.87 (1.43) above.
Further in the presence of substructure, the averaging approach suffers even further.
As a simple estimate of this if we took Maxwellian substructure, with the much smaller velocity dispersion $v_0 = 10$ km/s but the same boost velocity as the SHM, then even at the maximum the ratio is 5.42.

Using this maximum we can also determine the impact on limits.
Recalling the definition of the test statistic for upper limits in~\eqref{eq:TSlim}, we find the condition for a 95\% limit is determined when
\begin{equation}
\tilde{A}_{95\%} = A_t + \sqrt{2.71 \frac{m_a \lambda_B^2}{T \pi}} \left(\int_0^{v_{\rm max}} dv\, \frac{f(v)}{v_{\rm max}} \right)^{-1}\,.
\end{equation}
To compare this to case discussed in the main body, we take the simplifying values of $A_t = 0$ and again the default SHM speed distribution.
Doing so we find
\begin{align}
\frac{\tilde{A}_{95\%}^{\rm full}}{\tilde{A}_{95\%}^{\rm av.}} = &\left(\int_0^{v_{\rm max}} dv\, \frac{f(v)}{v_{\rm max}} \right) \left( \frac{1}{2} \int dv\,\frac{f(v)^2}{v} \right)^{-1/2} \nonumber\\
\approx & 0.94\,,
\end{align}
which corresponds to a ratio of the axion electromagnetic couplings of 0.97 ($A \propto g_{a \gamma \gamma}^2$).
This value shows that the full framework sets similar, but slightly stronger, constraints.

Accordingly, in all cases the framework described in the main body outperforms the averaged-power technique described in this appendix.
For the case of the SHM, when that technique is optimized the improvements are marginal.
Nevertheless in the presence of substructure, or if the optimal signal window is not chosen, then the gain from resolving the individual frequency bins can be much more substantial. 
Moreover, it is very difficult to constrain aspects of the DM phase-space distribution with the power-averaged technique, since the frequency dependence of the signal is not resolved.

\section{Verifying the Asimov Derivation of Upper Limit Bands}\label{app:MC_Expectations}

Using the Asimov dataset analysis, in Sec.~\ref{sec:95limits} we were able to calculate the expected 95\% limit on the signal strength $A$ at a given $m_a$.
We were also able to calculate the 1 and 2$\sigma$ containment bands around the expected 95\% limit without recourse to Monte Carlo simulations.
In this appendix we confirm that these results, presented in~\eqref{eq:A95} and~\eqref{eq:CLSB}, match those derived using Monte Carlo methods.

\begin{figure}[t]
\includegraphics[scale=0.41]{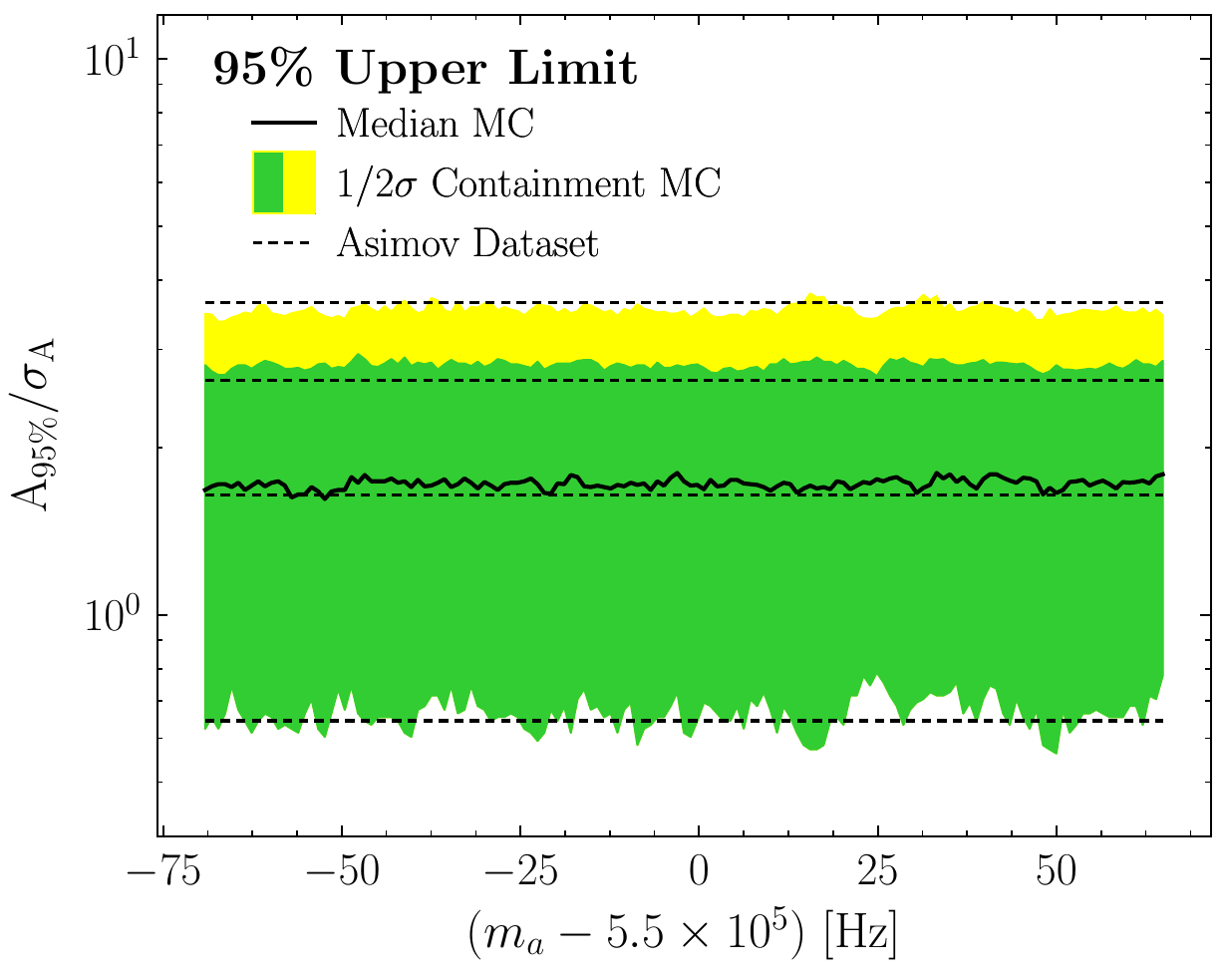} 
\caption{
A comparison between the variation in the 95\% upper limit found in Monte Carlo (MC) simulations to that derived analytically with the Asimov dataset.
As shown the two are in good agreement.
}
\label{fig:MC_Constraints}
\vspace{-0.4cm}
\end{figure}

For this procedure, we generate 1000 background-only datasets over frequencies in a $22$Hz window centered at $550$kHz and then scan these PSDs for a bulk SHM model. 
According to our estimate in~\eqref{eq:Nmaapprox}, we expect there to be approximately $55$ independent mass points for which we can scan contained within this frequency data.
However, for the sake of precision, we will arbitrarily increase our resolution to scan over $150$ mass points, between which there may be some degeneracy. 
At each mass point, we scan over $A$ values between $-5\sigma_A$ and $10 \sigma_A$ calculated according to~\eqref{eq:SigmaA}.
We emphasize again that it is necessary that we allow $A$ to take on negative values despite that, by its definition, $A$ must be nonnegative. 
In practice, this is resolved by imposing a power-constrained limit such that constraints on $A$ are placed no lower than $1\sigma$ below the expected constraint as calculated by~\eqref{eq:CLSB}.
In Fig.~\ref{fig:MC_Constraints} we show the median 95\% upper limit as well as the 1 (shaded green) and 2$\sigma$ (shaded yellow) containment intervals constructed from the ensemble of Monte Carlo simulations.
Note that we only show the upper 2$\sigma$ region, since we anticipate neglecting fluctuations below 1$\sigma$ with the power-constrained method.
Additionally,  we indicate the same quantities predicted by our Asimov analysis with dashed lines.
As the figure demonstrates, the Monte Carlo and Asimov results are generally in good agreement.\footnote{While there may be a small systematic offset, as visible in Fig.~\ref{fig:MC_Constraints}, the agreement is likely satisfactory for use at direct detection experiments.
However, if required the containment intervals could be further tuned to agree with Monte Carlo simulations like those presented here.}

\section{Asymptotic Distribution for the Discovery Test Statistic}\label{app:asymTS}

In this appendix we will explicitly calculate, from our likelihood, the survival function for the local ${\rm TS}$ under the null hypothesis.
We will then show that asymptotically the ${\rm TS}$ is $\chi^2$-distributed, and therefore there is a simple connection with the significance, $Z$, given by $Z = \sqrt{\rm TS}$.
Doing so will verify~\eqref{eq:survival}, presented in Sec.~\ref{sec:discoveryreach}.
Note that this appendix is in many ways an explicit illustration of Wilks' theorem.

\begin{figure}[t]
\includegraphics[scale=0.38]{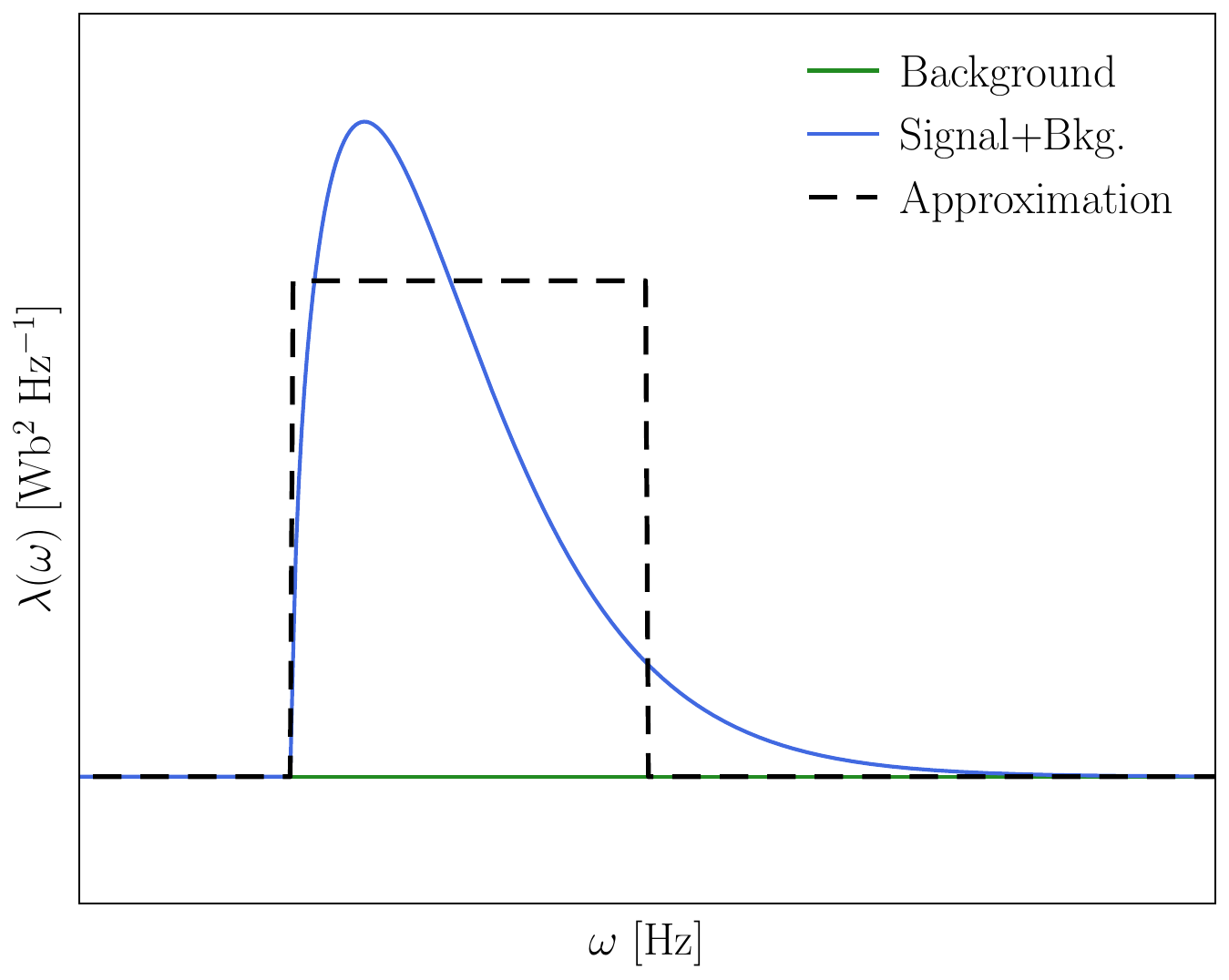} 
\caption{
Schematic depiction of the approximation made to the model used to derive ${\rm TS}_{\rm thresh}$.
Specifically we assume that the signal model is non-zero only within a finite frequency range, and equal to the background outside this, and within this range the combined signal and background is flat.
}
\label{fig:BoxApprox}
\vspace{-0.4cm}
\end{figure}

To begin with, the situation to keep in mind is that we have a dataset that is drawn from the background only distribution, where in some frequency range there is an upward fluctuation that can be well described by a model including the signal.
From this picture, in order to derive our result we will make two simplifying assumptions:
\begin{enumerate}
\item that the signal model we use is only non-zero in a set of $n_S$ frequency bins, the set of which we denote $\Omega_S$, and outside this $\lambda_k = \lambda_B$; and
\item that in these $n_S$ bins the background and model predictions are both frequency independent, so to avoid confusion we denote our signal prediction in this range as the $k$-independent $\lambda_S$.
\end{enumerate}
Taken together these assumptions imply we are approximating our model for this upward fluctuation in the background as a step function, similar to what is shown in Fig.~\ref{fig:BoxApprox}.
In that figure, which is intended to be schematic, we have shown a flat background model, and added on top of this the signal distribution as expected from~\eqref{eq:PSDasExponential}, and also shown the shape of the full model approximation we will use.
Note that nothing in our first approximation or the derivation below requires $n_S \ll N$, however for this approximation to be realistic this will usually be the case.

Our aim now is to determine how the discovery test statistic is distributed under these assumptions.
Combining these assumptions with the form of $\Theta$ given in~\eqref{eq:Thetadef}, and then choosing the $A$ that maximizes this quantity, we arrive at:
\begin{equation}
\widetilde{\rm TS} = \left\{
\begin{array}{lc}
2 n_S \left[ \frac{\overbar{S}_{\Phi\Phi}}{\lambda_B} - 1 - \ln \frac{\overbar{S}_{\Phi\Phi}}{\lambda_B} \right] & \overbar{S}_{\Phi\Phi} > \lambda_B\,, \\
0 & \overbar{S}_{\Phi\Phi} \leq \lambda_B\,,
\end{array}
\right.
\label{eq:TSmaxapprox}
\end{equation}
where we have defined the average data PSD in this range:
\begin{equation}
\overbar{S}_{\Phi\Phi} \equiv \frac{1}{n_S} \sum_{k \in \Omega_S} S^k_{\Phi\Phi}\,.
\label{eq:Sav}
\end{equation}
Note that this should be distinguished from the subinterval averaged PSD in~\eqref{eq:StackedPSD}. 
Note also as written this result is independent of $m_a$, so we have suppressed the dependence on the mass.

Now recall that as each of our PSD measurements are exponentially distributed, the average PSD, $\overbar{S}_{\Phi\Phi}$, will follow an Erlang distribution.
In detail, we have
\begin{equation}
P[\overbar{S}_{\Phi\Phi}] = \frac{n_S^{n_S}}{(n_S-1)!} \frac{\left(\overbar{S}_{\Phi\Phi}\right)^{n_S-1}}{\lambda_B^{n_S}} e^{-n_S\overbar{S}_{\Phi\Phi}/\lambda_B}\,.
\label{eq:AvPSD}
\end{equation}
We emphasize again that we are taking the data to follow the background distribution, as in calculating ${\rm TS}_{\rm thresh}$ we are interested in the distribution of the discovery test statistic under the null hypothesis.
This explains why the mean in the above distribution is simply $\lambda_B$.

Now we want to use this to derive the distribution for $\widetilde{\rm TS}$.
Before doing so, we need to take a brief aside.
Observe that the distribution for the average PSD given in~\eqref{eq:AvPSD} is correctly normalized for $\overbar{S}_{\Phi\Phi} \in \left[ 0, \infty \right)$.
Nevertheless, from~\eqref{eq:TSmaxapprox}, we see that we only get a non-zero test statistic for $\overbar{S}_{\Phi\Phi} > \lambda_B$, thus in the probability distribution for $\widetilde{\rm TS}$ there will be a pileup of probability at zero accounting for the fact that any time the average PSD is less than the background value, the maximum discovery test statistic will be zero.
We can determine the probability of that occurring as:
\begin{equation}
\int_0^{\lambda_B} d\overbar{S}_{\Phi\Phi}\,P[\overbar{S}_{\Phi\Phi}] = 1 - \frac{\Gamma(n_S,n_S)}{(n_S-1)!}\,,
\label{eq:TSzero}
\end{equation}
where $\Gamma(n_S,n_S)$ is the upper incomplete gamma function.
Keeping this additional probability in mind, we determine the distribution for $\widetilde{\rm TS}$ from our distribution for $\overbar{S}_{\Phi\Phi}$ via a change of variables.
As an intermediate step, observe that we can invert that equation for $\overbar{S}_{\Phi\Phi}$ in terms of TS using
\begin{equation}
\overbar{S}_{\Phi\Phi} = - \lambda_B W_{-1} \left( - \exp \left[-1-\frac{\widetilde{\rm TS}}{2n_S} \right] \right)\,,
\end{equation}
where $W_{-1}$ is the lower branch of the Lambert $W$ function.
This function provides an inverse to equations of the form $y=x e^x$, such that $x = W(y)$.
As $W$ is multivalued, we choose the lower branch $W_{-1}$, where $W < -1$, which implies that $\overbar{S}_{\Phi\Phi} \geq \lambda_B$.
This shows that the change of variables will not cover the situation where the average PSD is less than the background, which we account for using the result of~\eqref{eq:TSzero}.
Using this change of variables, we then arrive at
\begin{align}
P[\widetilde{\rm TS}] &= \frac{n_S^{n_S}}{2n_S!} \frac{w^{n_S}e^{-n_S w}}{w-1} + \left[ 1 - \frac{\Gamma(n_S,n_S)}{(n_S-1)!} \right] \delta(\widetilde{\rm TS})\,, \nonumber \\
w &\equiv - W_{-1} \left( - \exp \left[-1-\frac{\widetilde{\rm TS}}{2n_S} \right] \right)\,.
\end{align}

At this stage we can move to the asymptotic form of this result.
To invoke Wilk's theorem, we need to take the large sample size limit.
Here this is controlled by $n_S$, and so we take $n_S \to \infty$, and in particular $n_S \gg {\rm TS}$.
Taking these limits and keeping just the leading term, we obtain
\begin{equation}\begin{aligned}
P[\widetilde{\rm TS}] &= \frac{e^{-\widetilde{\rm TS}/2}}{\sqrt{8 \pi\,\widetilde{\rm TS}}} + \frac{1}{2} \delta(\widetilde{\rm TS}) \,.
\label{eq:ApproxDiscTS}
\end{aligned}\end{equation}

This equation represents the asymptotic form of the discovery test statistic distribution under the background only hypothesis.
We can now directly integrate this distribution to get the survival function, in detail to find the probability of a background fluctuation yielding a test statistic greater than some value:
\begin{equation}\begin{aligned}
S[\widetilde{\rm TS}] \equiv \int_{\widetilde{\rm TS}}^{\infty} d \widetilde{\rm TS}'\,P[\widetilde{\rm TS}'] 
= &\frac{1}{2} {\rm erfc} \left( \sqrt{\frac{\widetilde{\rm TS}}{2}} \right) \\
= &1 - \Phi \left(\sqrt{\widetilde{\rm TS}} \right)\,,
\label{eq:survivalApp}
\end{aligned}\end{equation}
where ${\rm erfc}$ is the complementary error function and again $\Phi$ is a zero mean, unit variance Gaussian.
This result verifies~\eqref{eq:survival}.

\section{Sensitivity Scaling for $T < \tau$}\label{app:RootTScaling}

The main results from the Asimov dataset analysis performed in Sec.~\ref{sec:sensitivity} demonstrated that our sensitivity increased with collection time as $T^{1/4}$, which is manifest in both~\eqref{eq:g95} and~\eqref{eq:gthresh}.
Nevertheless in deriving both of these results, we assumed that $T$ was large enough that frequency bins fully resolved variations in the signal; explicitly, we assumed that $T \gg \tau$, where $\tau$ represents the coherence time of the signal.
This assumption was used in~\eqref{eq:AsimovThetaInt} so that we could rewrite the sum over frequency modes as an integral.
As commented in Sec.~\ref{sec:SN1}, we would expect that for $T < \tau$ the sensitivity should instead scale as $T^{1/2}$~\cite{Budker:2013hfa}.
In this appendix we repeat our analysis, now assuming the collection time is less than the coherence time, and demonstrate we recover this scaling also.

To do so, we start with $\Theta$, from which we can derive 95\% limits and the TS of an excess, as described in Sec.~\ref{sec:sensitivity}.
In particular, we begin with~\eqref{eq:AsimovThetaSum} which is the furthest we advanced in the Asimov analysis of $\Theta$ before invoking the assumption of $T \gg \tau$.
Repeating that result for convenience, we have
\begin{equation}
\tilde{\Theta}(A) = 2 \sum_{k=1}^{N-1} \left[ \lambda_k^t \left( \frac{1}{\lambda_B} - \frac{1}{\lambda_k} \right) - \ln \frac{\lambda_k}{\lambda_B} \right]\,,
\end{equation}
where again $\lambda_k^t$ is the expected signal plus background, but with the signal set to its true value.

In the case where $T < \tau$, where we cannot resolve the signal, we can approximate it as being confined to a single $k$ mode, say $k=k_S$.
We are effectively approximating $T \ll \tau$ here, much as we did $T \gg \tau$ in the main body, simply to expose the scaling.
This allows us to rewrite the above as
\begin{equation}
\tilde{\Theta}(A) = 2 \left[ \lambda_{k_S}^t \left( \frac{1}{\lambda_B} - \frac{1}{\lambda_{k_S}} \right) - \ln \frac{\lambda_{k_S}}{\lambda_B} \right]\,,
\end{equation}
as for all other modes $\lambda_k^t = \lambda_k = \lambda_B$, and so the contributions vanish.
As in the main body, if we again consider the case of an emerging signal, then we can assume that $A \pi f(v) / (m_a v) \sim A_t \pi f(v) / (m_a v) \ll \lambda_B$, which to lowest order simplifies our result as
\begin{equation}
\tilde{\Theta}(A) = 2A(A_t - A) \left( \frac{\pi f(v)}{m_a v \lambda_B} \right)^2\,.
\end{equation}
Note the velocity appearing in this result is fixed by the value of $k_S$.

By relating the collection time to the width of our frequency bins and hence velocity, we have again that $1/T = m_a v \Delta v/(2\pi)$, where recall $\Delta v$ is the width with which we can probe in velocity space.
Accordingly we arrive at
\begin{equation}
\tilde{\Theta}(A) = \frac{1}{2} T^2 A (A_t - A) \left( \frac{f(v) \Delta v}{\lambda_B} \right)^2\,.
\end{equation}
Importantly, note that as $f(v)$ is a normalized pdf and $\Delta v$ is roughly the range over which it varies, we have $f(v) \Delta v \sim \mathcal{O}(1)$.
The exact numerical value is irrelevant: the key observation is that the combination is no longer dependent on $T$.
As such we see in this limit $\tilde{\Theta} \propto T^2$, which should be contrasted with~\eqref{eq:Thetavelint}, where $\tilde{\Theta} \propto T$.
As $A \propto g_{a\gamma \gamma}^2$, when we use $\tilde{\Theta}$ to derive the TS or 95\% limit as we did in the main body we will find they both scale as $T^{-1/2}$, as expected.

\bibliography{AxionLikelihood}

\end{document}